\documentclass[11pt]{article}
\usepackage{amsmath, amssymb, amsfonts, amsthm}
\usepackage{epsfig}

\newtheorem{theorem}{Theorem}[section]
\newtheorem{proposition}[theorem]{Proposition}

\newtheorem{lemma}[theorem]{Lemma}

\newcommand{\tri}{| \! | \! |}
\newcommand{\rd}{{\rm d}}
\newcommand{\be}{\begin{equation}}
\newcommand{\ee}{\end{equation}}
\newcommand{\bey}{\begin{eqnarray}}
\newcommand{\eey}{\end{eqnarray}}

\newcommand{\la}{\langle}
\newcommand{\ra}{\rangle}

\newcommand{\eps}{\varepsilon}

\newcommand{\bp}{{\bf p}}
\newcommand{\bq}{{\bf q}}
\newcommand{\br}{{\bf r}}

\newcommand{\bx}{{\bf x}}

\newcommand{\ph}{\varphi}

\newcommand{\boeta}{{\boldsymbol \eta}}
\renewcommand{\a}{\alpha}

\newcommand{\e}{\varepsilon}
\newcommand{\si}{\sigma}

\newcommand{\cU}{{\cal U}}
\newcommand{\cJ}{{\cal J}}

\newcommand{\bR}{{\mathbb R}}

\newcommand{\bN}{{\mathbb N}}

\newcommand{\tr}{\mbox{Tr}}

\newcommand{\wt}{\widetilde}
\newcommand{\wh}{\widehat}

\newcommand{\const}{\mathrm{const}}
\newcommand{\cG}{{\cal G}}
\newcommand{\cS}{{\cal S}}

\newcommand{\cF}{{\cal F}}
\newcommand{\cA}{{\cal A}}

\newcommand{\cK}{{\cal K}}
\newcommand{\cH}{{\cal H}}
\newcommand{\cL}{{\cal L}}

\input epsf

\newcommand{\donothing}[1]{}

\begin{document}

\title{Derivation of the Cubic Non-linear Schr\"odinger
 Equation from Quantum Dynamics of  Many-Body Systems}
\author{L\'aszl\'o Erd\H os${}^2$\thanks{Partially
supported by  EU-IHP Network ``Analysis and Quantum''
HPRN-CT-2002-0027.}\;, Benjamin Schlein${}^1$\thanks{Supported by
NSF postdoctoral fellowship.}\; and Horng-Tzer
Yau${}^1$\thanks{Partially supported by NSF grant
DMS-0307295 and MacArthur Fellowship.} \\ \\
Department of Mathematics, Harvard University\\ Cambridge,
MA 02138, USA${}^1$ \\ \\
Institute of Mathematics, University of Munich, \\
Theresienstr. 39, D-80333 Munich, Germany${}^2$
\\}

\date{Jun 12, 2006}

\maketitle

\begin{abstract}
We prove rigorously that the one-particle density matrix of
three dimensional
interacting Bose systems with a short-scale repulsive pair
interaction converges to the solution of the cubic non-linear
Schr\"odinger equation in a  suitable scaling limit.
The result is extended to
$k$-particle density matrices for all positive integer $k$.
\end{abstract}

\bigskip

AMS Classification Number (2000):  35Q55, 81Q15, 81T18, 81V70.

Running title: Derivation of the cubic NLS equation.

Keywords:  Feynman diagrams, BBGKY hierarchy, dispersive
estimates, propagation of chaos.

\section{Introduction}\label{sec:intro}
\setcounter{equation}{0}

The fundamental principle of quantum mechanics states that a
quantum  system of $N$ particles is described by a wave function
of $N$ variables satisfying a Schr\"odinger equation. In realistic
systems, $N$ is so large that a direct solution of the
Schr\"odinger equation for interacting systems is clearly an
impossible task. Many-body dynamics is thus  traditionally
approximated by simpler effective dynamics where only the time
evolution of a few cumulative degrees of freedom is monitored. In
the simplest case only the one-particle marginal densities are
considered. The many-body pair interaction is then replaced by an
effective non-linear mean-field potential and higher order quantum
correlations are neglected.

For Bose systems, there are three important examples:
 i) the Hartree equation describing
 Bose systems  with soft or Coulomb interactions in the mean field limit; ii) the
cubic non-linear Schr\"odinger (NLS) equation for  Bose systems
with short range interactions in suitable scaling limits and iii)
the Gross-Pitaevskii equation for the condensate of Bose systems
with short range interactions (including the hard core
interaction) in a certain scaling limit. All three equations are
non-linear Schr\"odinger equations and the Gross-Pitaevskii
equation is itself a cubic non-linear Schr\"odinger (NLS)
equation. The fundamental reason why nonlinear Schr\"odinger
equation describes these three systems is the following
observation: the effective dynamics of a given particle is
governed by the density times the two-body scattering process. In
particular, all three-body scattering processes  are negligible in
the limit; a fact that should be established rigorously. We now
introduce a model covering all these cases.

Before giving the precise definitions, we explain the physical
differences among these three effective theories. Let $\sum_{1\leq
i<j\leq N}U(x_i-x_j)$ be the  many-body pair interaction, where
$x_j$ is the position of the $j$-th particle. The largest
lengthscale of the problem is the size of the system and it is
typically comparable with the variation scale of the particle
density $\varrho(x)$. We will set this scale to be $O(1)$. Thus
the density of the system, $\varrho$, is of order $N$. The
potential $U$  defines two lengthscales: the range of $U$ and the
scattering length of $U$ (see Appendix for a definition), denoted
by $r_U$ and $a_U$, respectively. The scattering length determines
the effective lengthscale of the two particle correlations. We
shall set the scattering length $a_U \sim O(N^{-1})$ so that
$\varrho a_U $  is order one. This is needed to obtain an
effective one-particle dynamics in our model.

The case i),  the Hartree equation,  is obtained when $r_U\sim
O(1)$, i.e. the range of $U$ is comparable with spatial variation
of the density.  In this case, the effective two-body scattering
process is of mean-field type, and each particle is subject to an
effective potential $U\ast \varrho$. If $a_U\ll r_U\ll 1$, i.e.
the range of $U$ is much shorter than the spatial variation of the
density but bigger than the scattering length, the effective
two-body scattering process is its Born approximation and the
effective potential is $ \varrho \int U$. Mathematically, this can
be explained by the fact that the weak limit of $U(x)$ in this
scaling is $\delta (x) \int U$. Since $\varrho(x)$ is quadratic in
the quantum mechanical wave function, one obtains a cubic
nonlinear Schr\"odinger equation for the one-particle orbitals.
Finally, the Gross-Pitaevskii equation arises when  the potential
$U$ is so localized that its range is comparable with its
scattering length, $a_U\sim r_U$ (in particular, this is the case
of the hard-core interaction). In this case, the effective
two-body scattering is the full two-body scattering process and
the effective potential is $8\pi \varrho a_U$.

\bigskip

We now give the precise definition of this model. We consider a system of $N$
interacting bosons in $d=3$  dimensions. The state space of
the $N$-boson system is  $L^2_s (\bR^{3N},\rd \bx)$,  the subspace
of $L^2 (\bR^{3N}, \rd \bx)$ containing all functions symmetric
with respect to permutations of the $N$ particles. Let $V$ be a
smooth, compactly supported, non-negative and symmetric ( i.e.
$V(x)=V(-x)$) function. Define the rescaling of $V$ by
\be V_N
(x): =N^{3\beta} V(N^{\beta} x) \; ,\label{def:VN}
\ee
where
$\beta$ is a nonnegative parameter. The Hamiltonian with pair
interaction $\frac{1}{N}V_N(x_i-x_j)$ is given by the nonnegative
self-adjoint operator
\begin{equation}\label{eq:ham}
H_N = - \sum_{j=1}^N \Delta_j + \frac{1}{N} \sum_{i<j} V_N (x_i
-x_j)\;
\end{equation}
acting on  $L^2_s (\bR^{3N}, \rd \bx)$. Denote by $\psi_{N,t}$ the
wave function at time $t$; it satisfies the Schr\"odinger equation
\begin{equation}\label{eq:schr}
i\partial_t \psi_{N,t} = H_N \psi_{N,t} \, ,
\end{equation}
with  initial condition $\psi_{N,0}$. The Schr\"odinger equation
conserves the energy, the $L^2$-norm and the permutation symmetry
of  the wave function.

Instead of describing the system through its wave function
$\psi_{N,t}$, we can introduce the corresponding density matrix
$\gamma_{N,t}$, defined as the orthogonal projection onto
$\psi_{N,t}$ in the space  $L^2 (\bR^{3N}, \rd \bx)$,  i.e.,
$\gamma_{N,t} = \pi_ {\psi_{N,t}}$. The Schr\"odinger equation
(\ref{eq:schr}) assumes then the Heisenberg form
\begin{equation}\label{eq:heis}
i\partial_t \gamma_{N,t} = [ H_N , \gamma_{N,t} ]\, ,  \quad [A,
B]:=AB-BA \, .
\end{equation}
Since $\| \psi_{N,t} \| =1$, it follows that $\tr \, \gamma_{N,t}
=1$.

For any integer $k=1, \dots , N$, we define the {\it $k$-particle
marginal density}, $\gamma_{N,t}^{(k)}$, by taking the partial
trace of $\gamma_{N,t}$ over the last $N-k$ particles. If
$\gamma_{N,t} (\bx ; \bx'): = \psi_{N,t} (\bx)
\overline{\psi_{N,t}} (\bx')$ denotes the kernel of
$\gamma_{N,t}$, then the kernel of $\gamma^{(k)}_{N,t}$ is given
by
\begin{equation}\label{eq:gamk}
\gamma^{(k)}_{N,t} (\bx_k;\bx'_k) = \int_{\bR^{3(N-k)}} \rd
\bx_{N-k} \, \gamma_{N,t} (\bx_k, \bx_{N-k} ; \bx'_k , \bx_{N-k})
\, .
\end{equation}
Here and henceforth we use the notation $\bx = (x_1, \dots ,
x_N)$, $\bx_k = (x_1, \dots,x_k)$, $\bx_{N-k} = (x_{k+1} , \dots,
x_{N})$, and similarly for the primed variables. Due to the
permutational symmetry  of $\psi_{N, t}$ in all variables, the
marginal densities are also symmetric in the sense that \be
    \gamma^{(k)}_{N,t} (x_1, x_2, \ldots, x_k ;x'_1, x_2',\ldots , x_k')
  =  \gamma^{(k)}_{N,t} \Big(x_{\sigma(1)}, x_{\sigma(2)},
\ldots, x_{\sigma(k)} ;x_{\sigma(1)}', x_{\sigma(2)}', \ldots,
x_{\sigma(k)}'\Big) \label{symm} \ee for any permutation $\sigma
\in \cS_k$ ($\cS_k$ denotes the set of permutations on $k$
elements). Denote by  $\Theta_{\sigma}$ the unitary operator \be
\label{eq:Theta} (\Theta_{\sigma} \psi) (x_1, \dots ,x_k) = \psi
(x_{\sigma(1)} , \dots , x_{\sigma(k)}) \,. \ee Then  (\ref{symm})
is equivalent to \be \label{symm2} \Theta_{\sigma} \gamma^{(k)}
\Theta_{\sigma^{-1}} = \gamma^{(k)} \ee for all $\sigma \in
\cS_k$. By definition, {\it density matrices} are non-negative
trace class operators, $\gamma^{(k)}\ge 0$, acting on
$L^2(\bR^{3k})$ and with permutational symmetry (\ref{symm}).
With a slight abuse of notation we will use the same notation for
the operators and their kernels.

\medskip

The two-body potential of this model is $U=N^{-1} V_N$.
By scaling, the scattering length $a_U= O(N^{-1})$. The range of interaction,
$r_U$, is of order $r_U = O(N^{-\beta})$.
Cases i), ii) and iii) correspond to the cases $\beta =0$,
$0<\beta <1$ and $\beta =1$, respectively.
The focus of this paper is to study case ii).
Our main result, the following Theorem \ref{thm:main},  states that
the time evolution of the one-particle
density matrix  is given
by a cubic non-linear Schr\"odinger equation, provided $0<\beta <1/2$.
The same result is expected to hold for all $0<\beta <1$;
the regime $\beta \ge 1/2$ is an open problem.
The topology and spaces
used in this theorem, e.g., the weak* topology of $\cL^1_k$,  will
be defined in next section. We state the theorem only for the
dimension $d=3$, but the proof can be extended to $d\le 2$ as
well.

\begin{theorem}\label{thm:main}
Fix $\ph \in H^1 (\bR^3)$, with $\| \ph \|=1$. Assume the
Hamiltonian $H_N$ is defined as in (\ref{eq:ham})  with $0 < \beta
< 1/2$. Suppose that the unscaled potential $V$ is smooth,
compactly supported,
 positive and symmetric, i.e. it satisfies $V(x)=V(-x)$.
 Let $\psi_N (\bx) := \prod_{j=1}^N \ph (x_j)$, and
suppose $\psi_{N,t}$ is the solution of the Schr\"odinger equation
(\ref{eq:schr}) with initial data $\psi_{N,t=0} = \psi_N$. Let
$\Gamma_{N,t} = \{ \gamma^{(k)}_{N,t} \}_{k =1}^N$ be the family
of marginal distributions associated to $\psi_{N,t}$. Then, for
every fixed $t \in \bR$ and integer $k \geq 1$ we have
\begin{equation}
\gamma_{N,t}^{(k)} \to \gamma^{(k)}_t \quad \text{as } N \to
\infty
\end{equation}
with respect to the weak* topology of $\cL^1_k$. Here
\begin{equation}\label{eq:factorized}
\gamma^{(k)}_t (\bx ; \bx') := \prod_{j=1}^k \ph_t (x_j)
\overline{\ph_t} (x'_j)\; ,
\end{equation}
where $\ph_t$ is the solution of the nonlinear Schr\"odinger
equation
\begin{equation}\label{eq:gpe}
i\partial_t \ph_t = - \Delta \ph_t + b_0 |\ph_t|^2 \ph_t
\end{equation}
with initial data $\ph_{t=0} = \ph$ and $b_0=\int V(x)\rd x$.
\end{theorem}

\bigskip

For dimension $d=1$, Adami, Golse and Teta recently obtained a
result similar to Theorem \ref{thm:main} in \cite{AGT} (certain
partial results, for the one-dimensional case, were already
obtained by these authors, together with Bardos, in \cite{ABGT}).

\medskip

Theorem \ref{thm:main} holds also for $\beta=0$. The nonlinear
equation (\ref{eq:gpe}) then becomes the Hartree equation
\begin{equation}\label{eq:hartree}
i\partial_t u_t = - \Delta u_t + (V * |u_t|^2)u_t \, .
\end{equation}
In this special case, the result was rigorously proven by Hepp
\cite{He} for smooth potentials $V (x)$ and by Spohn \cite{Sp} for
bounded potentials. Ginibre and Velo \cite{GV} treated integrable
potentials, but they required the initial state to be coherent. In
particular, in the approach of \cite{GV}, the number of particles
cannot be fixed. For Coulomb potential,  partial results were
obtained in \cite{BGM} and a complete proof was given in
\cite{EY}. If the particles have a relativistic dispersion then
(\ref{eq:hartree}) has to be replaced by
\[i\partial_t u_t = (1 -  \Delta)^{1/2} u_t + (V * |u_t|^2)u_t \,
.\] This equation was recently derived in \cite{ES}, starting from
many-body dynamics, for the case of a Coulomb potential. A concise
overview on results and open problems related to the Hartree
equation in physical context is found in \cite{FL}.

\bigskip

The basic strategy to prove Theorem \ref{thm:main} is the same as
in \cite{EY}: We first write down the BBGKY hierarchy
\eqref{eq:BBGKY1} for the marginal densities (\ref{eq:gamk}). Then
we take the limit $N \to \infty$ to obtain an infinite hierarchy
of equations \eqref{eq:GPH}. Since this hierarchy was first
mentioned in the context of the Gross-Pitaevskii scaling,
$\beta=1$, we will continue to call it {\it Gross-Pitaevskii
hierarchy} even when the coefficient of the nonlinear term is
given by the Born approximation of the scattering length. Finally
we prove that the Gross-Pitaevskii hierarchy has a unique solution
in a suitable space. Since tensor products of solutions to the
non-linear Schr\"odinger equation (\ref{eq:gpe}) are trivial
solutions to the hierarchy, this identifies the limit and thus
proves Theorem \ref{thm:main}.

The key steps in this approach are an a-priori estimate, the
convergence of the BBGKY hierarchy to the infinite hierarchy and
the proof of the uniqueness of the infinite hierarchy. The first
part was already proved in \cite{EESY}. The convergence has to be
proven in a somewhat stronger sense than in \cite{EESY}.
 The key point of the present paper is the
uniqueness result stated as Theorem \ref{thm:unique}. This theorem
is indeed the well-posedness of the Gross-Pitaevskii hierarchy.
Since solutions of the cubic non-linear Schr\"odinger equation
naturally generate solutions of the hierarchy by taking tensor
products, Theorem \ref{thm:unique} can be viewed as an extension
of the well-posedness theorem of the NLS equation to infinite
dimensions. Therefore, we have to either extend the fundamental
tool in the well-posedness of the nonlinear Schr\"odinger
equation,  the Stricharz inequality, to infinite dimensions or
find a method avoiding it (for a review on the Stricharz
inequality and the well-posedness of nonlinear Schr\"odinger
equations see, e.g., \cite{B}, \cite{C}, \cite{T}). Apart from
this issue, we have to control correlation effects of
many-particle systems which are absent in the nonlinear equation
in  Euclidean space. Our method, based on the analysis of Feynman
graphs, provides a solution to both problems and contains in
certain sense a version of Stricharz inequality in infinite
dimension---albeit we deal only with cubic nonlinearity.
We will explain this issue in the remarks after Theorem
\ref{thm:bounds}.

We emphasize that our uniqueness result is valid for any coupling
constant in the non-linear Schr\"odinger equation, in particular
it is valid also for the special case of the Gross-Pitaevskii
equation. Thus the main part of the paper actually  is independent
of the scaling in the $N$-body model, in particular it is
independent of the choice of $\beta$. The restriction $\beta <1/2$
is only used to obtain the a-priori bound and therefore the
convergence to the infinite hierarchy.

\medskip

As mentioned earlier,
the coupling constant $b_0 = \int V$
in the non-linear Schr\"odinger equation in
Theorem \ref{thm:main} is the Born approximation to
the physically correct
coupling constant: $8\pi$ times the scattering length $a_N$ of the
potential $\frac{1}{N}V_N$. To see this,
denote
the scattering length of  $V$ by
$a_0$. Then we have
\be
 \lim_{N\to\infty} N a_N=  \begin{cases}
  b_0 /8\pi   & \text{if} \quad 0 < \beta < 1, \\
   a_0  &\text{if} \quad  \beta=1.
  \end{cases}
\label{liman} \ee The limit in the first case is proved in Lemma
\ref{lm:scatt} for radial potential; the second one is just a
rescaling. Notice that at $\beta=1$, the coupling constant is the
scattering length of the unscaled potential, which indicates that
the full two-body scattering process needs to be taken into
account. For stationary problem in this case,  the connection
between the ground state energy of the Bose system and the
Gross-Pitaevskii energy functional was rigorously established by
Lieb and Yngvason \cite{LY}. Furthermore, the existence of
Bose-Einstein condensation in this limit was proved by Lieb and
Seiringer \cite{LS}. An excellent overview on the recent
development in these problems, see \cite{LSSY}. For the dynamical
problem, we have proved in \cite{ESY} that the family of the
reduced density matrices converges to a solution of the
Gross-Pitaevskii hierarchy with the correct coupling constant $8
\pi a_0$. However, the a priori estimate obtained in \cite{ESY}
was not strong enough to apply the uniqueness theorem in this
paper, Theorem \ref{thm:unique}. Furthermore, the pair
interactions in \cite{ESY} were cutoff whenever many particles are
in a very small physically unlikely region. To complete the
project for $\beta=1$, we still have to remove this cutoff and
establish the a priori estimate needed for Theorem
\ref{thm:unique}.

\bigskip

{\it Acknowledgement.} We would like to thank the referees for their
helpful comments on how to improve the presentation of the results
in the manuscript.

\section{The BBGKY Hierarchy}
\setcounter{equation}{0}

The marginal density matrices (\ref{eq:heis}) satisfy the BBGKY
hierarchy:
\begin{equation}\label{eq:BBGKY1}
\begin{split}
i\partial_t &\gamma^{(k)}_{N,t} (\bx_k; \bx'_k) = \sum_{j=1}^k
\left( -\Delta_{x_j} + \Delta_{x'_j} \right)  \gamma^{(k)}_{N,t}
(\bx_k ; \bx'_k)  \\ &+ \frac{1}{N} \sum_{1\leq i < j \leq k}
\left( V_N (x_i - x_j) - V_N (x'_i - x'_j) \right)
\gamma^{(k)}_{N,t} (\bx_k ; \bx'_k) \\
&+ \Big(1 -\frac{k}{N}\Big) \sum_{j=1}^k \int \rd x_{k+1} \,
\left( V_N (x_j - x_{k+1}) - V_N (x'_j - x_{k+1}) \right) \,
\gamma^{(k+1)}_{N,t} (\bx_k , x_{k+1} ; \bx'_k , x_{k+1} ) \;
\end{split}
\end{equation}
for $k =1 ,2 , \dots , N$. The first term on the right hand side
of the hierarchy describes the kinetic energy of the first $k$
particles; the second term is associated with the interactions
among the first $k$ particles, and the last term corresponds to
interactions between the first $k$ particles and the other $N-k$
particles.

Recall that $b_0 = \int \rd x \, V(x)$ is the $L^1$ norm of the
non-negative potential $V$. Since $V_N (x) \to b_0 \delta (x)$ as
$N \to \infty$, the BBGKY hierarchy (\ref{eq:BBGKY1}) converges
formally, as $N \to \infty$, to the following {\it
Gross-Pitaevskii hierarchy} of equations:
\begin{equation}\label{eq:GPH}
\begin{split}
i\partial_t \gamma^{(k)}_{t} (\bx_k ; &\bx'_k) = \sum_{j=1}^k
\left(-\Delta_{x_j} + \Delta_{x'_j} \right) \gamma^{(k)}_{t}
(\bx_k ; \bx'_k) \\
&+ b_0 \sum_{j=1}^k \int \rd x_{k+1} \, \left( \delta (x_j -
x_{k+1}) - \delta (x'_j - x_{k+1}) \right) \gamma^{(k+1)}_{t}
(\bx_k , x_{k+1} ; \bx'_k , x_{k+1}) \, ,
\end{split}
\end{equation}
for any $k \geq 1$. It is easy to see that the family of
factorized densities $\gamma^{(k)}_t = \prod_{j=1}^k \ph_t (x_j)
\overline{\ph_t} (x'_j)$ is a solution of (\ref{eq:GPH}) if and
only if $\ph_t$ is a solution of the  cubic nonlinear
Schr\"odinger equation
\begin{equation}\label{eq:GPE}
i\partial_t \ph_t = -\Delta \ph_t + b_0 |\ph_t|^2 \ph_t \;.
\end{equation}
If we can establish the uniqueness of the Gross-Pitaevskii
hierarchy, then for every fixed $k \geq 1$,
\begin{equation}\label{eq:claim1}
\gamma^{(k)}_{N,t} \to \gamma^{(k)}_t= \prod_{j=1}^k \ph_t (x_j)
\overline{\ph_t} (x'_j) \; , \quad \text{as } N\to \infty
\end{equation}
with respect to some suitable topology.

As a technical point, we remark that the action of the
delta-function on general density matrices in (\ref{eq:GPH}) is
not well defined. However, in our case, the density matrices are
in $\cH_k$. For such density matrices  (\ref{eq:GPH}) can be
defined through an appropriate limiting procedure, see Section
\ref{sec:conv}.

\section{Banach Spaces of Density Matrices}\label{sec:topo}
\setcounter{equation}{0}

In this section we define some Banach spaces which will be useful
in order to take the limit $N \to \infty$ of the marginal
densities $\gamma_{N,t}^{(k)}$. For $k \geq 1$, we denote by
$\cL_k^1$ and by $\cK_k$ the space of trace class and,
respectively, of compact operators on the $k$-particle Hilbert
space $L^2 (\bR^{3k}, \rd \bx_k)$. We have
\begin{equation}\label{eq:L1K}
(\cL^1_k, \| \, . \, \|_1) = (\cK_k , \| \, . \, \|)^*\; ,
\end{equation}
where $\| \, . \, \|_1$ is the trace norm, and $\|\, .\, \|$ is
the operator norm (see, for example, Theorem VI.26 in \cite{RS}).
The density matrices are the nonnegative elements of $\cL^1_k$
with permutational symmetry (\ref{symm2}).

\bigskip

For $\gamma^{(k)} \in \cL^1_k$, we define the norm
\begin{equation*}
\| \gamma^{(k)} \|_{\cH_k} = \tr \,| S_1 \dots S_k \,\gamma^{(k)}
S_k \dots S_1|
\end{equation*}
and the corresponding Banach space \[ \cH_k = \{ \gamma^{(k)} \in
\cL^1_k : \| \gamma^{(k)} \|_{\cH_k} < \infty \}. \] Moreover we
define the space of operators
\[ \cA_k = \{ T^{(k)} = S_1 \dots S_k K^{(k)} S_k \dots S_1 : K^{(k)} \in
\cK_k \} \] with the norm
\[ \| T^{(k)} \|_{\cA_k} = \| S_1^{-1} \dots S_k^{-1} T^{(k)}
S_k^{-1} \dots S_1^{-1} \| \] where $\| \, . \, \|$ is the
operator norm. Then, analogously to (\ref{eq:L1K}), we have
\begin{equation}\label{eq:dual}
(\cH_k , \| \, . \, \|_{\cH_k}) = ( \cA_k, \| \, . \,
\|_{\cA_k})^*
\end{equation}
for every $k \geq 1$. This induces a weak* topology on $\cH_{k}$
(for a proof of (\ref{eq:dual}) see Lemma 3.1 in \cite{EY}). Since
$\cA_k$ is separable, we can fix a dense countable subset of the
unit ball of $\cA_k$: we denote it by $\{J^{(k)}_i\}_{i \ge 1} \in
\cA_k$, with $\| J^{(k)}_i \|_{\cA_k} \leq 1$ for all $i \ge 1$.
Using the operators $J^{(k)}_i$ we define the following metric on
$\cH_k$: for $\gamma^{(k)}, \bar \gamma^{(k)} \in \cH_k$ we set
\begin{equation}\label{eq:rho}
\rho_k (\gamma^{(k)}, \bar \gamma^{(k)}) : = \sum_{i=1}^\infty
2^{-i} \left| \tr \; J^{(k)}_i \left( \gamma^{(k)} - \bar
\gamma^{(k)} \right) \right| \, .
\end{equation}
Then the topology induced by the metric $\rho_k ( \cdot ,\cdot)$
and the weak* topology are equivalent on the unit ball of $\cH_k$
 (see \cite{Ru}, Theorem 3.16) and hence on
any ball of finite radius as well. In other words, a uniformly
bounded sequence $\gamma_N^{(k)} \in \cH_k$ converges to
$\gamma^{(k)} \in \cH_k$ with respect to the weak* topology, if
and only if $\rho_k (\gamma^{(k)}_N , \gamma^{(k)}) \to 0$ as $N
\to \infty$.

\bigskip

For a fixed $T \geq 0$, let $C ([0,T], \cH_k)$ be the space of
functions of $t \in [0,T]$ with values in $\cH_k$ which are
continuous with respect to the metric $\rho_k$. On $C ([0,T],
\cH_k)$ we define the metric
\begin{equation}\label{eq:hatrho}
\widehat \rho_k (\gamma^{(k)} (\cdot ) , \bar \gamma^{(k)} (\cdot
)) := \sup_{t \in [0,T]} \rho_k (\gamma^{(k)} (t) , \bar
\gamma^{(k)} (t))\,.
\end{equation}

Finally, we define the space $\cH$ as the direct sum over $k \geq
1$ of the spaces $\cH_k$, that is
\begin{equation*}
\cH = \bigoplus_{k \geq 1} \cH_k = \left\{ \Gamma = \{
\gamma^{(k)} \}_{k \geq 1} : \gamma^{(k)} \in \cH_k, \quad \forall
\, k \geq 1 \right\} \, ,
\end{equation*}
and, for a fixed $T \geq 0$, we consider the space \[ C([0,T] ,
\cH) = \bigoplus_{k \geq 1} C([0,T], \cH_k),\] equipped with the
product of the topologies induced by the metric $\widehat
\rho^{(k)}$ on $C([0,T],\cH_k)$. Let $\tau$ denote this topology.
That is, for $\Gamma_{N,t}= \{ \gamma_{N,t}^{(k)} \}_{k \geq 1}$
and $ \Gamma_t = \{ \gamma^{(k)}_t \}_{k \geq 1}$ in $C ([0,T],
\cH)$, we have $\Gamma_{N,t} \overset{\tau}{\to} \Gamma_t$ for $N
\to \infty$ if and only if, for every fixed $k \geq 1$,
\begin{equation*}
\widehat \rho_k (\gamma_{N,t}^{(k)},\gamma_t^{(k)}) = \sup_{t \in
[0,T]} \rho_k (\gamma_{N,t}^{(k)}, \gamma_t^{(k)}) \to 0
\end{equation*}
as $N \to \infty$.

\medskip

{\it Notation.} As in (\ref{eq:gamk}), we use $\bx = (x_1,\dots
,x_N) \in \bR^{3N}$, $\bx_k = (x_1,\dots,x_k) \in \bR^{3k}$,
$\bx_{N-k} = (x_{k+1},\dots,x_{N}) \in \bR^{3(N-k)}$, and
analogously for the primed variables. We also use the notation
$\la x \ra = (1 +x^2)^{1/2}$, for all $x \in \bR^d$. We also set
$S_j = \la p_j \ra = (1-\Delta_j)^{1/2}$, for all integer $j \geq
1$ ($p_j = -i \nabla_{x_j}$ is the momentum of the $j$-th
particle). Moreover $\tr_j$ will indicate for the partial trace
over the $x_j$-th variable. The norm notation
 without subscript, $\| \cdot \|$, will always refer to the $L^2$-norm
for functions and to the operator norm in case of operators.
Unless stated otherwise, all integrals are over $\bR^3$, or on
$\bR^{3k}$ if the measure is $\rd \bx_k, \rd \bp_k$ etc. Universal
constants will be denoted by $\mbox{(const.)}$. Constants, that
may depend on other parameters, will be denoted by $C$. The
dependence is indicated in the statements but not always in the
proofs. Typically $C$ depends on the initial function $\varphi$
and on the potential $V$. To compare positive numbers $A, B$, we
also use $A\lesssim B$ to indicate that there is a universal
constant $\mbox{(const.)}>0$ with $A\leq \mbox{(const.)} B$. The
fact that $A\lesssim B$ and $B\lesssim A$ is denoted by $A\sim B$.

\section{Outline of the Proof of the Main Theorem }
\setcounter{equation}{0}

The proof of Theorem \ref{thm:main} is divided into several steps.
In the following we fix $T > 0$ arbitrary.

\bigskip

{\it Step 1: Regularization of the initial data.} Fix $\kappa
>0$ and $\chi \in C_0^{\infty} (\bR)$, with $0\leq \chi\leq 1$,
$\chi (s) = 1$, for $0 \leq s \leq 1$, and $\chi (s) =0$ if $s
\geq 2$. Define the regularized initial function
\[
\psi_N^{\kappa} := \frac{ \chi (\kappa H_N /N) \psi_N }{ \| \chi
(\kappa H_N /N) \psi_N \|} ,
\]
and  we denote by $\psi_{N,t}^{\kappa}$ the solution of the
Schr\"odinger equation (\ref{eq:schr}) with initial data
$\psi_N^{\kappa}$.  Denote   by $\wt \Gamma_{N,t} = \{ \wt
\gamma_{N,t}^{(k)} \}_{k=1}^N$ the family of marginal densities
associated with $\psi_{N,t}^{\kappa}$. The tilde in the notation
indicates dependence on the cutoff parameter $\kappa$.

This regularization cuts off the high energy part of $\psi_N$ and
it allows us to obtain the strong a-priori estimate
\begin{equation}\label{eq:apri2}
\tr \; (1-\Delta_1) \dots (1- \Delta_k) \wt \gamma^{(k)}_{N,t}
\leq \wt C^k
\end{equation}
for sufficiently large $N$ and uniformly in $t$ (see Theorem
\ref{thm:compact}). Here the constant $\wt C$ depends on the
cutoff $\kappa >0$. We thus have the compactness stated in the
following step.

\bigskip

{\it Step 2: Compactness of $\wt \Gamma_{N,t}$.} For fixed $k \geq
1$, it follows from Theorem \ref{thm:compact} that the sequence
$\wt \gamma^{(k)}_{N,t} \in C([0,T], \cH_k)$ is compact with
respect to the topology induced by the metric $\wh \rho_k$. By a
standard Cantor diagonalization argument, this implies that $\wt
\Gamma_{N,t}$ is a compact sequence in $C([0,T], \cH)$ with
respect to the $\tau$-topology. It also follows from Theorem
\ref{thm:compact}, that, if $\wt \Gamma_{\infty,t} = \{ \wt
\gamma^{(k)}_{\infty,t} \}_{k\geq 1} \in C([0,T], \cH)$ denotes an
arbitrary limit point of $\wt \Gamma_{N,t}$, then $\wt
\gamma^{(k)}_{\infty,t}$ is non-negative, symmetric w.r.t.
permutations in the sense of (\ref{symm2}), and satisfies
\begin{equation}\label{eq:aprigam}
\tr \; | S_1 \dots S_k \wt \gamma^{(k)}_{\infty,t} S_k \dots S_1|
\leq \wt C^k \,
\end{equation}
for all $k \geq 1$ and $t \in [0,T]$. We put a tilde in the
notation for $\wt \Gamma_{\infty,t}$ and $\wt C$, because a-priori
they may depend on the cutoff $\kappa$, which is kept fixed here.
Notice that $\wt C$ is independent of $k$. From Step 3 and Step 4
below it will follow that $\wt \Gamma_{\infty,t}$ is actually
independent of $\kappa$.

\bigskip

{\it Step 3: Convergence to solutions of the Gross-Pitaevskii
hierarchy.}  Define  the free evolution operator \be \cU_0^{(k)}
(t) \gamma^{(k)} = \exp \left( - i t \sum_{j=1}^k (-\Delta_j)
\right) \gamma^{(k)} \exp \left(i t \sum_{j=1}^k(- \Delta_j)
\right)\; . \label{eq:freeev} \ee Theorem \ref{thm:convergence}
states that an arbitrary limit point $\wt \Gamma_{\infty,t} \in
C([0,T], \cH)$ of the sequence $\wt \Gamma_{N,t}$ satisfies the
infinite Gross-Pitaevskii hierarchy in the integral form
\begin{equation}\label{eq:BBGKYint1}
\wt \gamma_{\infty,t}^{(k)} = \cU_0^{(k)} (t) \wt
\gamma_{\infty,0}^{(k)} -i b_0\sum_{j=1}^k \int_0^t \rd s \;
\cU_0^{(k)} (t-s) \tr_{k+1} \; [ \delta (x_j -x_{k+1}), \wt
\gamma^{(k+1)}_{\infty, s}] \, ,
\end{equation}
with initial data
\begin{equation}\label{eq:init}
\wt \gamma_{\infty,t=0}^{(k)} (\bx_k ; \bx'_k) = \gamma_0^{(k)}
(\bx_k ; \bx'_k) = \prod_{j=1}^k \ph (x_j) \overline{\ph} (x'_j) .
\end{equation}
Notice that (\ref{eq:init}) implies that the initial data $\wt
\Gamma_{\infty,t=0}$ is independent of the cutoff $\kappa$.

\bigskip

{\it Step 4: Uniqueness of the solutions of the Gross-Pitaevskii
hierarchy.} In Theorem \ref{thm:unique} we prove that there is at
most one solution  $\Gamma_t = \{ \gamma^{(k)}_t \}_{k \geq 1}$ of
(\ref{eq:BBGKYint1}) in the space $C([0,T], \cH)$, such that, for
all $k \geq 1$ and $t \in [0,T]$, $\gamma^{(k)}_t$ is
non-negative, symmetric w.r.t. permutation (in the sense
(\ref{symm2})) and satisfies (\ref{eq:aprigam}) and
(\ref{eq:init}). Hence the family of factorized densities
$\Gamma_t = \{ \gamma^{(k)}_t \}_{k\geq 1}$ defined in
(\ref{eq:factorized}) is the unique nonnegative symmetric solution
of the Gross-Pitaevskii hierarchy (\ref{eq:BBGKYint1}). Thus  $\wt
\Gamma_{N,t} \to \Gamma_t$ as $N \to \infty$ in the
$\tau$-topology and this holds for any fixed $\kappa$, so the
limit is independent of $\kappa$. Since $\wt\gamma_{N,t}^{(k)}$ is
bounded in the $\cH_k$ norm, uniformly in $N$, the convergence in
the metric $\varrho_k$ and the weak* convergence of $\cH_k$ are
equivalent. It therefore follows that for every fixed $k \geq 1$,
$t \in [0,T]$ and $\kappa >0$, we have $\wt \gamma^{(k)}_{N,t} \to
\gamma_t^{(k)}$ as $N \to \infty$ with respect to  the weak*
topology of $\cH_k$. Convergence in weak* $\cL_k^1$ then trivially
follows.

\bigskip

{\it Step 5: Removal of the cutoff and the  conclusion of the
proof.}
 It follows from Proposition
\ref{prop:initialdata}, part ii), that
\[ \| \psi_{N,t} - \psi_{N,t}^{\kappa} \| = \| \psi_{N} -
\psi_{N}^{\kappa} \| \leq C \kappa^{1/2} \; ,\] where the constant
$C$ is independent of $N$ and $\kappa$. This implies that, for
every $J^{(k)} \in \cK_k$, we have
\begin{equation}\label{eq:remove}
\Big| \tr \; J^{(k)} \left( \gamma^{(k)}_{N,t} - \wt
\gamma^{(k)}_{N,t} \right) \Big| \leq C \kappa^{1/2}
\end{equation}
where the constant $C$ depends on $J^{(k)}$, but is independent of
$N$,  $k$ or $\kappa$.

For fixed $k \geq 1$ and $t$, we choose $J^{(k)} \in \cK_k$ and
$\e >0$. Then for any $\kappa >0$, we have from (\ref{eq:remove})
that
\begin{equation}\label{eq:lastproof}
\begin{split}
\Big| \tr \; J^{(k)} \left( \gamma^{(k)}_{N,t} - \gamma^{(k)}_t
\right) \Big| \leq &\; \Big| \tr \; J^{(k)} \left(
\gamma^{(k)}_{N,t} - \wt \gamma^{(k)}_{N,t} \right) \Big| + \Big|
\tr \; J^{(k)} \left( \wt \gamma^{(k)}_{N,t} - \gamma^{(k)}_{t}
\right) \Big| \\
\leq & \; C \kappa^{1/2} + \Big| \tr \; J^{(k)} \left( \wt
\gamma^{(k)}_{N,t} - \gamma^{(k)}_{t} \right) \Big| \, .
\end{split}
\end{equation}
Since $\wt \gamma^{(k)}_{N,t} \to \gamma_t^{(k)}$  with respect to
the weak* topology of $\cL^1_k$, the last term vanishes in the
limit $N \to \infty$. Since $\kappa >0$ is arbitrary, the r.h.s.
of (\ref{eq:lastproof}) is smaller as $\e$ for $N$ large enough.
This completes the proof of the theorem.

\medskip

{\it Remark.} Note that the main body of the proof, Steps 1--4,
would have proven weak* convergence in $\cH_k$. It is in the
removal of the cutoff, Step 5, that we can prove only under a
weaker convergence. This situation is similar to the Coulomb
interaction paper, \cite{EY}, where the removal of the cutoffs
resulted eventually in a weaker sense of convergence.

\section{Cutoff of the initial wave function}
\setcounter{equation}{0}

In this section we show how to regularize the initial wave
function $\psi_N (\bx) = \prod_{j=1}^N \ph (x_j)$. The aim is to
find an approximate wave function $\psi_N^{\kappa}$, depending on
a cutoff parameter $\kappa >0$, so that, on the one hand, the
expectation of $H_N^k$ in the state $\psi_N^{\kappa}$ is of the
order $N^k$, and, on the other hand, the difference between
$\psi_N$ and $\psi_N^{\kappa}$ converges to zero, as $\kappa \to
0$, uniformly in $N$.

\begin{proposition}\label{prop:initialdata}
Let $\ph \in H^1 (\bR^3)$, with $\| \ph \|_{L^2} =1$. Assume $H_N$
is defined as in (\ref{eq:ham}): suppose that the unscaled
potential $V$ is smooth and positive, and that $0 < \beta < 2/3$.
We define $\psi_N (\bx) = \prod_{j=1}^N \ph (x_j)$, and, for
$\kappa >0$,
\begin{equation}\label{eq:psieps}
\psi^{\kappa}_N = \frac{\chi ( \kappa H_N/N ) \psi_N}{\| \chi
(\kappa H_N/N) \psi_N \|} \, .
\end{equation}
Here $\chi \in C^{\infty}_0 ( \bR )$ is a cutoff function
 such that $0\leq \chi\leq 1$,
 $\chi (s) =1$ for
$0 \leq s \leq 1$ and $\chi (s) =0$ for $s \geq 2$. We denote by
$\wt \gamma^{(k)}_{N}$, for $k =1, \dots, N$, the marginal
densities associated with $\psi_N^{\kappa}$.
\begin{enumerate}
\item[i)] For every integer $k \geq 1$ and for $\kappa>0$ small
enough, we have
\begin{equation}
(\psi_N^{\kappa} , H_N^k \psi_N^{\kappa} ) \leq \frac{2^k
N^k}{\kappa^k} \frac{1}{1- C \kappa^{1/2}}
\end{equation}
where the constant $C$
 only depends on the
$H^1$-norm of $\ph$ and on the unscaled potential $V(x)$.
\item[ii)] We have
\[ \| \psi_N - \psi_N^{\kappa} \| \leq C \kappa^{1/2}
\] uniformly in $N$ ($C$ only depends on the $H^1$ norm of $\ph$
and on the unscaled potential $V$). \item[iii)] Let
\begin{equation}\label{eq:gamma0}
\gamma^{(k)}_0 (\bx_k ; \bx'_k) = \prod_{j=1}^k \ph (x_j)
\overline{ \ph} (x'_j).
\end{equation}
Then, for every fixed $k \geq 1$ and $J^{(k)} \in \cK_k$, we have
\begin{equation}
\tr \; J^{(k)} \left( \wt \gamma^{(k)}_{N} - \gamma^{(k)}_0
\right) \to 0
\end{equation}
as $N \to \infty$ (the convergence is uniform in $\kappa > 0$, for
$\kappa$ small enough).
\end{enumerate}
\end{proposition}

\begin{proof}
First  we compute
\begin{equation}
\| \chi  (\kappa H_N/N) \psi_N - \psi_N \|^2 =  \Big(\psi_N , ( 1 -
\chi (\kappa H_N /N) )^2 \psi_N\Big) \leq \Big(\psi_N, {\bf 1} ( \kappa H_N
\geq N) \psi_N\Big)\;,
 \end{equation}
where ${\bf 1} (s \geq \lambda)$ is the characteristic function of
$[\lambda , \infty)$. Next we use that $\chi (s \geq 1) \leq s$,
for all $s \geq 0$. Therefore
\begin{equation}
\begin{split}
\| \chi  (\kappa H_N/N) \psi_N - \psi_N \|^2 &\leq
\frac{\kappa}{N} (\psi_N , H_N \psi_N) \\ &= \frac{\kappa}{N}
\left( N \| \nabla \ph \|^2 + \frac{ (N-1)}{2} N^{3\beta} (\psi_N ,
V (N^{\beta} (x_1 -x_2)) \psi_N) \right)
\\ &\leq \mbox{(const.)}\kappa \left( \| \ph \|_{H^1}^2 + \| V \|_{L^1}
  \| \ph \|_{H^1}^2 \right) \; ,
\end{split}
\end{equation}
where we used that $N^{3\beta} V(N^{\beta} (x_1 -x_2)) \leq
\mbox{(const.)}\| V \|_{L^1} (1 - \Delta_1) (1-\Delta_2)$ (see
Lemma \ref{lm:sob} from Appendix~\ref{app}). Hence
\begin{equation}\label{eq:eps1}
\| \chi (\kappa H_N/N) \psi_N - \psi_N \| \leq C \kappa^{1/2}
\end{equation}
for a constant $C$ only depending on the $H^1$ norm of $\ph$ and
on the unscaled potential $V$. Using (\ref{eq:eps1}) we obtain
\begin{equation*}
(\psi^{\kappa}_N, H_N^k \wt \psi^{\kappa}_N) = \frac{(\psi_N,
\chi^2 (\kappa H_N / N) H_N^k \psi_N )}{\| \chi (\kappa H_N / N)
\psi_N \|^2} \leq \frac{2^k N^k}{\kappa^k} \frac{1}{1 -2 \| \chi
(\kappa H_N /N)\psi_N - \psi_N \|} \leq \frac{2^k N^k}{\kappa^k}
\frac{1}{1- C \kappa^{1/2}}
\end{equation*}
for all $0 < \kappa < 1/ C^2$, which proves i).

Part ii) follows very easily from (\ref{eq:eps1}) because, using
the shorthand notation $\chi = \chi (\kappa H_N /N)$  and
using that $\| \psi_N \|=1$:
\begin{equation}
\begin{split}
\Big\| \psi_N - \frac{ \chi \psi_N}{\| \chi \psi_N \|} \Big\| &\leq \|
\psi_N - \chi \psi_N \| + \Big\| \chi \psi_N -\frac{ \chi \psi_N}{\|
\chi \psi_N \|} \Big\| = \| \psi_N - \chi \psi_N \| + | 1 - \| \chi
\psi_N \| | \\ &\leq 2 \| \psi_N - \chi \psi_N \| \, .
\end{split}
\end{equation}

Finally, we prove iii). For fixed $k \geq 1$, $J^{(k)} \in \cK_k$
and $\e >0$ we prove that
\begin{equation}\label{eq:cutoff}
\left|\tr \; J^{(k)} \left( \wt \gamma^{(k)}_N - \gamma^{(k)}_0
\right) \right| \leq \e
\end{equation}
if $N$ is large enough (uniformly in $\kappa$, for $\kappa$
sufficiently small). To this end, we choose $\ph^* \in H^2
(\bR^3)$ with $\| \ph^* \| =1$, such that $\| \ph - \ph^* \| \leq
\e/ (24k \| J^{(k)} \|)$. Then we define $\psi_{N,*} (\bx) =
\prod_{j=1}^k \ph^* (x_j) \prod_{j=k+1}^N \ph (x_j)$, and $\wt
\psi_{N,*} = \chi (\kappa H_N /N) \psi_{N,*} / \|\chi (\kappa H_N
/N) \psi_{N,*} \|$. Moreover we set
\begin{equation}
\wt \gamma^{(k)}_{N,*} (\bx_k ; \bx'_k) = \int \rd \bx_{N-k} \,
\wt \psi_{N,*} (\bx_k , \bx_{N-k}) \overline{\wt \psi_{N,*}
(\bx'_k, \bx_{N-k})} .
\end{equation}
Note that even though $\wt \psi_{N,*}$ is not symmetric with
respect to permutation of the $N$ particles, it is still symmetric
in the first $k$ and the last $N-k$ variables; hence $\wt
\gamma^{(k)}_{N,*}$ is a symmetric density matrix. Next we define
the Hamiltonian
\[ \wh H_N =
- \sum_{j \geq k+1} \Delta_j + \frac{1}{N} \sum_{k +1 \leq i < j
\leq N}
 V_N (x_i -x_j) ,\] with $V_N (x) = N^{3\beta} V (N^{\beta} x)$.
We denote $\wh \chi = \chi (\kappa \wh H_N /N)$ and we set $\wh
\psi_N = \wh \chi \psi_N / \| \wh \chi \psi_N \|$ and $\wh
\psi_{N,*} = \wh \chi \psi_{N,*} / \| \wh \chi \psi_{N,*} \|$. We
also define
\begin{equation}\begin{split}
  \wh \gamma_{N}^{(k)}(\bx_k ; \bx_k') &:=
  \int \rd \bx_{N-k} \, \wh\psi_N (\bx_k, \bx_{N-k})\overline{
 \wh\psi_N (\bx_k', \bx_{N-k})}, \\
\wh \gamma_{N,*}^{(k)}(\bx_k ; \bx_k') &:=
  \int \rd \bx_{N-k} \, \wh\psi_{N,*} (\bx_k, \bx_{N-k})\overline{
 \wh\psi_{N,*} (\bx_k', \bx_{N-k})}\,.
 \end{split}
 \end{equation}
Although $\wh \psi_N$ and $\wh \psi_{N,*}$ are not symmetric with
respect to permutations of the $N$ particles, they are still
symmetric w.r.t. permutations of the first $k$ and the last $N-k$
particles; hence $\wh \gamma_N^{(k)}$ and $\wh \gamma_{N,*}^{(k)}$
are density matrices symmetric in all their variables in the sense
\eqref{symm}. Apart from the physical densities $\gamma^{(k)}_N$,
we introduced, starting from the wave functions $\psi_N$ and
$\psi_{N,*}$, two more sets of density matrices; the densities
$\wt \gamma^{(k)}_{N}$ and $\wt \gamma^{(k)}_{N,*}$, regularized
with the cutoff $\chi (\kappa H_N /N)$, and the densities $\wh
\gamma^{(k)}_N$ and $\wh \gamma^{(k)}_{N,*}$, regularized with the
cutoff $\wh \chi = \chi (\kappa \wh H_N / N)$. Observe that, since
the operator $\wh H_N$ acts trivially on the first $k$ variables
of $\psi_N$ and $\psi_{N,*}$, we have
\begin{equation}
\begin{split}
 \wh \psi_N = \ph^{\otimes k} \otimes \frac{ \wh \chi \ph^{\otimes
N-k}}{\| \wh \chi \ph^{\otimes N-k} \|}\quad \text{and } \quad
\wh \psi_{N,*} = (\ph^*)^{\otimes k} \otimes \frac{ \wh \chi
\ph^{\otimes N-k}}{\| \wh \chi \ph^{\otimes N-k} \|}
\end{split}
\end{equation}
 where
\[ \ph^{\otimes k
} = \underbrace{\ph \otimes \dots \otimes \ph}_{k \text{
factors}}\,. \] Hence $\wh \gamma^{(k)}_{N} = \gamma^{(k)}_0$ (see
(\ref{eq:gamma0})) and \[ \wh \gamma^{(k)}_{N,*} = |\ph^* \rangle
\langle \ph^*|^{\otimes k} = \underbrace{|\ph^* \rangle \langle
\ph^*| \otimes \dots \otimes |\ph^* \rangle \langle \ph^*|}_{k
\text{ factors}} \, ,\] for every $\kappa
>0$ and $N \geq k$. We estimate the l.h.s. of (\ref{eq:cutoff}) by
\begin{equation}\label{eq:cutoff2}
\begin{split}
\left|\tr \; J^{(k)} \left( \wt \gamma^{(k)}_N - \gamma^{(k)}_0
\right) \right| = \left|\tr \; J^{(k)} \left( \wt \gamma^{(k)}_N -
\wh \gamma^{(k)}_N \right) \right|  \leq \; &\left| \tr \; J^{(k)}
\left( \wt \gamma^{(k)}_N - \wt \gamma_{N,*}^{(k)} \right) \right|
+ \left|\tr \; J^{(k)} \left( \wt \gamma_{N,*}^{(k)} - \wh
\gamma_{N,*}^{(k)} \right) \right|
\\ &+ \left|\tr \; J^{(k)} \left(\wh \gamma_{N,*}^{(k)} - \wh
\gamma^{(k)}_N \right) \right| \, .
\end{split}
\end{equation}
The first term on the r.h.s. can be bounded by
\begin{equation}\label{eq:cut1}
\begin{split}
\left| \tr \; J^{(k)} \left( \wt \gamma^{(k)}_N - \wt
\gamma_{N,*}^{(k)} \right) \right| &\leq  2 \, \| J^{(k)} \| \,
\Big{\|} \frac{\chi \psi_{N}}{\| \chi \psi_N \|} - \frac{\chi
\psi_{N,*}}{\| \chi \psi_{N,*} \|} \Big{\|} \leq 4 \, \| J^{(k)}
\| \, \frac{ \| \chi (\psi_N - \psi_{N,*})\|}{\| \chi \psi_N \|}
\\ &\leq \frac{4 k \, \| J^{(k)} \|}{1- C \kappa^{1/2}} \, \| \ph -
\ph^* \| \leq \e /3
\end{split}
\end{equation}
uniformly in $\kappa$, for $\kappa$ small enough (recall that
$\chi = \chi (\kappa H_N /N)$). Here we used (\ref{eq:eps1}), $\|
\chi (\kappa H_N / N) \| \leq 1$, $\| \ph \| = \| \ph^* \| =1$,
and the choice $\| \ph - \ph^* \| \leq \e/ (24 k \| J^{(k)} \|)$.
Analogously, the third term on the r.h.s. of (\ref{eq:cutoff2}) is
bounded by
\begin{equation}\label{eq:cut3}
\left|\tr \; J^{(k)} \left(\wh \gamma_{N,*}^{(k)} - \wh
\gamma^{(k)}_N \right) \right| = \left|\tr \; J^{(k)} \left(
|\ph^* \rangle \langle \ph^* |^{\otimes k} - |\ph \rangle \langle
\ph |^{\otimes k} \right) \right| \leq 2 k \, \| J^{(k)}\| \, \|
\ph - \ph^* \| \leq \e /3\, .
\end{equation}
It remains to bound the second term on the r.h.s. of
(\ref{eq:cutoff2}). To this end, we note that
\begin{equation}\label{eq:g-g}
\begin{split}
\Big| \tr\; J^{(k)} \left( \wt \gamma^{(k)}_{N,*} - \wh
\gamma^{(k)}_{N,*} \right) \Big| &\leq 2 \| J^{(k)} \| \Big{\|}
\frac{\chi \psi_{N,*}}{\| \chi \psi_{N,*} \|} - \frac{\wh \chi \,
\psi_{N,*}}{ \| \wh \chi \, \psi_{N,*} \|} \Big\|
\\ &\leq 4\| J^{(k)} \| \frac{\| (\chi - \wh \chi) \psi_{N,*} \|}{\|
\chi \psi_{N,*} \|} \leq \frac{4\| J^{(k)} \|}{1- C \kappa^{1/2}}
\| (\chi - \wh \chi) \psi_{N,*} \|
\end{split}
\end{equation}
where we used (\ref{eq:eps1}). To estimate the last norm we expand
the function $\chi$ using the Helffer-Sj\"ostrand functional
calculus (see, for example, \cite{Dav}). Let $\wt \chi$ be an
almost analytic extension of the smooth function $\chi$ of order
two (that is $|\partial_{\overline z} \wt \chi (z)| \leq C |y|^2$,
for $y =\text{Im} z$ near zero): for example we can take $\wt\chi
(z=x+iy): = (\chi (x) + i y \chi' (x) + \chi'' (x) (iy)^2 /2)
\theta (x,y)$, where $\theta \in C_0^{\infty} (\bR^2)$ and $\theta
(x,y) = 1$ for $z =x +iy$ in some complex neighborhood of the
support of $\chi$. Then
\begin{equation}
\begin{split}
(\chi - \wh \chi) \psi_{N,*} &= -\frac{1}{\pi} \int \rd x \, \rd y
\, \partial_{\bar z} \wt \chi (z) \, \left( \frac{1}{z - (\kappa
H_N/N)} - \frac{1}{z - (\kappa \wh H_N/N)} \right) \psi_{N,*} \\
&= - \frac{\kappa}{N \pi} \int \rd x \, \rd y \,
\partial_{\bar z} \wt \chi (z) \, \frac{1}{z - (\kappa
H_N/N)} (H_N - \wh H_N) \frac{1}{z - (\kappa \wh H_N/N)}
\psi_{N,*} \, .
\end{split}
\end{equation}
Taking the norm we obtain
\begin{equation}\label{eq:Helffer}
\| (\chi - \wh \chi) \psi_{N,*} \| \leq \frac{C\kappa}{N} \int \rd
x \, \rd y \, \frac{|\partial_{\bar z } \wt \chi (z)|}{|y|} \,
\Big\| (H_N - \wh H_N ) \frac{1}{z - (\kappa \wh H_N/N)}\psi_{N,*}
\Big\| \, .
\end{equation}
Next we note that \[ H_N -\wh H_N = -\sum_{j=1}^k \Delta_j +
\frac{1}{N} \sum_{i\leq k, i < j \leq N} V_N (x_i -x_j) \, .\]
Therefore, using the symmetry of $\psi_{N,*}$ and of $\wh H_N$
w.r.t. permutations of the first $k$ and the last $N-k$ particles,
we obtain
\begin{equation}
\begin{split}
\Big\| (H_N -\wh H_N) \frac{1}{z - (\kappa \wh H_N / N)}
\psi_{N,*} \Big\| \leq \; &k \, \Big\| \Delta_1 \frac{1}{z-(\kappa
\wh H_N / N)} \psi_{N,*} \Big\| \\ &+ \frac{k^2}{N}
 \Big\| V_N (x_1 - x_2) \frac{1}{z - (\kappa \wh H_N / N)} \psi_{N,*} \Big \| \\
&+ \frac{k (N-k)}{N}  \Big\| V_N (x_1 - x_{k+1}) \frac{1}{z-
(\kappa \wh H_N /N)} \psi_{N,*}  \Big\| \, .
\end{split}
\end{equation}
Using Lemma \ref{lm:sob}, we can bound
\begin{equation}
 \Big\| V_N (x_1 -x_j) \frac{1}{z- (\kappa \wh H_N/N)} \psi_{N,*} \Big \| \leq
\mbox{(const.)}N^{3\beta/2} \| V^2 \|_{L^1}  \Big\| (1 -\Delta_1)
\frac{1}{z-(\kappa \wh H_N/N)} \psi_{N,*}  \Big\|\, .
\end{equation}
Since $\wh H_N$ does not depend on the variable $x_1$, we can
commute the derivatives with respect to $x_1$ through the
resolvent $(z - (\kappa \wh H_N /N))^{-1}$ and we conclude that
\begin{equation}
 \Big\| (H_N -\wh H_N) \frac{1}{z - (\kappa \wh H_N /N)} \psi_{N,*}  \Big\| \leq
\frac{C N^{3\beta/2}}{|y|} \| \ph^* \|_{H^2}
\end{equation}
for a constant $C$ which depends on $k$, but is independent of $N$
and $\kappa$. Inserting this bound into (\ref{eq:Helffer}) we
obtain $\| (\chi - \wh \chi ) \psi_{N,*} \| \leq  C N^{3\beta/2 -
1}$ and thus, since we assumed $\beta <2/3$ , we have, by
(\ref{eq:g-g}),
\begin{equation}
\Big| \tr\; J^{(k)} \left( \wt \gamma^{(k)}_{N,*} - \wh
\gamma^{(k)}_{N,*} \right) \Big| \leq \e/3
\end{equation}
for $N$ sufficiently large (uniformly in $\kappa$). Together with
(\ref{eq:cut1}) and (\ref{eq:cut3}), this completes the proof of
part iii).
\end{proof}

\section{A-Priori Estimate}
\setcounter{equation}{0}

The aim of this section is to prove an a-priori bound for the
solution $\psi_{N,t}^{\kappa}$ of the Schr\"odinger equation
(\ref{eq:schr}) with initial data $\psi_N^{\kappa}$ (as defined in
Proposition \ref{prop:initialdata}). Introduce the operator
$$
   S_j : = (1-\Delta_j)^{1/2} \; .
$$
The a-priori bound is an
estimate of the form
\begin{equation}\label{eq:apripsi}
(\psi_{N,t}^{\kappa} , (1- \Delta_1) \dots (1- \Delta_k)
\psi_{N,t}^{\kappa} ) = \int \rd \bx \, |(1-\Delta_1)^{1/2} \dots
(1 -\Delta_k)^{1/2} \psi_{N,t}^{\kappa}|^2 \leq \wt C^k
\end{equation}
for all $k \geq 1$, uniformly in $t \in \bR$ and in $N$, for $N$
large enough (the constant $\wt C$ depends on $\kappa$, but is
independent of $N$, $t$ and $k$). With density matrix notation,
\ref{eq:apripsi} is equivalent to
$$
   \tr \; S_1 S_2\ldots S_k \wt\gamma_{N, t}^{(k)} S_k \ldots S_2 S_1
 \leq \wt C^k \; .
$$
To prove (\ref{eq:apripsi}), we make use of the following energy
estimate, which gives an upper bound on the mixed derivative
operator in terms of higher powers of the Hamiltonian $H_N$.

\begin{proposition}[Energy Estimate]\label{prop:lower}
Let $H_N$ be defined as in (\ref{eq:ham}), with $V$ smooth and
positive, and with $0 < \beta <3/5$.  Define
\[ \bar{H}_N := \sum_{j=1}^N S_j^2 + \frac{1}{N} \sum_{\ell \neq
m} V_N (x_{\ell} - x_m) = H_N + N .\]
 Fix $k\in
\bN$ and $0 < C <1$. Then there is $N_0 = N_0 (k)$ such that
\begin{equation}
   (\psi, S_1^2 S_2^2
\dots S_k^2 \psi)\leq \frac{2^k}{N^k}\, (\psi, \bar{H}_N^k \psi )
\end{equation}
for all $N > N_0$ and all $\psi \in L_s^2 (\bR^{3N})$ (the
subspace of $L^2 (\bR^{3N})$ containing all permutation symmetric
functions).
\end{proposition}
\begin{proof} The proof of this proposition can be found in
\cite{EESY}. The constant $2$ could
 be replaced by any constant bigger than 1 at the expense of increasing $N_0$.
\end{proof}

Using this energy estimate, the conservation of the energy along
the time evolution, and the fact that at time zero,
$(\psi_N^{\kappa} , H_N^k \psi_N^{\kappa}) \leq \wt C^k N^k$ by the
choice of $\psi_N^{\kappa}$ (see Proposition
\ref{prop:initialdata}), we obtain, in the next theorem, the bound
(\ref{eq:apripsi}).

\begin{theorem}[A-Priori Estimate]\label{thm:apriori}
Let $H_N$ be defined as in (\ref{eq:ham}), with $V$ smooth and
positive and with $0 < \beta <3/5$. Fix $\ph \in H^1 (\bR^3)$
with $\| \ph \| =1$ and
$\kappa
>0$ sufficiently small, $\kappa\leq \kappa_0(\|\varphi\|_{H^1},
V)$. Let $\psi_N (\bx) = \prod_{j=1}^N \ph (x_j)$ and
$\psi_N^{\kappa} = \chi (\kappa H_N /N)\psi_N / \|\chi (\kappa H_N
/N)\psi_N \|$. Suppose that $\psi_{N,t}^{\kappa}$ is the solution
of the Schr\"odinger equation (\ref{eq:schr}) with initial data
$\psi_N^{\kappa}$. Then, for every $k \geq 1$, there exists $N_0 =
N_0 (k)$, with
\begin{equation}\label{eq:aprioripsi}
(\psi^{\kappa}_{N,t} , \, S_1^2 S_2^2 \dots S_k^2 \,
\psi^{\kappa}_{N,t}) \leq \wt C^k
\end{equation}
for all $N \geq N_0$, and for all $t \in \bR$. The constant $\wt
C$ depends on the unscaled potential $V(x)$, on the $H^1$-norm of
$\ph$, and on the cutoff $\kappa >0$ (actually, $\wt C$ is
proportional to $1/\kappa$ for small $\kappa$), but it is
independent of $t \in \bR$, $N$ and $k$. If we denote by $\wt
\gamma_{N,t}^{(k)}$ the marginal densities associated with the
wave function $\psi_{N,t}^{\kappa}$, then (\ref{eq:aprioripsi}) is
equivalent to the bound
\begin{equation}\label{eq:apriori-gamma}
\tr \; | S_1 \dots S_k \wt\gamma_{N,t}^{(k)} S_k \dots S_1 | \leq
\wt C^k
\end{equation}
for all $N \geq N_0$.
\end{theorem}
\begin{proof}
For fixed $k \geq 1$, using $\bar H_N^k\leq 2^k(H_N^k + N^k)$, it
follows from Proposition \ref{prop:lower} that
\begin{equation}
(\psi^{\kappa}_{N,t}, S_1^2 \dots S_k^2 \,
\psi^{\kappa}_{N,t}) \leq \frac{4^k}{N^k} (\psi^{\kappa}_{N,t}, H_N^k
\psi^{\kappa}_{N,t}) + 2^k \|\psi^{\kappa}_{N,t}\|^2
 =  \frac{4^k}{N^k} ( \psi^{\kappa}_{N}, H_N^k
\psi^{\kappa}_{N}) + 2^k
\end{equation}
for all $N \geq N_0(k)$. Here we used that the energy and the
$L^2$-norm are conserved along the time evolution. {F}rom
Proposition \ref{prop:initialdata}, we obtain
\begin{equation}
 ( \psi^{\kappa}_{N,t}, S_1^2 \dots S_k^2 \,
\psi^{\kappa}_{N,t}) \leq \frac{8^k }{\kappa^k}
\frac{1}{1-C\kappa^{1/2}} + 2^k
\end{equation}
which completes the proof of the theorem (assuming that $0 <
\kappa < 1/C^2$).
\end{proof}

\section{Compactness of $\wt \gamma^{(k)}_{N,t}$}
\setcounter{equation}{0}

In this section we keep the cutoff $\kappa >0$ fixed. We prove
that, for fixed $k \geq 1$, $\wt \gamma^{(k)}_{N,t}$ defines a
compact sequence in $C( [0,T], \cH_k)$ (recall that $\wt
\gamma^{(k)}_{N,t}$ is the $k$-particle marginal density
associated with the wave function $\psi_{N,t}^{\kappa}$). To
establish this result we prove the equicontinuity of $\wt
\gamma^{(k)}_{N,t}$ in $t \in [0,T]$, and then we apply the
Arzela-Ascoli theorem. To prove the equicontinuity of $\wt
\gamma^{(k)}_{N,t}$ we use a simple criterium, stated at the end
of this section, in Lemma \ref{lm:equi}.

\begin{theorem}\label{thm:compact}
Fix $k \geq 1$, $T \geq 0$, and $\kappa >0$ small enough. Let $\wt
\gamma^{(k)}_{N,t}$ be the $k$-particle marginal distribution
associated with the solution $\psi_{N,t}^{\kappa}$ of the
Schr\"odinger equation (\ref{eq:schr}), with regularized initial
data $\psi_N^{\kappa}$ (see Proposition \ref{prop:initialdata} for
the definition of $\psi_N^{\kappa}$). Then we have $\wt
\gamma_{N,t}^{(k)} \in C([0,T], \cH_k)$ for all $N \geq N
(k,\kappa)$ large enough. Moreover, the sequence $\wt
\gamma_{N,t}^{(k)}$ is compact in $C([0,T], \cH_k)$. If $\wt
\gamma^{(k)}_{\infty,t} \in C([0,T],\cH_k)$ is an arbitrary limit
point of $\wt \gamma_{N,t}^{(k)}$, then $\wt
\gamma_{\infty,t}^{(k)}$ is non-negative, symmetric with respect
to permutations (in the sense (\ref{symm2})) and satisfies
\begin{equation}\label{eq:apribound}
\| \wt \gamma_{\infty,t}^{(k)} \|_{\cH_k} = \tr\;  | S_1 \dots S_k
\wt \gamma_{\infty,t}^{(k)} S_k \dots S_1 | \leq \wt C^k
\end{equation}
for all $t \in [0,T]$ and $k \geq 1$ (the constant $C$ is the same
as in (\ref{eq:apriori-gamma}), and depends on the unscaled
potential $V(x)$, on the $H^1$ norm of $\ph$ and on the cutoff
$\kappa >0$, but it is independent of $k$).
\end{theorem}
\begin{proof}
We prove that the sequence $\wt \gamma_{N,t}^{(k)}$ is
equicontinuous in $t$, for $t\in [0,T]$. To this end we define the
following dense subset of $\cA_k$:
$$
 \cJ_k: = \{ J^{(k)} \in \cK_k\; : \; \|S_i
S_j J^{(k)} S_i^{-1} S_j^{-1}\|<\infty, \; 1 \leq i<j\leq k\} \; .
$$
We will prove that there exists a threshold $N(k,\kappa)$ such
that  for every $\eps >0$ and for every $J^{(k)}\in \cJ_k$ there
exists $\delta >0$ such that
\begin{equation}\label{eq:claim}
\sup_{N\ge N(k, \kappa)}
\Big| \tr \, J^{(k)} \left(\wt \gamma_{N,t}^{(k)} - \wt
\gamma_{N,s}^{(k)}\right) \Big| \leq \eps
\end{equation}
for all $t,s \in [0,T]$ with $|t-s| \leq \delta$. Combining this
with  Lemma \ref{lm:equi} below, we will obtain the
equicontinuity.

In order to prove (\ref{eq:claim}), we use the BBGKY hierarchy
(\ref{eq:BBGKY1}), rewritten in the integral form
\begin{equation}
\begin{split}
\wt \gamma_{N,t}^{(k)} - \wt \gamma_{N,s}^{(k)} = &\; -i
\sum_{j=1} \int_s^t \rd \tau \, [-\Delta_j , \wt
\gamma_{N,\tau}^{(k)}] -\frac{i}{N} \sum_{i<j}^k \int_s^t \rd \tau
\, [ V_N (x_i -x_j) , \wt \gamma_{N,\tau}^{(k)}]  \\ &-i
\left(1-\frac{k}{N}\right) \sum_{j=1}^k \int_s^t \rd \tau \,
\tr_{k+1} [V_N (x_j - x_{k+1}) , \wt \gamma^{(k+1)}_{N,\tau} ]
\end{split}
\end{equation}
where we recall the notation $V_N (x) = N^{3\beta} V(N^{\beta}
x)$. Multiplying last equation with $J^{(k)}$ and taking the trace
we get the bound
\begin{equation}
\begin{split}
\Big| \tr \, J^{(k)} &\left( \wt \gamma_{N,t}^{(k)} - \wt
\gamma_{N,s}^{(k)} \right) \Big| \leq\sum_{j=1}^k \int_s^t \rd
\tau \, \Big| \tr \; \left( S_j^{-1} J^{(k)} S_j - S_j J^{(k)}
S_j^{-1} \right) \, S_j \wt \gamma_{N,\tau}^{(k)} S_j \Big| \\ &+
\frac{1}{N} \sum_{i<j}^k \int_s^t \rd \tau \, \Big| \tr \; \left(
S^{-1}_i S^{-1}_j J^{(k)} S_i S_j S_i^{-1} S_j^{-1} V_N (x_i -x_j)
S_i^{-1} S_j^{-1} S_i S_j \wt \gamma_{N,\tau} S_i S_j \right. \\
&\hspace{3cm} \left. - S_i S_j J^{(k)} S^{-1}_i S^{-1}_j S_i S_j
\wt \gamma_{N,\tau} S_i
S_j S_i^{-1} S_j^{-1} V_N (x_i -x_j) S_i^{-1} S_j^{-1} \right) \Big| \\
&+\left(1-\frac{k}{N} \right) \sum_{j=1}^k \int_s^t \rd \tau \,
\Big| \tr \; \left( S_j^{-1} J^{(k)} S_j S_j^{-1} S_{k+1}^{-1} V_N
(x_j - x_{k+1}) S_{k+1}^{-1} S_j^{-1} S_{k+1} S_j \wt
\gamma_{N,\tau}^{(k+1)} S_j S_{k+1} \right. \\ &\hspace{3cm}
\left. - S_j J^{(k)} S_j^{-1} S_j S_{k+1} \wt
\gamma_{N,\tau}^{(k)} S_{k+1} S_j S_j^{-1} S_{k+1}^{-1} V_N (x_j
-x_{k+1}) S_{k+1}^{-1} S_j^{-1}\right) \Big|
\end{split}
\end{equation}
using that $S_{k+1}$ commutes with $J^{(k)}$. Using that
$\|S_i^{-1} S_j^{-1} V_N (x_i -x_j) S_i^{-1} S_j^{-1} \|$ is
finite, uniformly in $N$ (see Lemma \ref{lm:sob}), and the
assumption that $S_i S_j J^{(k)} S_i^{-1} S_j^{-1}$ is bounded,
for every $i,j \leq k$, we find
\begin{equation}
\begin{split}
\Big| \tr \, J^{(k)} \left( \wt \gamma_{N,t}^{(k)} - \wt
\gamma_{N,s}^{(k)} \right) \Big| \leq \; &2k |t-s| \sup_{j\leq k}
\| S_j^{-1} J^{(k)} S_j \| \; \sup_{j\leq k, \tau \in [s,t]} \tr
\; | S_j \wt \gamma_{N,\tau}^{(k)} S_j| \\ &+ k^2 N^{-1} |t-s| \;
\sup_{i,j \leq k} \| S_i S_j J^{(k)} S_i^{-1} S_j^{-1} \| \,
\sup_{i < j \leq k, \tau \in [s,t]} \tr \; | S_i S_j
\wt\gamma^{(k)}_{N,\tau} S_j S_i| \\ &+ 2k \Big(1-\frac{k}{N}\Big) \sup_{j\leq k}
\| S_j^{-1} J^{(k)} S_j \| \; \sup_{j \leq k, \tau \in [s,t]} \tr
\; |S_j S_{k+1} \wt \gamma^{(k+1)}_{N,\tau} S_{k+1} S_j|
\end{split}
\end{equation}
and thus, by Theorem \ref{thm:apriori},
\begin{equation}
\Big| \tr \, J^{(k)} \left( \wt \gamma_{N,t}^{(k)} - \wt
\gamma_{N,s}^{(k)} \right) \Big| \leq C_k \, |t-s|
\end{equation}
for a constant $C_k$ depending on $k$ and $J^{(k)}$ (but
independent of $t,s$ and $N$) and for all $N$ large enough
(depending on $k$ and on the cutoff $\kappa>0$). This implies
(\ref{eq:claim}) and, by Lemma \ref{lm:equi}, it implies that the
sequence $\wt \gamma^{(k)}_{N,t} \in C([0,T], \cH_k)$ is
equicontinuous in $t$ (with respect to the metric $\rho_k$ defined
on $\cH_k$). Since moreover the sequence $\wt \gamma^{(k)}_{N,t}$
is uniformly bounded in $\cH_k$ (for $N$ sufficiently large, by
Theorem \ref{thm:apriori}: note that here $\kappa >0$ is fixed),
it follows by the Arzela-Ascoli Theorem that it is compact.

\medskip

To prove that an arbitrary limit point $\wt
\gamma^{(k)}_{\infty,t}$ of the non-negative sequence $\wt
\gamma_{N,t}^{(k)}$ is also non-negative, we observe that, for an
arbitrary $\ph \in L^2 (\bR^{3k})$ with $\| \ph \| =1$, the
orthogonal projection $P_{\ph} = |\ph \rangle\langle \ph |$ is in
$\cA_k$ and therefore we have
\begin{equation}
\begin{split}
\langle \ph , \wt \gamma^{(k)}_{\infty,t} \ph \rangle = \tr \;
P_{\ph} \wt \gamma^{(k)}_{\infty,t} = \lim_{j \to \infty} \tr \;
P_{\ph} \wt \gamma^{(k)}_{N_j,t} = \lim_{j \to \infty} \langle
\ph, \wt \gamma^{(k)}_{N_j ,t} \ph \rangle \geq 0 \, ,
\end{split}
\end{equation}
for an appropriate subsequence $N_j$ with $N_j \to \infty$ as $j
\to \infty$. Similarly, the symmetry of $\wt \gamma^{(k)}_{\infty,t}$
w.r.t. permutations is inherited from the symmetry of $\wt
\gamma^{(k)}_{N,t}$ for finite $N$. In fact, for an arbitrary
$J^{(k)} \in \cA_k$ and a permutation $\pi \in \cS_k$, we have
\begin{equation}
\begin{split}
\tr \; J^{(k)} \wt \gamma^{(k)}_{\infty,t} &= \lim_{j \to \infty}
\; J^{(k)} \wt \gamma^{(k)}_{N_j ,t} = \lim_{j \to \infty} \tr \;
J^{(k)} \Theta_{\pi} \wt \gamma^{(k)}_{N_j ,t} \Theta_{\pi^{-1}} =
\lim_{j \to \infty} \tr \; \Theta_{\pi^{-1}} J^{(k)} \Theta_{\pi}
\wt \gamma_{N_j ,t}^{(k)} \\ &= \tr \; \Theta_{\pi^{-1}} J^{(k)}
\Theta_{\pi} \wt \gamma_{\infty,t}^{(k)} = \tr \; J^{(k)}
\Theta_{\pi} \wt \gamma_{\infty,t}^{(k)} \Theta_{\pi^{-1}}
\end{split}
\end{equation}
where we used that, since $J^{(k)} \in \cA_k$, also
$\Theta_{\pi^{-1}} J^{(k)} \Theta_{\pi} \in \cA_k$, because
\be
\begin{split} \|\Theta_{\pi^{-1}} J^{(k)} \Theta_{\pi} \|_{\cA_k}
&= \| S_1^{-1} \dots S_k^{-1} \Theta_{\pi^{-1}} J^{(k)}
\Theta_{\pi} S_k^{-1} \dots S_1^{-1} \| \\ &= \| \Theta_{\pi^{-1}}
S_1^{-1} \dots S_k^{-1} J^{(k)} S_k^{-1} \dots S_1^{-1}
\Theta_{\pi^{-1}} \| \\& = \| S_1^{-1} \dots S_k^{-1} J^{(k)}
S_k^{-1} \dots S_1^{-1} \|= \| J^{(k)} \|_{\cA_k} \,.
\end{split}\ee
 Finally, the bound (\ref{eq:apribound}) follows
because in the weak limit the norm can only decrease.
\end{proof}

\begin{lemma}\label{lm:equi}  Fix $k$.
A sequence of time-dependent density matrices $\gamma_{N,
t}^{(k)}$, $N=1,2, \ldots$, defined for  $t \in [0,T]$ and
satisfying
 \begin{equation}\label{eq:rec}
\sup_{N\ge 1}\sup_{t\in [0,T]} \|
\gamma_{N,t}^{(k)} \|_{\cH_k} \leq C \; ,
\end{equation}
 is equicontinuous in $C([0,T], \cH_k)$ with respect to the metric
$\rho_k$ (defined in (\ref{eq:rho})), if and only if there exists
a dense subset $\cJ_k$ of $\cA_k$ such that for any $J^{(k)}\in
\cJ_k$ and for every $\eps
>0$ there exists a $\delta > 0$ such that
\begin{equation}\label{eq:equi02}
\sup_{N\ge 1}\Big| \tr \;  J^{(k)} \left( \gamma_{N,t}^{(k)} -
\gamma_{N,s}^{(k)} \right) \Big| \leq \eps
\end{equation}
for all $t,s \in [0,T]$ with $|t -s| \leq \delta$.
\end{lemma}

\begin{proof}
The proof of this lemma is similar to the proof of Lemma 9.2 in
\cite{ESY}; the main difference is that here we keep $k$
fixed, while in \cite{ESY} we considered equicontinuity in the
direct sum $C([0,T], \cH) = \oplus_{k \geq 1} C([0,T], \cH_k)$
over all $k \geq 1$.
\end{proof}

\section{Convergence to solutions of the Gross-Pitaevskii hierarchy}
\label{sec:conv}\setcounter{equation}{0}

{F}rom Theorem \ref{thm:compact} and from the Cantor
diagonalization argument explained in Step 2 of the proof of
Theorem \ref{thm:main}, we know that the the sequence $\wt
\Gamma_{N,t} = \{ \wt\gamma^{(k)}_{N,t}\}_{k=1}^N$ has at least
one limit point $\wt \Gamma_{\infty,t} = \{ \wt
\gamma_{\infty,t}^{(k)} \}_{k \geq 1}$ in $C([0,T], \cH)$ with
respect to the $\tau$-topology.
 In the next theorem, we show that any
such limit point is a solution of the infinite Gross-Pitaevskii
hierarchy (\ref{eq:GPH}) in the integral form
(\ref{eq:BBGKYint1}). The analogous theorem from \cite{EESY}
cannot be directly applied since here we work in $\bR^3$ in
contrast to the compact configuration space of \cite{EESY}.
Moreover, the infinite hierarchy (\ref{eq:BBGKYint1}) is defined
somewhat differently than (1.8) from  \cite{EESY}.

\begin{theorem}\label{thm:convergence}
Assume $H_N$ is defined as in (\ref{eq:ham}), with $0< \beta <
1/2$. For a fixed $\kappa >0$, let $\psi_{N,t}^{\kappa}$ be the
solution of the Schr\"odinger equation (\ref{eq:schr}), with
initial data $\psi_N^{\kappa}$ (defined as in Proposition
\ref{prop:initialdata}), and let $\wt \Gamma_{N,t} = \{ \wt
\gamma_{N,t}^{(k)} \}_{k=1}^N$ be the marginal densities
associated with $\psi_{N,t}^{\kappa}$. Suppose $\wt
\Gamma_{\infty,t} = \{ \wt \gamma_{\infty,t}^{(k)} \}_{k\geq 1}
\in C([0,T], \cH)$ is a limit point of the sequence $\wt
\Gamma_{N,t}$ with respect to the $\tau$-topology.
Then $\wt \Gamma_{\infty,t}$ is a solution of the infinite
hierarchy
\begin{equation}\label{eq:GPH2}
\wt \gamma_{\infty,t}^{(k)} = \cU_0^{(k)} (t) \wt
\gamma_{\infty,0}^{(k)} -ib_0 \sum_{j=1}^k \int_0^t \rd s \;
\cU_0^{(k)} (t-s) \tr_{k+1} \; [ \delta (x_j -x_{k+1}), \wt
\gamma^{(k+1)}_{\infty, s}] \, ,
\end{equation}
with initial data
\begin{equation}\label{eq:initial}
\wt \gamma_{\infty,t=0}^{(k)} (\bx_k ; \bx'_k) = \gamma_0^{(k)}:=
\prod_{j=1}^k \ph (x_j) \overline{\ph} (x'_j) ,
\end{equation}
for all $k \geq 1$.
\end{theorem}

The action of the delta-function in the second term on the r.h.s.
of (\ref{eq:GPH2}) is defined through a limiting procedure. We
define the operator $B^{(k)}$, acting on densities
$\gamma^{(k+1)}$ with smooth kernel, $\gamma^{(k+1)}(\bx_{k+1};
\bx_{k+1}') \in \cS( \bR^{6(k+1)})$ by
\be
B^{(k)} \gamma^{(k+1)} = -ib_0 \sum_{j=1}^k  \tr_{k+1} \; [
\delta (x_j -x_{k+1}), \gamma^{(k+1)}] \, .
\label{def:Bker}
\ee
If we interpret this definition formally for arbitrary density
matrices, then
 the infinite
hierarchy (\ref{eq:GPH2}) can be rewritten in the more compact
form \begin{equation}\label{eq:GPH3}
 \wt \gamma_{\infty,t}^{(k)} =
\cU_0^{(k)} (t) \wt \gamma_{\infty,0}^{(k)} + \int_0^t \rd s \;
\cU_0^{(k)} (t-s) B^{(k)} \wt \gamma^{(k+1)}_{\infty, s} \,.
\end{equation}
The action of $B^{(k)}$ on kernels is formally
given by
\begin{equation}
(B^{(k)} \gamma^{(k+1)} ) (\bx_k;\bx'_k) = -ib_0 \sum_{j =1}^k \int
\rd x_{k+1} \, \left( \delta (x_j -x_{k+1}) - \delta (x'_j
-x_{k+1}) \right) \gamma^{(k+1)} (\bx_k ,x_{k+1} ; \bx'_k ,
x_{k+1})\,.
\end{equation}

For the more precise definition of $B^{(k)}$, we choose a positive
smooth function $h \in C^{\infty} (\bR^3)$, with compact support
and such that $\int \rd x \, h(x) =1$. For $\a >0$, we put
$\delta_{\a} (x) = \a^{-3} h (\a^{-1} x)$. Then, for
$\gamma^{(k+1)} \in \cH_{k+1}$, we put
\begin{equation}\label{eq:Blim}
\begin{split} \left(B^{(k)}
\gamma^{(k+1)}\right) (\bx_k; \bx'_k): = -ib_0 &\lim_{\a_1,\a_2
\to 0} \sum_{j=1}^k \int \rd x_{k+1} \rd x_{k+1}' \, \delta_{\a_2}
(x_{k+1} -x'_{k+1}) \\ &\times
 \left( \delta_{\a_1} (x_j -x_{k+1})  - \delta_{\a_1}
(x'_j -x_{k+1}) \right)\gamma^{(k+1)} (\bx_k ,x_{k+1}
; \bx'_k , x'_{k+1})  .\end{split}
\end{equation}
Lemma \ref{lm:sobsob} below will show that $B^{(k)}$ is well
defined for  any $\gamma^{(k+1)}\in \cH_{k+1}$. We introduce the
norm \be\label{eq:Jnorm} \tri J^{(k)} \tri_{j} := \sup_{\bx_k,
\bx'_k} \la x_1 \ra^4 \dots \la x_k \ra^4 \la x'_1 \ra^4 \dots \la
x'_k \ra^4  \left( |J^{(k)} (\bx_k ; \bx'_k)| + |\nabla_{x_j}
J^{(k)} (\bx_k;\bx'_k)| + |\nabla_{x'_j} J^{(k)} (\bx_k;\bx'_k)|
\right) \, \ee for any $j\leq k$ and for any function
$J^{(k)}(\bx_k ; \bx'_k)$.

\begin{lemma}\label{lm:sobsob}   Suppose that
$\delta_\alpha(x)$ is  a function satisfying
 $0 \leq \delta_{\alpha} (x) \leq C \alpha^{-3} {\bf 1} (|x| \leq
\alpha)$ and $\int \delta_{\alpha} (x) \rd x=1$ (for example
$\delta_{\alpha} (x) = \alpha^{-3} h (x/\alpha)$, for a bounded
probability density $h(x)$ supported in $\{ x : |x| \leq 1\}$).
Then if $\gamma^{(k+1)} (\bx_{k+1};\bx'_{k+1})$ is the kernel of a
density matrix on $L^2 (\bR^{3(k+1)})$, we have, for any $j\leq
k$,
\begin{multline}\label{eq:gammaintbound}
\Big| \int \rd \bx_{k+1} \rd \bx'_{k+1} \, J^{(k)} (\bx_k ;
\bx'_k) \left(\delta_{\alpha_1} (x_{k+1} - x'_{k+1})
\delta_{\alpha_2} (x_j -x_{k+1}) - \delta (x_{k+1} -x'_{k+1})
\delta (x_j - x_{k+1})\right)\\ \times \gamma^{(k+1)} (\bx_{k+1} ;
\bx'_{k+1}) \Big| \\ \leq (\const.)^k \, \tri J^{(k)} \tri_j
\left( \alpha_1 + \sqrt{\alpha_2}\right) \, \tr \, | S_j S_{k+1}
\gamma^{(k+1)} S_j S_{k+1}|\;.
\end{multline}
Recall here that $S_{\ell} = (1-\Delta_{x_{\ell}})^{1/2}$. Exactly
the same bound holds if $x_j$ is replaced with $x_j'$ in
(\ref{eq:gammaintbound}) by symmetry.
\end{lemma}

 This lemma is similar to Proposition 8.1 in \cite{ESY}, with
the difference that here we work in the infinite space $\bR^3$
instead of a compact set $\Lambda$ as in \cite{ESY}. For
completeness we give a proof of Lemma \ref{lm:sobsob} at the end
of Appendix \ref{app}.

\medskip

It follows from the this lemma that the limit (\ref{eq:Blim})
exists for $\gamma^{(k+1)}\in \cH_{k+1}$,
 in an appropriate weak topology, and that it is
independent of the choice of $h \in C^{\infty} (\bR^3)$. Here,
with a slight abuse of the notation,  $\cH_{k+1}$ is used both for
the space of densities defined in Section \ref{sec:topo}, and for
the space of kernels associated with these densities. Hence the
operator $B^{(k)}$, originally defined on Schwarz functions, can
be extended to a bounded operator from the whole $\cH_{k+1}$ and
with values in some sufficiently large Banach space determined by
the conditions on the test function $J^{(k)}$. Moreover, the
following bound holds \be \Bigg| \int \rd \bx_{k} \rd\bx_{k}'
 J^{(k)}(\bx_k;\bx_k') \tr_{k+1} [\delta(x_j
-x_{k+1}),\gamma^{(k+1)}]\Bigg| \leq C^k \tri J^{(k)} \tri_j \,
 \tr |S_jS_{k+1} \gamma^{(k+1)}S_{k+1}S_j|
\label{Bbound} \ee for each term in (\ref{def:Bker}), therefore a
similar bound holds for the operator $B^{(k)}$ as well.

The equality in
(\ref{eq:GPH3}) is then interpreted in the sense that there exists
a representation of \[ \int_0^t \rd s \; \cU_0^{(k)} (t-s) B^{(k)}
\gamma^{(k+1)}_{\infty, s}\] which lies in $\cH_k$ and such that
(\ref{eq:GPH3}) holds. This follows from the fact that both
$\gamma_{\infty,t}^{(k)}$ and $\cU_0^{(k)} (t)
\gamma_{\infty,0}^{(k)}$ are in $\cH_k$ and the equality can be
checked in a weak sense.

\begin{proof}[Proof of Theorem \ref{thm:convergence}.]
Without loss of generality we can assume that $\wt \Gamma_{N,t}$
converges to $\wt \Gamma_{\infty,t}$ with respect to the
$\tau$-topology. This implies that, for every fixed $k\geq 1$ and
$t \in [0,T]$ we have
\[ \wt \gamma_{N,t}^{(k)} \to \wt \gamma^{(k)}_{\infty,t} \] with
respect to the weak* topology of $\cH_k$. That is, for every
$J^{(k)} \in \cA_k$ we have
\begin{equation}\label{eq:conv1}
\tr \; J^{(k)} \, \left(\wt \gamma^{(k)}_{N,t} - \wt
\gamma_{\infty,t}^{(k)} \right) \to 0
\end{equation}
for $N \to \infty$.

Let
$$
   \Omega_k : =  \prod_{j=1}^k \left( \la
x_j \ra + S_j \right)   \quad \quad (\text{with } S_j
= (1-\Delta_{x_j})^{1/2} ) \, .
$$
In the following we assume that the observable $J^{(k)} \in \cK_k
\subset \cA_k$ is such that
\begin{equation}\label{eq:assJ} \Big\| \Omega_k^7
 J^{(k)}\Omega_k^7 \Big\|_{\text{HS}} < \infty ,
\end{equation}
where  $\| A \|_{\text{HS}}$ denotes the Hilbert-Schmidt norm of
the operator $A$, that is $\| A \|^2_{\text{HS}} = \tr A^* A$.
 Note that the set of observables $J^{(k)}$
satisfying the condition (\ref{eq:assJ}) is a dense subset of
$\cA_k$.

It is straightforward to check that
\begin{equation}\label{eq:assJsimple}
\| S_1 \dots S_k \, J^{(k)} \| < \Big\| \Omega_k^7
 J^{(k)}\Omega_k^7 \Big\|_{\text{HS}}, \quad \text{ and }\quad
\| J^{(k)} S_k \dots S_1 \| < \Big\| \Omega_k^7
 J^{(k)}\Omega_k^7 \Big\|_{\text{HS}}  \; .
\end{equation}
Moreover,  for any $j\leq k$
\be \label{HStoj}
    \tri J^{(k)}\tri_j \leq (\mbox{const.})^k \Big\| \Omega_k^7
 J^{(k)}\Omega_k^7 \Big\|_{\text{HS}} ,
\ee where the norm $\tri . \tri_j$ is defined in (\ref{eq:Jnorm}).
This follows from the standard Sobolev inequality $\| f\|_\infty
\leq (\const.) \,  \| f \|_{2,2}$ in three dimensions applied to
each variable separately in the form
\begin{equation*}
\begin{split}
\Big(\sup_{x,x'} \; \langle x \rangle^4 \langle x'\rangle^4
|\nabla_x J(x,x')|\Big)^2 &\leq (\const.) \int \rd x \rd x' \Big|
  (1-\Delta_x) \Big[\langle x \rangle^4
 \big(\nabla_x J(x,x') \big)\langle x'\rangle^4\Big]\Big|^2
\\ &\leq (\const.) \; \tr \; (1-\Delta)  \langle x \rangle^4 \nabla \; J
 \, \langle x \rangle^8 \, J^* \;  \nabla^*  \;
 \langle x \rangle^4 \, (1-\Delta)
\\& \leq (\const.) \; \tr \; \Omega^7 J\Omega^{14} J^* \Omega^7
\end{split}
\end{equation*}
with $\Omega = \langle x \rangle + (1-\Delta)^{1/2}$. Similar
estimates are valid for each term in the definition of $\tri \cdot
\tri_j$, for $j \leq k$. Here we  commuted derivatives and the
weights $\langle x \rangle$; the commutators can be estimated
using Schwarz inequalities.

\medskip

Rewriting the BBGKY hierarchy (\ref{eq:BBGKY1}) in integral form,
and multiplying it with $J^{(k)}$ we obtain
\begin{equation}\label{eq:BBGKYint}
\begin{split}
\tr^{(k)} \; J^{(k)} \wt \gamma_{N,t}^{(k)} = \; &\tr^{(k)} \;
J^{(k)} \cU^{(k)} (t) \wt \gamma_{N,0}^{(k)}\\ & -
i\left(1-\frac{k}{N}\right) \sum_{j=1}^k \int_0^t \rd s \,
\tr^{(k)} \, J^{(k)} \cU^{(k)} (t-s) \tr_{k+1} \; [ V_N (x_j
-x_{k+1}) , \wt\gamma_{N,s}^{(k+1)}] \end{split}
\end{equation}
where we recall the notation $V_N (x) = N^{3\beta} V(N^{\beta} x)$
and
\begin{equation}
\cU^{(k)} (t) \gamma^{(k)} = e^{-iH^{(k)}_N t} \gamma^{(k)}
e^{iH^{(k)}_N t}
\end{equation}
with \[ H^{(k)}_N = - \sum_{j=1}^k \Delta_j + \frac{1}{N}
\sum_{i<j}^k V_N (x_i -x_j)\,. \] Here we use the notation
$\tr^{(k)}$ instead of $\tr$ to explicitly stress that we take the
trace over the degrees of freedom of $k$ particles.

The l.h.s. of (\ref{eq:BBGKYint})
clearly converges, as $N \to \infty$, to $\tr^{(k)} \; J^{(k)} \wt
\gamma_{\infty,t}^{(k)}$ (by (\ref{eq:conv1}) and because $J^{(k)}
\in \cK_k \subset \cA_k$ by assumption). As for the first term on
the r.h.s. of (\ref{eq:BBGKYint}) we have
\begin{equation}\label{eq:conv2}
\tr^{(k)} \; J^{(k)} \left( \cU^{(k)} (t) \wt \gamma_{N,0}^{(k)} -
\cU_0^{(k)} (t) \wt \gamma_{\infty,0}^{(k)} \right) \to 0
\end{equation}
for $N \to \infty$. The definition of $\cU_0^{(k)}$ is recalled
from (\ref{eq:freeev}). To prove (\ref{eq:conv2}) we note that
\begin{equation}\label{eq:conv3}
\begin{split}
\tr^{(k)} \; J^{(k)} \left( \cU^{(k)} (t) \wt \gamma_{N,0}^{(k)} -
\cU_0^{(k)} (t) \wt \gamma_{\infty,0}^{(k)} \right) = \;
&\tr^{(k)} \; J^{(k)} \left( \cU^{(k)} (t) - \cU_0^{(k)} (t)
\right) \wt \gamma_{N,0}^{(k)} \\ &+ \tr^{(k)} \; J^{(k)}
\cU_0^{(k)} (t) \left( \wt \gamma_{N,0}^{(k)} - \wt
\gamma_{\infty,0}^{(k)} \right)\,.
\end{split}
\end{equation}
The second term converges to zero, for $N \to \infty$, because, if
$J^{(k)} \in \cA_k$, then also $\cU_0^{(k)} (-t) J^{(k)} \in
\cA_k$, and hence
\[\tr \; J^{(k)} \cU_0^{(k)} (t) \left( \wt
\gamma_{N,0}^{(k)} - \wt \gamma_{\infty,0}^{(k)} \right) = \tr \;
(\cU_0^{(k)} (-t) J^{(k)} ) \left(\wt \gamma_{N,0}^{(k)} - \wt
\gamma_{\infty,0}^{(k)} \right) \to 0 \] as $N \to \infty$. As for
the first term on the r.h.s. of (\ref{eq:conv3}) we have
\begin{equation}
\tr^{(k)} \; J^{(k)} \left( \cU^{(k)} (t) - \cU_0^{(k)} (t)
\right) \wt \gamma_{N,0}^{(k)} =\frac{-i}{N} \sum_{i<j}^k \int_0^t
\rd s \, \tr^{(k)} \, J^{(k)} \cU^{(k)} (t-s) V_{N} (x_i -x_j) \,
\cU_0^{(k)} (s) \wt \gamma_{N,0}^{(k)} \, .
\end{equation}
This implies that
\begin{equation*}
\begin{split}
\left|\tr^{(k)} \, J^{(k)} \, \left( \cU^{(k)} (t) - \cU_0^{(k)}
(t) \right) \wt \gamma^{(k)}_{N,0}\right| &\leq \frac{k^2 t \|
J^{(k)}\|}{N} \sup_{i<j \leq k} \|V_N (x_i -x_j) S_i^{-1}
S_j^{-1}\| \sup_{i<j \leq k} \tr \; |S_i S_j \wt\gamma_{N,0}^{(k)}| \\
&\leq \frac{k^2 t \| J^{(k)} \|}{N^{1-(3\beta/2)}} \tr \; |S_1 S_2
\wt\gamma_{N,0}^{(k)} S_2 S_1| \; , \end{split}
\end{equation*}
where we used $\tr \; |S_i S_j\gamma|\leq \tr \; |S_i S_j\gamma
S_j S_i|$, the permutation symmetry of $\wt\gamma_{N,0}^{(k)}$ and
the bound $\| V_N (x_i -x_j) S_i^{-1} S_j^{-1} \| \leq C
N^{3\beta/2}$ with a constant $C$ that only depends on the
unscaled potential $V(x)$ (see Lemma \ref{lm:sob}). Since $\beta <
1/2<2/3$, we get, from Theorem \ref{thm:apriori},
\begin{equation}
\left|\tr^{(k)} \, J^{(k)} \, \left( \cU^{(k)} (t) - \cU_0^{(k)}
(t) \right) \wt \gamma^{(k)}_{N,0}\right| \to 0
\end{equation}
as $N \to \infty$.

Next we consider the second term on the r.h.s. of
(\ref{eq:BBGKYint}). More precisely we prove that the difference
\begin{multline}
\left(1-\frac{k}{N}\right) \sum_{j=1}^k \int_0^t \rd s \,
\tr^{(k)} \, J^{(k)} \cU^{(k)} (t-s) \tr_{k+1} \; [ V_N (x_j
-x_{k+1}) , \wt\gamma_{N,s}^{(k+1)}] \\
- b_0 \sum_{j=1}^k \int_0^t \rd s \, \tr^{(k)} \, J^{(k)}
\cU_0^{(k)} (t-s) \tr_{k+1} \; [ \delta (x_j -x_{k+1}) , \wt
\gamma_{\infty,s}^{(k+1)}]
\end{multline}
converges to zero, as $N \to \infty$. To this end we write this
difference as the following sum of four terms
\begin{equation}\label{eq:difference}
\begin{split}
&-\frac{k}{N} \sum_{j=1}^k \int_0^t \rd s \, \tr^{(k)} \, J^{(k)}
\cU^{(k)} (t-s) \tr_{k+1} \; [ V_N (x_j -x_{k+1}) ,
\wt\gamma_{N,s}^{(k+1)}] \\
&+ \sum_{j=1}^k \int_0^t \rd s \, \tr^{(k)} \, J^{(k)} \left(
\cU^{(k)} (t-s) - \cU^{(k)}_0 (t-s) \right) \tr_{k+1} \; [ V_N
(x_j -x_{k+1}) , \wt\gamma_{N,s}^{(k+1)}] \\
&+ \sum_{j=1}^k \int_0^t \rd s \, \tr^{(k)} \, J^{(k)} \cU_0^{(k)}
(t-s) \tr_{k+1} \; [ \left( V_N (x_j -x_{k+1}) - b_0 \delta (x_j
-x_{k+1}) \right), \wt\gamma_{N,s}^{(k+1)}] \\
&+ b_0 \sum_{j=1}^k \int_0^t \rd s \, \tr^{(k)} \, J^{(k)}
\cU_0^{(k)} (t-s) \tr_{k+1} \; \Big[ \delta (x_j -x_{k+1}) ,
\left( \wt\gamma_{N,s}^{(k+1)} -\wt \gamma_{\infty,s}^{(k+1)}
\right)\Big]
\end{split}
\end{equation}
and we prove that each one of these terms converges to zero when
$N \to \infty$.

Using that $S_{k+1}$ commutes with $J^{(k)}\cU^{(k)}$, the first
term can be bounded in absolute value by
\begin{multline}
\frac{k}{N} \sum_{j=1}^k \int_0^t \rd s \, \Big| \tr^{(k+1)}
J^{(k)} \cU^{(k)} (t-s) [ S_{k+1}^{-1} V_N (x_j -x_{k+1})
S_{k+1}^{-1} , S_{k+1} \wt \gamma_{N,s}^{(k+1)} S_{k+1}] \Big|
\\ \leq \frac{2 k^2 t \| J^{(k)} \|}{N} \| S_{k+1}^{-1} V_N (x_j
-x_{k+1}) S_{k+1}^{-1} \| \, \sup_{s \in [0,t]} \tr \; S_{k+1} \wt
\gamma_{N,s}^{(k+1)} S_{k+1} \leq \wt C N^{-1 + \beta} \to 0
\end{multline}
as $N \to \infty$. Here we used Theorem \ref{thm:apriori} and that
$\| S_{k+1}^{-1} V_N (x_j -x_{k+1}) S_{k+1}^{-1} \| \leq C
N^{\beta}$, for a constant $C$ which only depends on the unscaled
potential $V(x)$ (by the second statement in Lemma \ref{lm:sob}).
The constant $\wt C$ on the r.h.s. also depends on the cutoff
$\kappa$.

The second term in
(\ref{eq:difference}), can be rewritten as
\begin{equation}
\begin{split}
\sum_{j=1}^k \int_0^t \rd s \, \tr^{(k)} \, J^{(k)} &\left(
\cU^{(k)} (t-s) - \cU^{(k)}_0 (t-s) \right) \tr_{k+1} \; [ V_N
(x_j -x_{k+1}) , \wt\gamma_{N,s}^{(k+1)}] \\ &= -i N^{-1}
\sum_{j=1}^k \sum_{\ell<m}^k \int_0^t \rd s \int_0^{t-s} \rd \tau
\, \tr^{(k+1)} \, J^{(k)} \, \cU_0^{(k)} (t-s-\tau) \\
& \hspace{2cm} \times\left[ V_N (x_{\ell} - x_m), \cU^{(k)} (\tau)
\left[ V_N (x_{j} - x_{k+1}), \wt
\gamma^{(k+1)}_{N,s}\right]\right] \,.
\end{split}
\end{equation}
Expanding the two commutators, we find that the absolute value of
the r.h.s. of the last equation can be estimated by
\begin{equation}
\begin{split}
C N^{-1} \sum_{j=1}^k \sum_{\ell<m}^k \int_0^t \rd s \int_0^{t-s}
\rd \tau \, & \left( \| J^{(k)} S_{\ell} S_m \|  + \| S_{\ell} S_m
J^{(k)} \| \right) \, \| S_{\ell}^{-1} S_m^{-1} V_N (x_{\ell} - x_m) \| \\
&\times \| S_{k+1}^{-1} V_N (x_j - x_{k+1}) S_{k+1}^{-1} S_j^{-1}
\| \, \tr\; |S_j S_{k+1} \wt \gamma^{(k+1)}_{N,s} S_{k+1}| \, .
\end{split}
\end{equation}
 Notice that $\| S_{k+1}^{-1} V_N (x_j - x_{k+1})
S_{k+1}^{-1} S_j^{-1} \| \leq C N^{\beta/2}$ using both
inequalities from Lemma \ref{lm:sob} and a Schwarz estimate.
Combining this with (\ref{eq:assJsimple}), with $\| S_{\ell}^{-1}
S_m^{-1} V_N (x_{\ell} - x_m) \| \leq C N^{3\beta/2}$
 and
\begin{equation}
\tr \; |S_j S_{k+1} \wt \gamma^{(k+1)}_{N,s} S_{k+1}| \leq \tr \;
S_1 S_2 \wt \gamma^{(k+1)}_{N,s} S_2 S_1
\end{equation}
for all $j=1,\dots,k$, by the symmetry of $\wt
\gamma^{(k+1)}_{N,s}$, we find
\begin{equation}
\Big| \sum_{j=1}^k \int_0^t \rd s \, \tr^{(k)} \, J^{(k)} \left(
\cU^{(k)} (t-s) - \cU^{(k)}_0 (t-s) \right) \tr_{k+1} \; [ V_N
(x_j -x_{k+1}) , \wt\gamma_{N,s}^{(k+1)}] \Big| \leq \wt C N^{-1 +
2\beta}
\end{equation}
which converges to zero, as $N \to \infty$, for $\beta <1/2$. We
remark that this is the only step where the more restrictive
$\beta<1/2$ condition is used, the rest of the proof works for
$\beta<3/5$.

Next we consider the third term in (\ref{eq:difference}). Using
the kernel representation of $\gamma^{(k+1)}_{N,s}$, the absolute
value of this term can be estimated by Lemma \ref{lm:sobsob} (with
$\alpha_1 =0$ and $\alpha_2 = N^{-\beta}$) as
\begin{equation}\label{eq:lemmasob}
\begin{split}
b_0 \sum_{j=1}^k &\int_0^t \rd s \, \left\{ \Bigg| \int \rd
\bx_{k} \rd \bx'_k \rd x_{k+1}  \left( \cU_0^{(k)} (s-t) J^{(k)}
\right) (\bx_k ; \bx'_k) \; \right. \\ &\hspace{2.5cm}\times
\left(\frac{1}{b_0} V_N (x_j -x_{k+1}) - \delta (x_j -x_{k+1})
\right) \wt\gamma_{N,s}^{(k+1)} (\bx_k , x_{k+1} ; \bx'_k ,
x_{k+1}) \Bigg| \\ &\hspace{1.5cm}+ \Bigg| \int \rd \bx_{k} \rd
\bx'_k \rd x_{k+1}
\left( \cU_0^{(k)} (s-t) J^{(k)} \right) (\bx_k ; \bx'_k) \;  \\
&\hspace{2.5cm}\left.\times \left(\frac{1}{b_0} V_N (x'_j
-x_{k+1}) - \delta (x'_j -x_{k+1}) \right) \wt\gamma_{N,s}^{(k+1)}
(\bx_k , x_{k+1} ;
\bx'_k , x_{k+1}) \Bigg| \right\}\\
\leq \; &C  N^{-\beta/2} \sup_{j \leq k, s \in [0,t]} \tr \, | S_j
S_{k+1} \wt \gamma_{N,s}^{(k+1)} S_{k+1} S_j| \; \sum_{j=1}^k
\int_0^t \rd s \, \tri \cU^{(k)}_0 (s-t) J^{(k)} \tri_j \, ,
\end{split}
\end{equation}
for a constant $C$ depending on $k$. Here we recall the definition
of the norm $\tri \cdot \tri_j$ from (\ref{eq:Jnorm}). Using the
estimate (\ref{HStoj}), we have
$$
 \tri \cU^{(k)}_0 (s-t) J^{(k)} \tri_j
\leq C \big\| \Omega_k^7 \; \cU^{(k)}_0 (s-t) J^{(k)} \;
\Omega_k^7 \Big\|_{\text{HS}}
$$
with a $k$-dependent constant $C$. Since $e^{i(s-t)p_j^2} \la x_j
\ra^{m} e^{-i(s-t) p_j^2} = \la x_j +2 (s-t) p_j \ra^{m}$, for any
$j=1,\dots,k$, $m\in \bN$, with $p_j = -i\nabla_j$, we obtain that
$$
 \tri \cU^{(k)}_0 (s-t) J^{(k)} \tri_j
\leq C (1 + |t-s|^7)\big\| \Omega_k^7  \; J^{(k)} \;  \Omega_k^7
\Big\|_{\text{HS}} \; .
$$
{F}rom the assumption (\ref{eq:assJ}) on $J^{(k)}$ and the
a-priori control on $\wt \gamma_{N, s}^{(k+1)}$, we obtain that
the r.h.s. of (\ref{eq:lemmasob}) is bounded by $\wt C
N^{-\beta/2}$ with a constant $\wt C$ depending on $k$, on $t$, on
$J^{(k)}$, and on the cutoff $\kappa$. Hence, the third term on
the r.h.s. of (\ref{eq:difference}) converges to zero, as $N \to
\infty$.

Finally, we consider the fourth term in (\ref{eq:difference}). For
fixed $s \in [0,t]$, and $j \leq k$ we have
\begin{equation}
\tr^{(k+1)} \; J^{(k)} \cU_0 (t-s) \delta (x_j -x_{k+1})  \left(
\wt\gamma_{N,s}^{(k+1)} - \wt \gamma_{\infty,s}^{(k+1)} \right)
\to 0
\end{equation}
for $N \to \infty$, because $J^{(k)} \cU_0 (t-s) \delta (x_j
-x_{k+1}) \in \cA_{k+1}$. In fact
\begin{multline}
\| S_1^{-1} \dots S_{k+1}^{-1} J^{(k)}  \cU_0 (t-s) \delta (x_j
-x_{k+1}) S_{k+1}^{-1} \dots S_1^{-1} \| \\ \leq \| J^{(k)} S_j \|
\| S_j^{-1} S_{k+1}^{-1} \delta (x_j - x_{k+1}) S_{k+1}^{-1}
S_j^{-1} \|
\end{multline}
is finite. It follows that the integrand in the fourth term in
(\ref{eq:difference}) converges to zero, for every $s \in [0,t]$,
and every $j \leq k$. Since the integrand is uniformly bounded
using  the uniform (in $s$) apriori estimates on
$\wt\gamma_{N,s}^{(k+1)}$ and $\wt\gamma_{\infty,s}^{(k+1)}$,
 and the uniformity of the $\cA_{k+1}$-norm
of  $J^{(k)} \cU_0 (t-s) \delta (x_j -x_{k+1})$, it follows that
the fourth term converges to zero as well, for $N \to \infty$.
This proves that, for every $t \in [0,T]$, $k \geq 1$ and $J^{(k)}
\in \cK_k$ satisfying (\ref{eq:assJ}), we have
\begin{equation}
\begin{split}
\tr^{(k)} \; J^{(k)} \wt \gamma^{(k)}_{\infty,t} = \; &\tr^{(k)}
\; J^{(k)} \cU_0^{(k)} (t) \wt \gamma^{(k)}_{\infty,0} \\ &-i b_0
\sum_{j=1}^k \int_0^t \rd s \, \tr^{(k)} \; J^{(k)} \cU_0^{(k)}
(t-s) \tr_{k+1} [\delta (x_j - x_{k+1}) , \wt
\gamma_{\infty,s}^{(k+1)}]\,.
\end{split}
\end{equation}
This implies that
\begin{equation}\label{eq:infhier}
\wt \gamma^{(k)}_{\infty,t} = \cU_0^{(k)} (t)
\gamma^{(k)}_{\infty,0} - i b_0 \sum_{j=1}^k \int_0^t \rd s \,
\cU_0^{(k)} (t-s) \tr_{k+1} [ \delta (x_j -x_{k+1}) , \wt
\gamma_{\infty,s}^{(k+1)}] \,
\end{equation}
if we consider $\wt \gamma^{(k)}_{\infty,t}$ as elements of a
large space of density matrices, the dual space of the Banach
space consisting of all sufficiently smooth $J^{(k)}$ (such that
$J^{(k)}$ satisfies (\ref{eq:assJ})). Next since $\wt
\gamma^{(k)}_{\infty,t} \in \cH_k$ and $\cU_0^{(k)} (t) \wt
\gamma^{(k)}_{\infty,0} \in \cH_k$, it follows that also the
second term on the r.h.s. of (\ref{eq:infhier}) lies in $\cH_k$
(or at least it has a representation as element of $\cH_k$), and
that (\ref{eq:infhier}) holds as an equality on $\cH_k$.

\medskip

Finally we prove (\ref{eq:initial}). For arbitrary $N \geq k$ and
$J^{(k)} \in \cK_k$, we have
\begin{equation}\label{eq:initialdata}
\begin{split}
\tr \; J^{(k)} \left( \wt \gamma_{\infty,0}^{(k)} -
\gamma_{0}^{(k)} \right) = \; &\tr \; J^{(k)} \left( \wt
\gamma_{\infty,0}^{(k)} - \wt \gamma_{N,0}^{(k)} \right) +\tr \;
J^{(k)} \left(\wt \gamma_{N,0}^{(k)} - \gamma_{0}^{(k)} \right) \,
.
\end{split}
\end{equation}
{F}rom (\ref{eq:conv1}) (with $t=0$), the first term converges to
zero, for $N \to \infty$. The second term converges to zero, as $N
\to \infty$, by Proposition \ref{prop:initialdata}, part iii).
\end{proof}

\section{Uniqueness of the infinite hierarchy}\label{sec:unique}
\setcounter{equation}{0}

In this section we show the uniqueness of the solution of the
infinite hierarchy (\ref{eq:GPH}). The following theorem is the
main result of this section.

\begin{theorem}\label{thm:unique}
Fix $T > 0$ and $b_0>0$. Suppose $\Gamma_0 = \{ \gamma_0^{(k)}
\}_{k \geq 1}$  is such that $\gamma_0^{(k)}$ is non-negative
and symmetric with respect to permutations (in the sense of
(\ref{symm2})) and it satisfies
\begin{equation}\label{eq:initbound}
\| \gamma_0^{(k)} \|_{\cH_k} = \tr \, | S_1 \dots S_k
\gamma^{(k)}_0 S_k \dots S_1 | \leq C^k
\end{equation}
for all $k \geq 1$ with some constant $C$. Then the infinite
hierarchy
\begin{equation}\label{eq:BBGKYint2}
\gamma_{t}^{(k)} = \cU_0^{(k)} (t) \gamma_{0}^{(k)}
-i b_0\sum_{j=1}^k \int_0^t \rd s \; \cU_0^{(k)} (t-s) \tr_{k+1} \; [
\delta (x_j -x_{k+1}), \gamma^{(k+1)}_{ s}] \, ,
\end{equation}
has at most one solution $\Gamma_t = \{ \gamma^{(k)}_{t} \}_{k
\geq 1} \in C([0,T], \cH)$ with $\Gamma_{t=0} = \Gamma_0$,
such that $\gamma_t^{(k)}$ is non-negative, symmetric with respect
to permutations and satisfies the bound
\begin{equation}\label{eq:tbound}
\| \gamma^{(k)}_{t} \|_{\cH_k} \leq C^k
\end{equation}
for all $k \geq 1$, and $t \in [0,T]$.
\end{theorem}
{\it Remark:} In the proof we set $b_0=1$ for convenience. The
inclusion of $b_0$ modifies all bounds in a trivial way, but it
plays no role in the argument.

In order to prove this theorem, we will expand the solution in a
Duhamel-type series. Recall from Section \ref{sec:conv}, the
formal definition of the operator $B^{(k)}$, given by
\begin{equation}\label{def:B}
B^{(k)} \gamma^{(k+1)} = -i \sum_{j=1}^k \tr_{k+1} [ \delta (x_j
-x_{k+1}) , \gamma^{(k+1)} ] \, .
\end{equation}
On kernels in momentum space $B^{(k)}$ acts according to
\begin{equation}\label{eq:Bk}
\begin{split}
(B^{(k)} \gamma^{(k+1)} ) &(\bp_k;\bp'_k) =  (-i) \sum_{j=1}^k
\int \rd q_{k+1} \rd q'_{k+1} \left( \gamma^{(k+1)} ( p_1, .. ,
p_j - q_{k+1} + q'_{k+1}, .. , p_k, q_{k+1} ; \bp'_k , q'_{k+1})
\right.
\\ &\left. - \gamma^{(k+1)} ( \bp_k, q_{k+1} ; p'_1,\dots, p'_j + q_{k+1} -
q'_{k+1}, \dots, p_k', q'_{k+1}) \right)\, \\
= \; &(-i) \sum_{j=1}^k \int \rd \bq_{k+1} \rd \bq'_{k+1} \left(
\prod_{\ell \neq j}^k \delta (p_{\ell} -q_{\ell}) \delta
(p'_{\ell} - q'_{\ell}) \right) \gamma^{(k+1)} (\bq_{k+1} ; \bq'_{k+1})\\
&\times \left[ \delta (p'_j - q'_j) \delta (p_j - (q_j + q_{k+1} -
q'_{k+1})) - \delta (p_j -q_j) \delta (p'_j - (q'_j + q'_{k+1} -
q_{k+1})) \right]  .
\end{split}
\end{equation}
These definitions are formal: they can be made precise using Lemma
\ref{lm:sobsob}, as explained in Section \ref{sec:conv}. In the
current paper we will work in momentum space, i.e. we apply
(\ref{eq:Bk}) repeatedly and we will show that all integrals are
absolute convergent.

With these notations we can expand the solution $\{ \gamma^{(k)}_t
\}$ of (\ref{eq:BBGKYint2}) for any $n \geq 1$ as
\begin{equation}\label{eq:duhamel}
\begin{split}
\gamma^{(k)}_t = & \; \cU_0^{(k)} (t) \gamma^{(k)}_0 +
\sum_{m=1}^{n-1} \int_0^t \rd s_1 \int_0^{s_1} \rd s_2 \dots
\int_0^{s_{m-1}} \rd s_m  \\&\hspace{1cm} \times \cU_0^{(k)}
(t-s_1) B^{(k)} \cU_0^{(k+1)} (s_1 -s_2)
B^{(k+1)} \dots B^{(k+m-1)} \cU_0^{(k+m)} (s_m) \gamma^{(k+m)}_0 \\
&+ \int_0^t \rd s_1 \int_0^{s_1} \rd s_2 \dots \int_0^{s_{n-1}}
\rd s_n  \\ &\hspace{1cm} \times \cU_0^{(k)} (t-s_1) B^{(k)}
\cU_0^{(k+1)} (s_1 -s_2) B^{(k+1)} \dots \cU_0^{(k+n-1)} (s_{n-1}
-s_n) B^{(k+n-1)} \gamma^{(k+n)}_{s_n} \, .
\end{split}
\end{equation}
The terms in the summation will be called {\it fully expanded}
as they contain only the initial data. The last error term
involves the density matrix at an intermediate time $s_n$.

In Sections \ref{sec:graphs} and \ref{sec:DuhF} below we show how
the terms in this expansion can be written as a sum of
contributions of suitable Feynman graphs. In Section
\ref{sec:boundgraph}, we show how to bound the contributions of
the Feynman graphs. Then, in Section \ref{sec:proofunique}, we use
these bounds to prove Theorem \ref{thm:unique}. Some technical
estimates, used in Section \ref{sec:boundgraph} to bound the
contributions of the Feynman graphs, are shown in Section
\ref{sec:bounds}.

\medskip

{\it Notation.} For the rest of this paper we will mostly work
in Fourier (momentum) space.
We  use
the convention that variables $p, q, r$ always refer
to three dimensional Fourier variables, while
$x, y, z$ are reserved for configuration space  variables.
With this convention, the usual hat indicating the Fourier
transform will be omitted.
For example, the kernel of
a two-particle density matrix $\gamma_0^{(2)}$ in
position space is $\gamma_0^{(2)}(x_1, x_2;x_1',x_2')$;
in momentum space it is given by the Fourier transform
\[
\gamma_0^{(2)} (q_1,q_2; q'_1,q'_2) = \int \rd x_1 \rd x_2 \rd x'_1
\rd x'_2 \; \gamma_0^{(2)} (x_1,x_2;x'_1,x'_2) \, e^{-i (x_1 \cdot
p_1 + x_2 \cdot p_2)} e^{i(x'_1 \cdot p'_1 + x'_2 \cdot p'_2)} \; ,
\]
with the slight abuse of notation of omitting the hat on left hand side.

Furthermore, to avoid $(2\pi)$-factors in the Fourier transform,
we make the convention that the integration
measure for the three dimensional momentum variables $p, q, r$
are always divided by $(2\pi)^3$, i.e.
$$
   \rd p: = \frac{\rd_{\mbox{\tiny Leb}}p}{(2\pi)^3}  \qquad \mbox{for all
three dimensional momentum variables}
$$
where $\rd_{\mbox{\tiny Leb}}$ denotes the usual Lebesgue measure.
We will also use delta functions in momentum space, $\delta(p)$,
and they will correspond to the measure $\rd p$ above, i.e.
$$
    \int f(p)\delta(p-q) \rd p = f(q)
$$
for smooth functions. Delta functions in position space,
$\delta(x)$, remain subordinated to the usual Lebesgue measure.

Similar convention is used for the frequency variables (dual
variables to the time) that will always be denoted by $\alpha$:
$$
   \rd \a: = \frac{\rd_{\mbox{\tiny Leb}}\a}{2\pi}  \qquad \mbox{for all
one dimensional frequency variables}
$$
and to the delta functions involving $\a$-variables.

\subsection{Graphs}
\label{sec:graphs}

Graphical representation is a convenient tool
to bookkeep various terms in the perturbation expansion
of many body systems; the graphs encode the collision
histories of the particles in  a concise form.
We start by introducing {\it classical graphs},
which would appear in a diagrammatic expansion of the solution of
the BBGKY hierarchy of a classical many particle system. Afterwards
we will define {\it quantum graphs}, which are suitable for the
diagrammatic representation of the expansion of a quantum system
like (\ref{eq:duhamel}). The quantum
nature of the expansion requires a doubling of each edge of the
classical graphs, corresponding to the fact that in the density
matrix description each particle is associated with two different
variables ($x_j$ and $x'_j$).  In this paper we will only
use  the quantum graphs. Classical graphs are included only
to facilitate the description of the quantum graphs and to derive
a combinatorial estimate on their numbers.

\subsubsection{Classical graphs}

\begin{figure}
\begin{center}
\epsfig{file=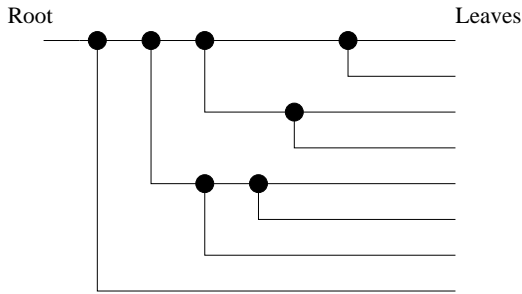,scale=.75}
\end{center}
\caption{Example of a rooted, marked, binary tree with $n=7$
vertices}\label{fig:classic}
\end{figure}

We consider rooted binary trees on $n$ (internal) vertices and
$n+1$ leaves. The root and the  leaves are not considered
vertices; instead we identify them with the unique edge they are
adjacent to. These edges are called {\it external} and we will use
the terminology root and leaves for the external edges. For $n=0$,
i.e. when there is no vertex, there is only one single edge, that
is the root and the single leaf at the same time. However, when
counting the external edges, we will count this edge twice.
 This tree will be called {\it trivial}.
At every vertex three edges meet; the one that is closest to the
root is called {\it father-edge}, the other two are called {\it
son-edges} of this vertex. At every internal vertex we mark one of
the son-edges. (In the expansion this son-edge will inherit the
father's identity.) For illustration,  one can draw such a graph so
that the marked son-edge  goes straight through, and the unmarked
edge joins from below (see Fig. \ref{fig:classic}). The set of
marked binary rooted trees of $n$
 vertices is denoted by
 $\cG_n$.

Note that the number of marked binary trees with $n$ vertices,
$C_n:=|\cG_n|$,  is given by the recursion
$$
  C_{n+1}= \sum_{k=0}^n C_k C_{n-k}, \qquad  C_0=1 \; .
$$
To show this formula, note that removing the vertex closest to the
root splits any marked binary tree into two smaller ones, with $k$
and $n-k$ vertices. This recursion defines the so-called Catalan
numbers. They are given by the closed formula
\[ C_n = \frac{1}{n+1} {2n \choose n} \, ,\]
and can be estimated by $C_n\leq 4^n$.

\begin{figure}
\begin{center}
\epsfig{file=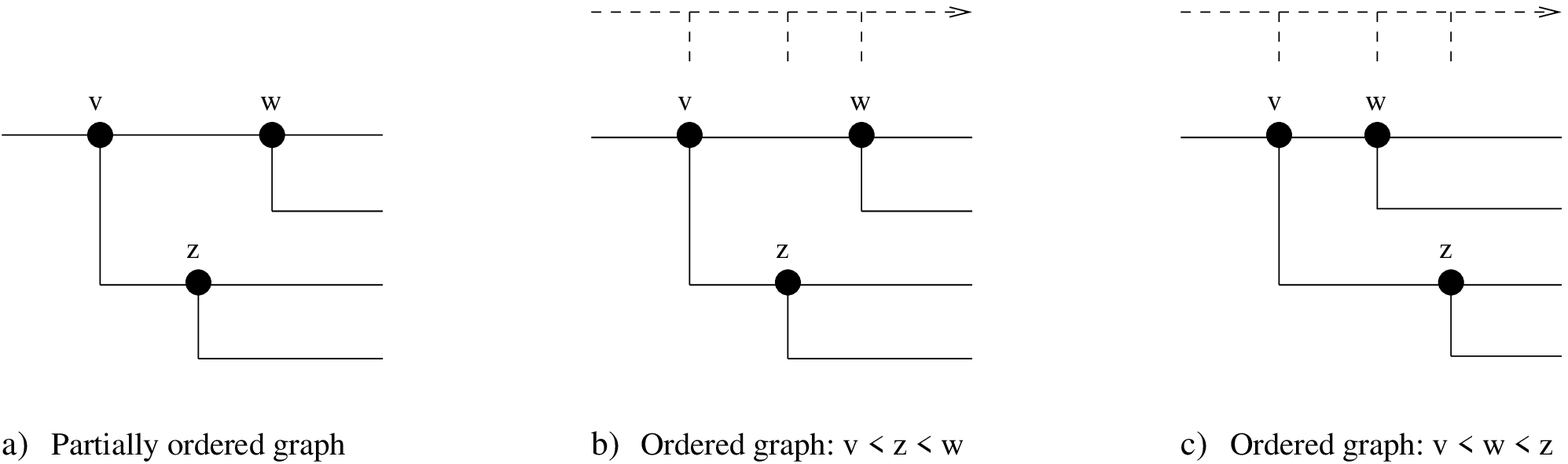,scale=.75}
\end{center}
\caption{Ordering of graphs in $\cG_n$}\label{fig:order}
\end{figure}

For any $G\in \cG_n$, the set of vertices is denoted by $V (G)$
and the set of edges is denoted by $E(G)$. The root is denoted by
$R=R(G)$, the set of $n+1$ leaves is denoted by $L=L(G)$. They
together form the set of external edges, $\text{Ext} (G):=
R(G)\cup L(G)$. The set of internal edges is defined as
$\text{Int} (G):=E(G)\setminus
 \text{Ext} (G)$. We will draw the graphs as in
Fig. \ref{fig:classic}, i.e., the root is on the left, and the
leaves are on the right of the graph. Note that the vertices $V(G)$
are partially ordered by their succession towards the root: for any
$v, v'\in V(G)$  we have $v\prec v'$ if $v$ lies on the (unique)
route from $v'$ to $R$. For any $G\in \cG_n$, we denote by $O(G)$
the set of complete orderings of the vertices $V(G)$ that are
compatible with the partial order $\prec$ of $V(G)$.
 In general, for
a given $G \in \cG_n$, there are several complete orderings which
are compatible with the partial order of $G$. The complete ordering
can be visualized by drawing the graph in such a way that the
horizontal coordinates of the vertices correspond to the ordering
(see Fig. \ref{fig:order}). Two rooted, binary, marked, and
completely ordered trees $G_1$ and $G_2$ are said to be equivalent
if there exists a one-to-one map between the edges and the vertices
of $G_1$ and $G_2$, such that all adjacency relations, all marks and
all labels of the ordered vertices are preserved. The total number
of inequivalent rooted, binary, marked trees with $n$ completely
ordered vertices is $n!$, i.e.
$$
    \sum_{G\in \cG_n} |O(G)|= n!  \; .
$$
This follows from the fact that one can build up such a graph on
$n$ vertices by successively adjoining new leaves: a new leaf will
be joined to an existing leaf in such a way that the existing leaf
keeps its ``father''-identity. In a graph with $k-1$ vertices and
$k$ leaves there are exactly $k$ possibilities to adjoin the new
leaf: one can create a new vertex on any of the existing leaves
and adjoin the new $(k+1)$-th leaf to it as unmarked.

\bigskip

\begin{figure}
\begin{center}
\epsfig{file=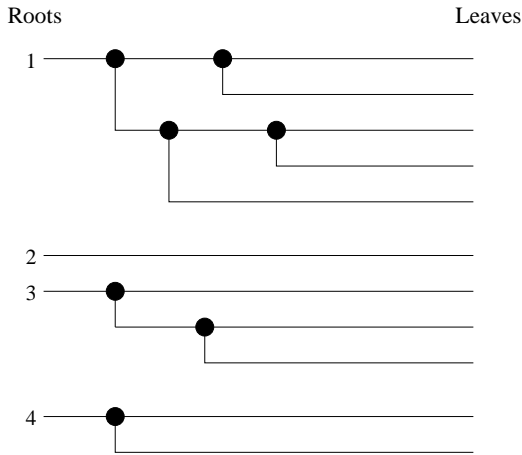,scale=.75}
\end{center}
\caption{Example of forest in $\cG_{n,k}$ with $n=7$ and
$k=4$}\label{fig:forest}
\end{figure}

For $k \geq 1$ and $n \geq 1$ consider forests consisting of $k$
rooted, marked,  binary trees, $G_1, G_2, \ldots, G_k$,
 so that the total
number of (internal) vertices is $n$. We assume that the trees,
i.e. their roots are labelled, i.e. the permutation of the trees
results in inequivalent forests. The set of such forests will be
denoted by $\cG_{n,k}$. In Fig. \ref{fig:forest}, we draw an
example of a graph in $\cG_{7,4}$. Note that the number of
inequivalent forests in $\cG_{n,k}$ is given by
$$
   \sum_{(n_1,\dots, n_k) : \sum_{i=1}^k n_i =n} C_{n_1} C_{n_2} \ldots C_{n_k}
$$
where the summation runs over all $k$-tuples of nonnegative
integers that add up to $n$. This number can be bounded by
$$
    |\cG_{n,k}|\leq 4^n \cdot {n+k-1\choose k-1} \leq 2^{3n+k} \,.
$$

Again, for $G \in \cG_{n,k}$, we will denote by $V(G)$ the set of
the vertices of $G$, by $E(G)$ the set of edges, by $\text{Int}
(G)$, and $\text{Ext} (G)$ the sets of internal and, respectively,
external edges. Moreover, $R(G)$ and $L(G)$ will be used for the
set of roots (there are $k$ roots in each forest), and for the set
of leaves (there are $n+k$ leaves) of $G$, respectively. The
vertices in $V(G)$ are again partially ordered by their succession
towards the roots: for any $v, v'\in V(G)$ within the same tree we
have $v\prec v'$ if $v$ lies on the (unique) route from $v'$ to
the root of the tree. There is no order relation between vertices
in different trees. For  a given $G \in \cG_{n,k}$, we define
$O(G)$ as the set of complete ordering of the $n$ vertices of $G$
which are compatible with the partial order of $G$: the number of
non-equivalent forests with $k$ trees and $n$ completely ordered
vertices is then
$$
   k(k+1)(k+2)\ldots (n+k-1)=\frac{(n+k-1)!}{(k-1)!} \; .
$$
This can again be seen by the successive build-up of the forest;
if one starts with $k$ trivial trees with no vertices, then the
first new leaf can be adjoined in $k$ different ways, the second one
in $(k+1)$ ways etc.

\bigskip

\subsubsection{Quantum (Feynman) Graphs}

For any $n \geq 0$, $k\geq 1$, we now define the set of Feynman
graphs $\cF_{n,k}$. An element $\Gamma\in \cF_{n,k}$ is a union of
$k$ labelled pairs $(T_j, T_j')$, where, for every $j =1,\dots
,k$, $T_j$ and $T'_j$ are two disconnected oriented marked rooted
trees. As in the classical graphs, the roots and the leaves of the
trees are not considered vertices; we identify them with the edges
they are adjacent to. These edges are called external. Each edge
in $T_j$ and $T'_j$ is oriented (indicated by an arrow) and the
edge is called {\it inward} or {\it outward} according to whether
the orientation points towards or away from the root. For each $j$
the root of $T_j$ is outward, the root of $T_j'$ is inward. Each
vertex is adjacent to four edges. At each vertex we require that
precisely two edges are incoming and two edges are outgoing. Note
that the concept of incoming/outgoing is relative to the adjacent
vertex and it does not coincide with the concept of inward/outward
that is solely a property of the edge.

Similarly to the classical
graphs, at each vertex we can introduce the notion of father-edge
and son-edges. The father-edge is on the route from the vertex to
the root, the other three edges are the son-edges. Moreover, two
son-edges have the same orientation as the father-edge and one of
them is distinguished and will be called the {\it marked son-edge}
(we will say that the marked son-edge inherits the identity of the
father-edge).

If the number of vertices of $T_j$ and $T_j'$ is $m_j$ and $m_j'$,
respectively, then the number of external edges is $2m_j+2$ and
$2m_j'+2$, respectively. Recall that for a trivial tree we count
the leaf and the root as separate external edges. We assume that
the total number of vertices is $\sum_{j=1}^k (m_j+m_j')=n$. The
vertices of $\Gamma$ are denoted by $V(\Gamma)$. The edges of
$\Gamma$ are denoted by $E(\Gamma)$, the internal edges are
$\text{Int} (\Gamma)$ and the external ones are $\text{Ext}
(\Gamma)$. For $e \in E(\Gamma)$ and $v \in V(\Gamma)$, the
notation $e \in v$ indicates that the edge $e$ is adjacent to the
vertex $v$. We denote by $R(\Gamma)$ the set of the root edges and
by $L(\Gamma)$ the set of leaf edges. Similarly to the classical
graphs, the fact that each component of $\Gamma$ has a root
induces a partial ordering among the vertices of $\Gamma$, this
will be denoted by $\prec$. Analogously, one can define a partial
ordering among the edges of $\Gamma$, which will also be denoted
by $\prec$.

 Note
that, because of the mark at every vertex, there is a natural
pairing of the leaves of $\Gamma$. To define the pairing, we
introduce the notion of the {\it ancestor} $a(\ell)$ of a leaf
$\ell$ as follows: if $\ell$ is an unmarked son-edge, then the
ancestor of $\ell$ is defined as $\ell$ itself, $a (\ell)=\ell$.
On the other hand, if $\ell$ is a marked son-edge, or if it is the
root, then the ancestor $a(\ell)$ is defined as the minimal edge
$\bar e$ (minimal with respect to the partial order $\prec$) on
the route from $\ell$ to the root, for which the set $\{ e \in
E(\Gamma): \bar e \prec e \prec \ell\}$ contains only marked
son-edges. If we say that at each vertex the marked son-edge
inherits the identity of the father-edge, then the ancestor
$a(\ell)$ is defined as the closest edge to the root whose
identity is inherited by the leaf $\ell$. Clearly, the
ancestor-map $a: L(\Gamma) \to E(\Gamma)$ is injective (any two
leaves have different ancestors) and it maps leaves of $T_j$ or
$T'_j$ into edges of $T_j$ or $T'_j$, respectively (the ancestor
of $\ell$ lies in the same tree as $\ell$). Moreover, for every
$j=1,\dots ,k$, the root of $T_j$ (and of $T'_j$) is always the
ancestor of exactly one leaf in $T_j$ (or $T'_j$, respectively).

Using the concept of ancestor, we define next the pairing of the
leaves of $\Gamma$. For fixed $j$, two leaves $\ell$ and $\ell'$
in $T_j$ (or in $T'_j$) are paired (we say they are {\it companion
leaves}) if $a(\ell)$ and $a(\ell')$ are the unmarked son-edges of
a vertex in $T_j$ (or in $T'_j$, respectively). Moreover, if the
ancestor of a leaf $\ell$ is the root of $T_j$, then we pair
$\ell$ with the unique leaf of $T'_j$ whose ancestor is the root
of $T'_j$. This completely defines a pairing of the leaves of $T_j
\cup T'_j$, and thus of the whole graph $\Gamma$. Note that, if
$T_j$ has $m_j$ vertices, then the number of leaves of $T_j$ is
$2m_j+1$: $2m_j$ leaves are paired within each other and one leaf,
the one with the identity of the root, is paired with the leaf of
$T'_j$ whose ancestor is the root of $T'_j$.

\medskip

For a given graph $\Gamma \in \cF_{n,k}$, we define the labelling
map, $\pi_1 : R(\Gamma) \to \{ 1, \dots ,k \}$, where $\pi_1(r) =
j$ if $r$ is the root of $T_j$ or $T_j'$. For $e \in E(\Gamma)$,
we also introduce the notation $\tau_e=1$ if $e$ is outward and
$\tau_e=-1$ if $e$ is inward. A root and the corresponding
component will be called {\it trivial} if it  contains no
(internal) vertex. For $\Gamma \in \cF_{n,k}$ we denote by $R_1
(\Gamma)$ the set of trivial roots, and we set $R_2 (\Gamma):
=R(\Gamma) \backslash R_1 (\Gamma)$. Let $k_1 (\Gamma) = |R_1
(\Gamma)|$ and $k_2 (\Gamma)= |R_2 (\Gamma)| = 2k - k_1 (\Gamma)$.
Moreover, we define by $E_2 (\Gamma): = E(\Gamma) \backslash R_1
(\Gamma)$, then $|E_2(\Gamma)|= 2k+3n- k_1 (\Gamma)$. Let
$L_1(\Gamma)$ be the set of leaves of the trivial components; this
set is naturally identified with $R_1(\Gamma)$. Let
$L_2(\Gamma):=L(\Gamma)\setminus L_1(\Gamma)$
 be the set of leaves of the non-trivial components, clearly
$|L_2(\Gamma)|= 2n+2k-k_1 (\Gamma)$.

\medskip

When we draw graphs in $\cF_{n,k}$, for every $j=1, \dots ,k$, we
superpose the two trees $T_j$ and $T'_j$; this makes the pairing
of the leaves clearer. As in the classical graphs, at each vertex
we draw the marked son-edge so that it goes straight through, while the
two unmarked son-edges join from below.

\bigskip

Next we will show that Feynman graphs can be constructed in a
natural way starting from the classical ones. Later we will
demonstrate that all quantum graphs can be obtained in this way. As
a by-product we will derive a bound for the number of elements in
$\cF_{n,k}$.

\begin{figure}
\begin{center}
\epsfig{file=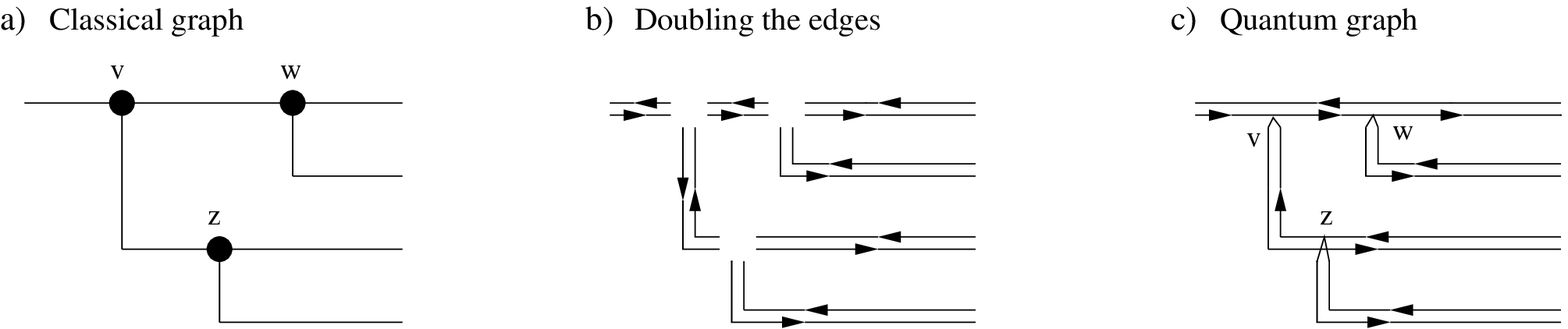,scale=.75}
\end{center}
\caption{{F}rom $\cG_{n,k}$ to $\cF_{n,k}$}\label{fig:quantum}
\end{figure}

For any $G\in \cG_{n,k}$ and $\si=\{ \sigma_v \in \{
\pm 1\} \; : \; v \in V(G)\}$ we define a Feynman graph,
$\Gamma=\Gamma(G, \si)\in \cF_{n,k}$, as follows: We double each
edge of $G$ and equip them with an opposite orientation (arrows).
At any  vertex of $G$ we define the new vertex of the six edges of
$\Gamma$ involved as follows. For $\sigma_v = 1$, the outward
father-edge is joined with both edges of the unmarked son-edge and
with the outward edge of the marked son-edge; this creates a
vertex of $\Gamma$ with four edges. The inward father-edge is
joined with the inward edge of the marked son-edge and we consider
it as a simple continuation, removing the (virtual) vertex.
Analogously, if $\sigma_v =-1$, the inward father-edge is joined
with both edges of the unmarked son-edge and with the inward edge
of the marked son-edge. The outward father-edge is joined with the
outward edge of the marked son-edge (and we consider it as a
simple continuation). In Fig. \ref{fig:quantum} we illustrate this
procedure with an example: given the graph $G \in \cG_{3,1}$ in
a), we first double each edge of $G$ and equip the new edges with
opposite orientation in b). Then we define the new vertices: for
example, if $(\sigma_v,\sigma_w,\sigma_z)= (1,1,-1)$ we obtain the
Feynman graph in drawn in c). Hence, for $G \in \cG_{n,k}$ and
$\si \in \{ \pm1 \}^n$, $\Gamma (G,\si)$ is a Feynman graph with
$n+k$ incoming and $n+k$ outgoing leaves (according to the arrows)
and with $k$ outgoing and $k$ incoming root edges. The vertices of
$\Gamma$, $V(\Gamma)$, have four edges: two incoming and two
outgoing. The set $V(\Gamma)$ is in a natural one-to-one
correspondence with $V(G)$.

\medskip

\begin{figure}
\begin{center}
\epsfig{file=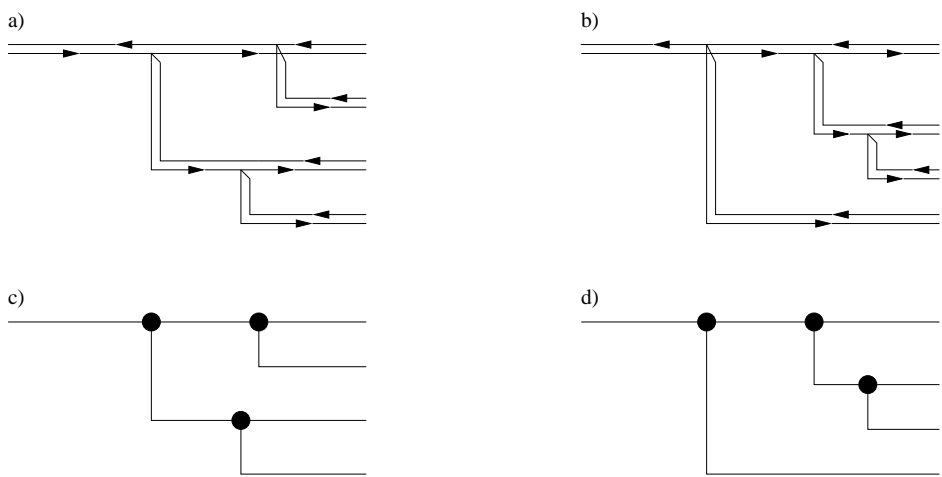,scale=.75}
\end{center}
\caption{The graphs a) and b) are the same element of $\cF_{3,1}$,
but c) and d) are different elements of $\cG_{3,1}$}\label{fig:F0}
\end{figure}

We will now show that every element $\Gamma \in \cF_{n,k}$ can be
obtained as $\Gamma (G,\si)$ for some $G\in \cG_{n,k}$ and $\si
\in \{ \pm 1 \}^n$. This representation is, however, not unique.
The ambiguity comes from the fact that although the vertices of
each component of $G$ are partially ordered (according to their
distance to the root), similarly for the components of $\Gamma$,
but the order $v\prec v'$ in the classical graph for $v, v'\in
V(G)$ is lost in the graph $\Gamma$ if $v$ and $v'$ are assigned
to different components of $\Gamma$. Therefore two different $G$
or $\sigma$ may lead to identical Feynman graphs (see Fig.
\ref{fig:F0}: the Feynman graphs a) and b) are the same element of
$\cF_{3,1}$, but they are obtained from two different classical
graphs in c) and d)).

We can equip each Feynman graph with an extra ordering  to make
this construction unique.
 First we introduce the notion of {\it
orderable pairs of vertices}. If $v, \bar v\in V(\Gamma)$ and they belong
to the same tree-pair but not to the same tree, i.e., $v\in T_j$
and $\bar v\in T_j'$ for some $j$, then the pair $(v, \bar v)$ is
called {\it orderable} if there is a leaf-pair $(\ell, \ell')$
such that $v$ is on the route from $\ell$ to the root of $T_j$ and
$\bar v$ is on the route from $\ell'$ to the root of $T_j'$.

For any $\Gamma\in \cF_{n,k}$ we define the set of orderings
$O(\Gamma)$ on the vertices $V(\Gamma)$ as follows. An ordering
$\prec $ belongs to $O(\Gamma)$ if:

\begin{enumerate}

\item whenever $v, \bar v\in V(\Gamma)$ belong to the same tree,
then $v\prec \bar v$ if and only if $v$ lies on the route from
$\bar v$ to the root of the tree.

\item If $v, \bar v\in V(\Gamma)$ belong to different tree-pairs,
then there is no order relation between them.

\item Any orderable pair $(v, \bar v)$ is ordered by $\prec$.

\end{enumerate}

It is easy to see that $O(\Gamma)$ is not empty; for example the
ordering $v\prec \bar v$ for every orderable pair $v\in T_j, \bar
v\in T_j'$ within the same tree-pair is compatible with the
requirements.

A Feynman graph $\Gamma\in \cF_{n,k}$ equipped with an ordering
$\prec \in O(\Gamma)$ is called {\it ordered Feynman graph}.

It is easy to see that every ordered Feynman graph
 $\Gamma\in\cF_{n,k}^o$ can be uniquely represented
as $\Gamma (G,\si)$ for some $G\in \cG_{n,k}$ and $\si \in \{ \pm
1 \}^n$. Moreover, since for any (unordered) Feynman graph
$\Gamma\in \cF_{n,k}$ the set of orderings $O(\Gamma)$ is not
empty, we obtain in particular, that every Feynman graph
$\Gamma\in \cF_{n,k}$ can be represented as $\Gamma = \Gamma(G,
\sigma)$.

The number of  Feynman graphs therefore
satisfy the bound
\be
|\cF_{n,k}|  \leq 2^n |\cG_{n,k}|\leq 2^{4n+k} \, .
\label{number}
\ee

\subsubsection{Amplitudes of Feynman graphs}

Each Feynman graph $\Gamma$ represents a map acting on density matrices;
it encodes how the initial density matrix changes
as the system undergoes a specific sequence of collisions.
In this section we describe the kernel of this map, commonly
known as the {\it amplitude} of the Feynman graph.

Given an arbitrary quantity $x_e$ defined for $e \in E(\Gamma)$,
and a vertex $v \in V(\Gamma)$, we will use the notation $\sum_{e
\in v} \pm \, x_e$ to indicate that $x_e$ is summed with a plus
sign if the edge $e$ is incoming (w.r.t $v$) while it is summed
with a minus sign if $e$ is outgoing.

For any $\Gamma \in \cF_{n,k}$, we choose a family $\boeta = \{
\eta_e \}_{e \in E(\Gamma)}$, with the property $\eta_e > 0$ for
all $e \in E(\Gamma)$, and such that, at every vertex $v \in
V(\Gamma)$,
\begin{equation}\label{eq:etarule}
\sum_{e \in v} \pm \tau_e \eta_e = 0 \, .
\end{equation}
(Recall that $\tau_e=1$ for outward and $\tau_e=-1$  for inward
edges.) It is easy to check that (\ref{eq:etarule}) is equivalent
to the requirement that the $\eta$ associated with any father-edge
equals the sum of the $\eta$ associated to its son-edges. In
particular, the values of $\eta$ on each of the $2(n+k)$ leaves
uniquely determine $\eta_e$ for all edges $e \in E(\Gamma)$. For a
given $\Gamma \in \cF_{n,k}$, and a given family $\boeta$, we
define the operator $K_{\Gamma,t,\boeta}$ through its kernel
\begin{equation}\label{eq:KGamma2}
\begin{split}
K_{\Gamma,t,\boeta} (\bq_k, &\bq'_k; \br_{n+k} , \br'_{n+k}): =
\frac{1}{(n+k)!} \sum_{\pi_2 \in S_{n+k}} \prod_{e \in R_1 (\Gamma)
= L_1(\Gamma)} (-i\tau_e) e^{-it \tau_e (q_{\pi_1 (e)}^{\sharp_e})^2
} \, \delta (q^{\sharp_e}_{\pi_1 (e)} - r^{\sharp_e}_{\pi_2 (e)}) \\
& \times\int \int_{\bR}\prod_{e \in E_2 (\Gamma)} \rd \alpha_e \rd
p_e \prod_{e \in R_2 (\Gamma)} \delta (p_e - q^{\sharp_e}_{\pi_1
(e)}) \prod_{e \in L_2 (\Gamma)} \delta (p_e -
r^{\sharp_e}_{\pi_2(e)}) \;e^{-it \sum_{e\in R_2 (\Gamma)}
\tau_e(\alpha_e+i\tau_e\eta_e)}
\\ &\times \prod_{e \in E_2 (\Gamma)} \frac{1}{\a_e - p_e^2 +
i\tau_e\eta_e} \prod_{v \in V(\Gamma)} \delta \Big(\sum_{e\in v}
\pm \alpha_e \Big) \delta\Big(\sum_{e\in v}\pm p_e \Big) \, ,
\end{split}
\end{equation}
where $q^{\sharp_e} = q$ if $e$ points away from the root and
$q^{\sharp_e} =q'$ if $e$ points toward the root and similar
notation is used for the $r$ variables. Here the map $\pi_1$
labelling the roots is fixed, since $\Gamma$ is considered as a
graph with labelled roots. The map $\pi_2 : L(\Gamma) \to \{
1,\dots ,n+k \}$ labels the leaves of $\Gamma$ in such a way that
the two elements of a leaf pair receives the same label.
 Since there is no natural order of the
leaves, we sum over all possible labelling $\pi_2$, and we divide
the result by the number of labelling $(n+k)!$. Notice that
$K_{\Gamma,t,\boeta}$ maps operators on $L^2_s( \bR^{3(n+k)})$ into
operators on $L^2_s( \bR^{3k})$ by the formula
$$
    \Big( K_{\Gamma,t,\boeta}\gamma^{(n+k)}\Big) (\bq_k ; \bq_k')=
    \int K_{\Gamma,t,\boeta} (\bq_k, \bq'_k; \br_{n+k} , \br'_{n+k})
    \gamma^{(n+k)}(\br_{n+k} ; \br'_{n+k}) \rd \br_{n+k}
  \rd  \br'_{n+k}
$$
where  $\gamma^{(n+k)}$ is an operator  on $L^2_s( \bR^{3(n+k)})$
given by the  kernel  $\gamma^{(n+k)}(\br_{n+k} ; \br'_{n+k})$ in
Fourier space.

\bigskip

Note that there are $ |R_2|+|L_2|+ |V| = 4k+3n-2k_1$ momentum
delta-functions involving $p_e$ variables and only
$|E_2|=2k+3n-k_1$ momentum integration variables. Together with
the $k_1$ direct delta-functions related to the roots in $R_1
(\Gamma)$, we see that the kernel $K_{\Gamma, t,\boeta}$ contains
$2k$ delta-functions among its $2n+4k$ variables. This corresponds
to the $2k$ momentum conservation in each of the $2k$ components
of $\Gamma$. It can easily be seen that all the $p_e$ momenta can
indeed be uniquely expressed through the external momenta
$\br_{n+k}, \br'_{n+k}, \bq_k, \bq'_k$.  In particular, the $\rd
p_e$ integrations are all well defined and they correspond to
substituting the appropriate linear combinations of the external
momenta into $p_e$.

\medskip

 We will  represent the
fully expanded terms in the Duhamel expansion (\ref{eq:duhamel}) as
a sum of contributions associated with the Feynman graphs. More
precisely, we will show in the next subsection that the $m$-th term
in the sum on the r.h.s. of (\ref{eq:duhamel}) can be rewritten as
the sum of $K_{\Gamma,t,\boeta} \gamma^{(k+m)}_0$ over all $\Gamma
\in \cF_{m,k}$ independent of the choice of $\boeta$.

On the intuitive level, it is possible to
recognize the origin of some of the factors in the formula
\eqref{eq:KGamma2} for the
kernel $K_{\Gamma,t,\boeta}$. The appearance of the one-dimensional
variables $\a_e$ and  the propagators $(\a_e - p_e^2 + i\tau_e
\eta_e)^{-1}$, for example, derives from the free evolution $e^{-it
\tau_e p_e^2}$, expressed as
\begin{equation}\label{1res}
e^{-it\tau_e p_e^2} = (i\tau_e) \int_\bR \rd \a_e \; \frac{e^{-it\tau_e ( \a_e
+i\tau_e \eta_e)}}{\a_e -p_e^2 + i\tau_e \eta_e}\,.
\end{equation}
The presence of the factor $\delta (\sum_{e \in v} \pm p_e)$
(conservation of momentum at every vertex), on the other hand, is
due to the translation invariance of the interaction, and it just
reproduces the kernel of the operators $B^{(k)}$ in momentum space
(see (\ref{eq:Bk})). Finally, one can understand the presence of the
factors $\delta (\sum_{e \in v} \pm \a_e)$ as the result of the
integration over the time-variables, after all the free evolutions
$e^{\pm itp_e^2}$ have been rewritten in terms of  resolvents
according to (\ref{1res}).

\medskip

\begin{figure}
\begin{center}
\epsfig{file=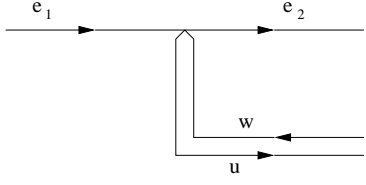,scale=.75}
\end{center}
\caption{Integration over $\a_e$} \label{fig:alpha1}
\end{figure}

The absolute convergence of the $\rd \alpha_e$ integrals
in \eqref{eq:KGamma2} can be
proven by induction on  $n$. For $n=0$ there is no such
integration. For $n=1$, the $\rd \alpha_e$ integrations are of the
form \be \int_\bR \frac{\rd\alpha_{e_1} \rd\alpha_{e_2}\rd\alpha_u
 \rd\alpha_w \;
\delta( \alpha_{e_2} +\alpha_{u}-\alpha_{e_1} - \alpha_{w})}
{(\a_{e_1} - p_{e_1}^2 +i\tau_{e_1}\eta_{e_1}) (\a_{e_2} -
p_{e_2}^2 +i\tau_{e_2}\eta_{e_2}) (\a_{u} - p_{u}^2
+i\tau_{u}\eta_{u}) (\a_{w} - p_{w}^2 +i\tau_{w}\eta_{w})}
\label{vert1} \ee with $\tau_{e_1}=\tau_{e_2}$ and
$\tau_u+\tau_w=0$. Here $e_2$ corresponds to the marked son-edge
of the father-edge $e_1$ in $\Gamma$, and $u$, $w$ correspond to
the doubling of the unmarked son-edge (see Fig. \ref{fig:alpha1}).
Recall that, by definition $\eta_{e_1} = \eta_{e_2} + \eta_w +
\eta_u$, and all the $\eta$'s are strictly positive.

We will use the following inequality with $\eta = \min (\eta_1,
\eta_2)$,
\begin{equation}\label{eq:ineq}
\begin{split}
\int_\bR  \frac{\rd \alpha}{|\alpha-a - i\eta_1| |\alpha-b +
i\eta_2|} &\leq \eta^{-2} \int_\bR  \frac{\rd \alpha}{|
(\alpha-a)/\eta - i| | (\alpha-b)/\eta + i|} \\ & \leq \eta^{-1}
\int_\bR  \frac{\rd y}{|y- \frac{a}{\eta} - i|^{1-\frac{\e}{3}}
|y- \frac{b}{\eta}|^{1-\frac{\e}{3}}}  \leq \frac{C \eta^{-\e}}{
|a - b - i\eta|^{1-\e}}
\end{split}
\end{equation}
for any $a, b\in \bR$, $\eta_1, \eta_2 > 0$ and $0 < \e < 1$ (in
the last inequality we applied Lemma \ref{lm:pre1}).

Using this inequality,  (\ref{vert1}) can be bounded by (with
$\eta: = \min (\eta_{e_2}, \eta_w, \eta_u)$)
\begin{equation}
\begin{split}
   C \eta^{-\e} &\int_\bR \frac{\rd\alpha_{e_1}\rd\alpha_{e_2}}
{|\a_{e_1} - p_{e_1}^2 +i \eta|\; |\a_{e_2} - p_{e_2}^2 +i\eta| \;
|\a_{e_2} -\a_{e_1} +p_u^2-p_w^2 + i\eta|^{1-\e}}\\
   &\leq C^2 \eta^{-2\e} \int_\bR \frac{\rd\alpha_{e_2}}
{|\a_{e_2} - p_{e_2}^2 +i\eta| \; |\a_{e_2} -p_{e_1}^2
+p_u^2-p_w^2 + i\eta|^{1-2\e}}\\
   &\leq  \frac{ C^3 \eta^{-3\e}}
  { |p_{e_2}^2 -p_{e_1}^2 +p_u^2-p_w^2 + i\eta|^{1-3\e}}
  \leq \frac{C}{
   \eta} \label{lasteta}
\end{split}
\end{equation}
where we applied a slightly modified version of (\ref{eq:ineq})
two more times, and where we assumed that $0 < \e <1/3$.

\bigskip

\begin{figure}
\begin{center}
\epsfig{file=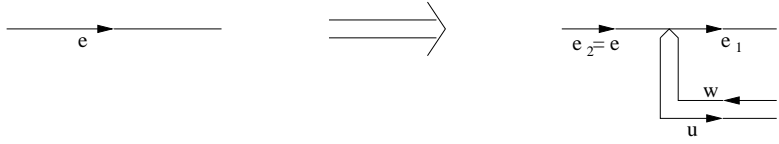,scale=.75}
\end{center}
\caption{Insertion of a new edge}\label{fig:alpha2}
\end{figure}

For a general $n$ we note that any Feynman graph can be built up
from the trivial graph with $2k$ roots by successively adjoining
new vertices to leaves. When we adjoin a new vertex to a Feynman
graph $\Gamma\in \cF_{n-1, k}$, we select a leaf, $e\in
L(\Gamma)$, split it into two edges, $e_1$ and $e=e_2$ ($e_1$
becomes the new leaf and it is equipped with
 the same orientation as $e$)
 and we adjoin two more edges, $u$ and $w$
 of opposite orientation (see Fig. \ref{fig:alpha2}).

We create three new denominators, three new $\alpha$ variables and
one new delta function among them. The additional integration is
\be \int_\bR \frac{\rd\alpha_{e_1} \rd\alpha_{u}
 \rd\alpha_{w}}{(\a_{e_1} - p_{e_1}^2 +i\tau_{e_1}\eta_{e_1})
(\a_{u} - p_{u}^2 +i\tau_{u}\eta_{u}) (\a_{w} - p_{w}^2
+i\tau_{w}\eta_w)} \delta\Big( \alpha_e +\alpha_{u}-\alpha_{e_1} -
\alpha_{w}\Big) \label{vert} \ee
 where the $\eta$'s are
chosen such that $\eta_{e_2} = \eta_{e_1} + \eta_w + \eta_u$. This
integral is absolutely convergent uniformly in $\alpha_e$ and for
any choice of $\eta_{e_1}, \eta_u, \eta_w >0$ and for any choice
of the $p$-variables. This guarantees that the absolute
convergence of all the $\rd\a_e$
 integrals in (\ref{eq:KGamma2})
can be checked with a step by step reduction.

\bigskip

Now we prove that the right hand side of (\ref{eq:KGamma2}) is
independent of the family $\boeta = \{ \eta_e \}_{e \in
E(\Gamma)}$. Notice that, because of the condition that at each
vertex the $\eta$ associated with the father-edge equals the sum
of the $\eta$'s associated with the son-edges, the only
independent $\eta$'s are the ones associated with the leaves of
$\Gamma$. It is moreover clear that, for every fixed $\bar e \in
L(\Gamma)$, $K_{\Gamma,t,{\boeta}}$, as a function of $\eta_{\bar
e}$, has an analytic extension in the whole half plane $\mbox{Re}
\; \eta_{\bar e} > 0$. It is therefore sufficient to show that
$K_{\Gamma,t,{\boeta}}$ is constant in a small neighborhood of a
given value $\eta_{\bar e}$ with $\mbox{Re} \; \eta_{\bar e} >0$
(while all the other $\eta_e>0$, $\bar e \neq e \in L (\Gamma)$,
are kept constant). After replacing $\eta_{\bar e}$ by $\eta_{\bar
e} + \xi$, we can shift the $\a_e$ variables as follows. For $e
\in E(\Gamma)$ on the route from $\bar e$ to the unique root
connected to $\bar e$, we shift $\a_e \to \tilde \a_e = \a_e + i
\tau_e \xi$, while for all other $e \in E(\Gamma)$ we leave
$\tilde \a_e =\a_e$. Here we assume that $|\mbox{Re} \, \xi| < \,
\min_{e \in E(\Gamma)} \mbox{Re}\; \eta_{e}$ to avoid deforming
the $\a_e$ integral contour through the pole at $\a_e = p^2_e
-i\tau_e \eta_e$. Then we can see that the integral remains
unchanged. This follows from observing that for all $v$
$$
   \sum_{e\in v} \pm \alpha_e =  \sum_{e\in v} \pm \tilde \alpha_e
$$
thanks to the definition of $\tau_e$ and the sign convention of
the summation. This proves the independence of (\ref{eq:KGamma2})
from the family $\boeta$. In particular we can use the simpler
notation $K_{\Gamma, t}$.

\bigskip

We note, that one may introduce dual variables $\alpha_e$ for the
trivial components $e\in R_1(\Gamma)$ as well. Using the identity
\eqref{1res},
one then may rewrite the
definition of $K_{\Gamma,t}$ into a more compact form
\begin{equation}\label{simpler}
\begin{split}
K_{\Gamma,t}&( \bq_k ,\bq'_k ; \br_{n+k}, \br'_{n+k}) =
\frac{1}{(n+k)!} \sum_{\pi_2 \in S_{n+k}} \\ &\times \int \int_\bR
\prod_{e\in E (\Gamma)} \rd \alpha_e \rd p_e \, \prod_{e \in
R(\Gamma)} \delta (p_e - q_{\pi_1 (e)}^{\sharp_e}) \prod_{e \in
L(\Gamma)} \delta (p_e - r_{\pi_2 (e)}^{\sharp_e}) \\
\ &\times e^{-it \sum_{e\in R(\Gamma)}
\tau_e(\alpha_e+i\tau_e\eta)} \prod_{e \in E(\Gamma)}
\frac{1}{\a_e - p_e^2 + i\tau_e\eta_e} \prod_{v \in V(\Gamma)}
\delta \Big(\sum_{e\in v}\pm \alpha_e \Big) \delta\Big(\sum_{e\in
v}\pm p_e \Big) \, .
\end{split}
\end{equation}
However, this definition will not result in absolute convergent
integrals, therefore (\ref{1res}) has to be used for each $e\in
R_1(\Gamma)$ before any estimates. With this remark in mind, we
can use the simpler formula (\ref{simpler}).

\bigskip

The amplitudes $K_{\Gamma,t}$ will describe
the terms of the summation in \eqref{eq:duhamel}.
The input of the last term in \eqref{eq:duhamel} is somewhat different
from the previous terms; it involves the
density matrix $\gamma^{(k+n)}$ at an intermediate time $s_n$
and the last free evolution is absent.
We thus introduce a slight modification of the
amplitude $K_{\Gamma,t}$ to represent this term.
We  define the operator $L_{\Gamma,t}$, for $\Gamma\in
\cF_{n,k}$, $n,k\ge1$, through
\begin{equation}\label{Lsimpler}
\begin{split}
L_{\Gamma,t} &( \bq_k ,\bq'_k ; \br_{n+k}, \br'_{n+k}) :=
\frac{1}{(n+k)!}  \sum_{\pi_2 \in S_{n+k}} \sum_{\bar v \in
M(\Gamma)} \sigma_{\bar v} \int \int_\bR\prod_{e\in E (\Gamma)\atop
e \not
\in S_{\bar v}} \rd \alpha_e  \prod_{e \in E (\Gamma)} \rd p_e \, \\
&\times \prod_{e \in R (\Gamma)} \delta (p_e - q_{\pi_1 (e)})
\delta (p_{e'} - q'_{\pi_1 (e)}) \prod_{ e \in L(\Gamma)} \delta
(p_e - r_{\pi_2 (e)}) \delta (p_{e'} - r'_{\pi_2 (e)}) \prod_{v
\in V(\Gamma)
}\delta\Big(\sum_{e\in v}\pm p_e \Big) \\
&\times \exp{\Big( -it \sum_{ e \in R(\Gamma)} \tau_e
(\alpha_e+i\tau_e\eta_e) \Big)} \prod_{e \in E(\Gamma)\atop e \not
\in  S_{\bar v}} \frac{1}{\a_e - p_e^2 + i\tau_e \eta_e}
\prod_{\bar v \neq v \in V(\Gamma)} \delta \Big(\sum_{e\in v}
\pm\alpha_e  \Big) \, ,
\end{split}
\end{equation}
where $M(\Gamma) \subset V(\Gamma)$ is the set of maximal vertices
of $\Gamma$, that is the set of all $v \in V(\Gamma)$ so that
there exists no $\tilde v$ with $\tilde v \succ v$ (recall that
$\prec$ was the partial order on $V(\Gamma)$ induced by the
distance to the root). Moreover, for a vertex $v \in V(\Gamma)$,
we denote by $S_v$ the set of son-edges of the vertex $v$
(clearly, $R(\Gamma) \cap S_{v} = \emptyset$, for any $v \in
V(\Gamma)$), and we set
$$
\sigma_v = \sum_{e \in S_v} \tau_e\; ,
$$
i.e., $\sigma_v =1$ if of the four edges adjacent to $v$, three are
outward and one is inward, $\sigma_v = -1$ otherwise.

 In other
words, $L_{\Gamma,t}$ is defined so that there are no propagators
associated with the son-edges of a $\bar v \in M(\Gamma)$. The
variables $\boeta = \{ \eta_e \}_{e\in E(\Gamma)}$ are chosen as
in (\ref{eq:etarule}): although $\eta_e$, for $e \in S_{\bar v}$,
does not appear directly in (\ref{Lsimpler}), the value of the
$\eta$-variable associated to the father-edge of $\bar v$ depends
on it. Analogously to $K_{\Gamma,t}$, it can be proven that
$L_{\Gamma,t}$ is independent of the choice of $\boeta$. In
(\ref{Lsimpler}) we define $L_{\Gamma,t}$ using a form analogous
to the simpler definition (\ref{simpler}); clearly $L_{\Gamma,t}$
can be rewritten into a form analogous to (\ref{eq:KGamma2}) as
well after integrating out all $\alpha_e$, $e\in R_1(\Gamma)$.

\subsection{Duhamel expansion in terms of Feynman graphs}\label{sec:DuhF}

In this section, we shall prove that the Duhamel expansion for
the solution (\ref{eq:duhamel}) can be expressed as sums over
Feynman graphs. Similar representations were used in the physics
literature and proofs are available for the expansions of the
imaginary time many-fermion Green functions via
the method of (Grassmannian)
functional integrals, see e.g., \cite{Man}. Our relatively
short proof is elementary and does not rely on functional integral.
It also allows for an explicit remainder term.

\begin{theorem}\label{thm:graphs}
Fix $k,n \geq 1$. Then, for any $\gamma_0^{(k+n)}$ that is
symmetric with
respect to permutations (in the sense of (\ref{symm2})), we have
\begin{equation}\label{eq:fullexp}
\begin{split}
\int_0^t \rd s_1 \int_0^{s_1} \rd s_2 \dots &\int_0^{s_{n-1}} \rd
s_n \; \cU_0^{(k)} (t-s_1) B^{(k)} \dots \cU_0^{(k+n-1)} (s_{n-1}
-s_n) B^{(k+n-1)} \cU_0^{(k+n)} (s_n) \gamma_0^{(k+n)} \\ &=
\sum_{\Gamma \in \cF_{n,k}} K_{\Gamma,t} \gamma_0^{(k+n)}\,.
\end{split}
\end{equation}
For $n=0$, $k\ge 1$ we have \be \cU_0^{(k)} (t) \gamma^{(k)}_0 =
\sum_{\Gamma \in \cF_{0,k}} K_{\Gamma,t} \gamma_0^{(k)} \label{n0}
\ee where the summation is only for the trivial graph.

Moreover, for any fixed $k,n \geq 1$, if $\gamma_t^{(k+n)} \in
C([0,T]; \cH_k)$ is symmetric with respect to permutations for all
$t \in [0,T]$, then we have
\begin{equation}\label{eq:error}
\begin{split}
\int_0^t \rd s_1 \int_0^{s_1} \rd s_2 &\dots \int_0^{s_{n-1}} \rd
s_n \; \cU_0^{(k)} (t-s_1) B^{(k)} \dots \cU_0^{(k+n-1)} (s_{n-1}
-s_n) B^{(k+n-1)} \gamma_{s_n}^{(k+n)} \\ &= -i \sum_{\Gamma \in
\cF_{n,k}} \int_0^t \rd s \,  L_{\Gamma,t-s} \gamma_s^{(k+n)}\,
\end{split}
\end{equation}
for all $t \in [0,T]$.
\end{theorem}

Before presenting the proof for the general case, it is
instructive to show how the structure of the
operators $K_{\Gamma, t}$ emerges from the Duhamel expansion.
We thus consider \eqref{eq:fullexp}
in the very simple case $n=1$, $k=1$, i.e.,
\begin{equation}\label{eq:ex-0}
 \int_0^t \rd s \, \cU_0^{(1)} (t-s) B^{(1)} \cU_0^{(2)} (s)
\gamma^{(2)}_0 = K_{\Gamma_1,t} \gamma_0^{(2)}  + K_{\Gamma_2,t}
\gamma_0^{(2)} , \end{equation}
where $\Gamma_1$ and $\Gamma_2$ are
the two elements of $\cF_{1,1}$ drawn in Figure \ref{fig:1,1}.
\begin{figure}
\begin{center}
\epsfig{file=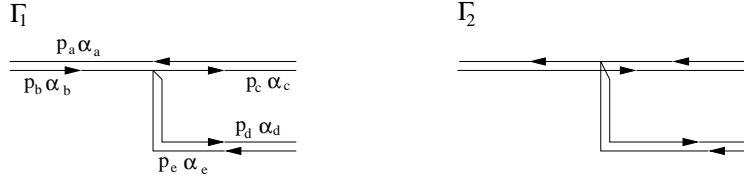,scale=.75}
\end{center}
\caption{The two Feynman graphs in $\cF_{1,1}$}\label{fig:1,1}
\end{figure}

By definition of the map $B^{(1)}$, the l.h.s. of (\ref{eq:ex-0}) is
given by
\begin{equation}\label{eq:ex-1}
-i \int_0^t \rd s \; \cU_0^{(1)} (t-s)  \tr_2 \, \delta (x_1 -x_2)
\, \cU_0^{(2)} (s) \gamma^{(2)}_0 +i \int_0^t \rd s \; \cU_0^{(1)}
(t-s) \tr_2 \,\cU_0^{(2)} (s) \gamma^{(2)}_0 \delta (x_1 -x_2)\; .
\end{equation}
We now show that the first term coincides with the contribution of
the Feynman diagram $\Gamma_1$ (analogously one can prove that the
second term equals the contribution of $\Gamma_2$).

Recall that $K_{\Gamma_1,t} \in \cF_{1,1}$  maps operators on
$L^2_s(\bR^3\times \bR^3)$ into operators on $L^2_s(\bR^3)$. In
momentum space, the kernel of $\gamma_0^{(2)}$ has four variables,
$\gamma_0^{(2)}(p_c, p_d; p_a, p_e)$. These are the momentum
variables on the right hand side of Feynman graph $\Gamma_1$. The
resulting operator, $K_{\Gamma,t}\gamma_0^{(2)}$, acts on
$L^2_s(\bR^3)$, and its kernel  has two variables, $(p_b ; p_a)$
 represented on the l.h.s. of the Feynman
graph.

The pairs of variables, $(p_c ; p_a)$ and $(p_d; p_e)$, in the
argument of $\gamma^{(2)}_0$ correspond to the input and output of
the first and the  second variable of the
$L^2_s(\bR^3\times \bR^3)$ space. Similarly, $p_b$ is the input and
$p_a$ is the output variable of the resulting operator
$K_{\Gamma,t}\gamma_0^{(2)}$: arrows in the Feynman graph pointing
away from the roots indicate input-variables, arrows pointing
towards the roots indicate output-variables of the corresponding
density matrix.

For any kernel $\gamma^{(2)}(p_1, p_2;p'_1, p'_2)$, the free time
evolution in momentum space acts as follows
$$
   \Big(\cU_0^{(2)} (s) \gamma^{(2)}\Big)(p_1, p_2; p_1', p_2')
   = e^{-is(p_1^2 + p_2^2)} e^{is([p_1']^2 + [p_2']^2)}
   \gamma^{(2)}_0(p_1, p_2; p_1', p_2')\,.
$$
The multiplication by $\delta (x_1 -x_2)$ corresponds to
convolution with $\delta(p_1+p_2)$ in Fourier space,
$$
    \Big( \delta(x_1-x_2) \gamma^{(2)}\Big)(p_1, p_2, p_1', p_2')
    = \int \rd r\;  \gamma^{(2)}(r, p_1-r+p_2; p_1', p_2')\; ,
$$
thus after taking the partial trace, we have
$$
   \Big[\tr_2 \Big( \delta(x_1-x_2) \gamma^{(2)}\Big)\Big](p_1; p_1')
    = \int \rd r \rd q \; \gamma^{(2)} (r, p_1-r+q; p_1', q)\,.
$$
Applying these elementary steps for the first term of \eqref{eq:ex-1},
we get
\begin{equation}\label{long}
\begin{split}
\Big[ -i &\int_0^t \rd s \; \cU_0^{(1)} (t-s)  \tr_2 \, \delta (x_1
-x_2) \, \cU_0^{(2)} (s) \gamma^{(2)}_0 \Big] (p_b; p_a)  \\
&= -i \int_0^t \rd s \, e^{-i(t-s)p_b^2} e^{i(t-s) p_a^2} \int \rd q
\rd r \,
\left(\cU_0^{(2)} (s) \gamma^{(2)}_0 \right) (r ,p_b -r+ q; p_a,q)\\
&= -i \int_0^t \rd s \, e^{-i(t-s)p_b^2} e^{i(t-s) p_a^2} \int \rd
p_c \rd p_d \rd p_e \, \delta(p_b+p_e-p_c-p_d) \left(\cU_0^{(2)} (s)
\gamma^{(2)}_0 \right) (p_c ,p_d; p_a, p_e)\,.
\end{split}
\end{equation}
In the last step we changed variables, which  now correspond
to the variables in the Feynman graph. In particular,
 the vertex involves a delta function expressing
the Kirchoff law (conservation of momentum at the vertex).

Now we show how the time integrals of the propagators can
be expressed in terms of auxiliary $\alpha$-integrals and
resolvents. Neglecting the momentum integrations,
the last term in \eqref{long} contains the following
propagators:
$$
 \Pi:=  \int_0^t \rd s\, e^{-i(t-s)p_b^2} e^{i(t-s) p_a^2}
   e^{-is(p_c^2+p_d^2)} e^{is(p_a^2+p_e^2)}\,.
$$
Using the identity \eqref{1res} for the propagator of
each momentum variable, we obtain
$$
  \Pi =  i \int_0^t \rd s
   \int_\bR \frac{\rd \alpha_a\rd \alpha_b \rd \alpha_c
\rd \alpha_d\rd \alpha_e\;
e^{it(\alpha_a-i\eta_a)-i(t-s)(\a_b+i\eta_b)
-is(\a_c+i\eta_c)-is(\a_d+i\eta_d)+is(\a_e-i\eta_e)}}
   {(\alpha_a - p_a^2 -i\eta_a)(\a_b-p_b^2+i\eta_b)
 (\a_c-p_c^2+i\eta_c)(\a_d-p_d^2+i\eta_d)(\a_e-p_e^2-i\eta_e)}
$$
(In our example, $\tau_b=\tau_c=\tau_d=1$ and $\tau_a=\tau_e=-1$
by the definition of $\tau$). Notice that the time integration
can be extended to $s\in (-\infty, \infty)$, since all $\eta$'s are positive
and  by residue calculation
$$
     \int_\bR  \frac{\rd \a \; e^{-is(\a+i\eta)}}{\a - p^2 + i\eta} =0
$$
if $s<0$ and $\eta>0$.

Using that $\eta_b=\eta_c+\eta_d+\eta_e$ and performing the $\rd s$
integration, we obtain the delta function in the $\a$ variables:
$$
  \Pi = i
   \int_\bR \frac{ e^{it(\a_a-i\eta_a)}e^{-it(\a_b+i\eta_b)}\;
\rd \alpha_a\rd \alpha_b \rd \alpha_c \rd \alpha_d\rd \alpha_e\;
\delta(\a_b-\a_c-\a_d+\a_e)}
   {(\alpha_a - p_a^2 -i\eta_a)(\a_b-p_b^2+i\eta_b)
 (\a_c-p_c^2+i\eta_c)(\a_d-p_d^2+i\eta_d)(\a_e-p_e^2-i\eta_e)}
$$
(recall that the $\delta$-function in the $\a$-variables is defined
w.r.t. to the measure $\rd \a= \rd_{\text{Leb}} \a /2\pi$).
Combining this formula with \eqref{long}, we arrive at
\begin{equation}
\begin{split}
\Big[ -i \int_0^t \rd s \, &\cU_0^{(1)} (t-s)  \tr_2 \, \delta (x_1
-x_2) \, \cU_0^{(2)} (s) \gamma^{(2)}_0 \Big] (p_b; p_a)  \\
 = & \int  \rd p_c \rd p_d \rd p_e
 \; \delta(p_b+p_e-p_c-p_d) \gamma_0^{(2)}(p_c, p_d; p_a, p_e) \\
&\times
\int  e^{it(\a_a-i\eta_a)}e^{-it(\a_b+i\eta_b)}
\Bigg[ \prod_{j=a,b,c,d,e} \frac{\rd \a_j}{\a_j -p_j^2 + i\tau_j\eta_j}\Bigg]
 \; \delta(\a_b+\a_e-\a_c-\a_d)
\end{split}
\end{equation}
which is exactly the action of the kernel $K_{\Gamma_1, t}$
(as defined in \eqref{simpler}) on $\gamma^{(2)}_0$.

\bigskip

Now we come to the proof of the  general theorem.

\medskip

{\it Proof of Theorem \ref{thm:graphs}.}  We start by proving
(\ref{eq:fullexp}) and (\ref{n0}). For $k \geq 1$, $n \geq 1$, and
$t\in [0,T]$, let
\begin{equation}
\begin{split}
\xi_{n,t}^{(k)} := \int_0^t \rd s_1 \int_0^{s_1} \rd s_2 \ldots
\int_0^{s_{n-1}} \rd s_n \; \cU_0^{(k)} (t-s_1) B^{(k)} \ldots
&  \; \cU_0^{(k+n-1)} (s_{n-1} -s_n) B^{(k+n-1)} \\ &\times \cU_0^{(k+n)}
(s_n) \gamma_0^{(k+n)} \, .
\end{split}
\end{equation}
For $k \geq 1$, $n =0$ and $ t \in [0,T]$, let
\begin{equation}
\xi_{0,t}^{(k)} := \cU_0^{(k)} (t) \gamma^{(k)}_0\; .
\end{equation}

We also define $\theta_{n,t}^{(k)}$
 for $k \geq 1$, $n \geq 0$, and $t \in [0,T]$
through its kernel given by
\begin{equation}\label{def:theta}
\theta_{n,t}^{(k)} (\bq_k ; \bq'_k): = \sum_{\Gamma \in \cF_{n,k}}
\int \rd \br_{n+k} \rd \br'_{n+k} \, K_{\Gamma,t} ( \bq_k , \bq'_k ;
\br_{n+k}, \br'_{n+k}) \gamma_0^{(n+k)} (\br_{n+k}, \br'_{n+k}) \, .
\end{equation}
We need to show that $\xi_{n,t}^{(k)}= \theta_{n,t}^{(k)}$ for all
$n\geq 0$, $k\geq 1$,  and for all $t \in [0,T]$. The proof is
based on induction over $n$. For $n=0$ the claim is trivial using
the identity (\ref{1res}), thus  from now on we assume that
$n\ge 1$. We prove that $\xi_{m,t}^{(k)} = \theta_{m,t}^{(k)}$
for $m=n$ and for all $k \geq 1$ and $t \in [0,T]$, assuming that
this is true for all $m <n$. We will first check that
$\xi_{n,0}^{(k)}= \theta_{n,0}^{(k)}$, then compare their time
derivatives.

At $t=0$ clearly $\xi_{n,0}^{(k)}=0$. To see $\theta_{n,0}^{(k)}=0$,
we check that the $K_{\Gamma, t}$ kernel vanishes at $t=0$, for
$\Gamma \in \cF_{n,k}$,  $n\ge 1$. We use the representation
(\ref{eq:KGamma2}). Since $n\ge 1$, we know that $E_2(\Gamma)\neq
\emptyset$, in particular there is at least one propagator factor
$(\a_{\bar e} - p_{\bar e}^2 + i\tau_{\bar e} \eta_{\bar e})^{-1}$
on the right hand side for some ${\bar e}\in L_2(\Gamma)$. Notice
that for $t=0$ the exponentially increasing factor $\prod_{e \in R_2
(\Gamma)} e^{\eta_e t}$ is not present. Using the absolute
convergence of all the vertex integrals (\ref{vert})  uniformly for
large $\boeta$ and the estimate (\ref{lasteta}) for the integration
of a maximal vertex, we can let $\eta_{e} \to \infty$ for all $e\in
L_2(\Gamma)$ (recall that the $\eta$'s on the leaves determine all
$\eta$ values), and conclude that $K_{\Gamma, t=0}(\bq_k , \bq'_k ;
\br_{n+k}, \br'_{n+k})$, for any fixed values of its argument, is
bounded by a quantity converging to zero as $\eta_{e}\to\infty$, for
all $e \in L_2 (\Gamma)$. Since $K_{\Gamma, t=0}$ is independent of
$\boeta$, we obtain that it is zero.

Next we prove that $\xi_{n,t}^{(k)}$ and $\theta_{n,t}^{(k)}$
both satisfy the same equations:
\begin{equation}\label{eq:xi-theta}
\begin{split}
i\partial_t \xi_{n,t}^{(k)} (\bq_k ; \bq'_k) &= \sum_{j=1}^k \left(
q_j^2 -(q'_j)^2 \right) \xi_{n,t}^{(k)} (\bq_k; \bq'_k) + i \left(
B^{(k)} \xi_{n-1,t}^{(k+1)} \right) (\bq_k ; \bq'_k)\, \\
i\partial_t \theta_{n,t}^{(k)} (\bq_k ; \bq'_k) &= \sum_{j=1}^k
\left(q_j^2 - (q'_j)^2 \right) \theta_{n,t}^{(k)} (\bq_k; \bq'_k) +
i \left( B^{(k)} \theta_{n-1,t}^{(k+1)}\right) (\bq_k ; \bq'_k).
\end{split}
\end{equation}
Since, by induction assumption, $\xi_{n,0}^{(k)} =
\theta_{n,0}^{(k)}$ and $\xi_{n-1,t}^{(k+1)} =
\theta_{n-1,t}^{(k+1)}$ for every $k \geq 1$ and every $t \in
[0,T]$,  it follows from (\ref{eq:xi-theta}) that $\xi_{n,t}^{(k)} =
\theta_{n,t}^{(k)}$ for all $k \geq 1$ and $t \in [0,T]$. It remains
to prove (\ref{eq:xi-theta}).

We start by deriving the equation for $\xi_{n,t}^{(k)}$. To this end
we compute the derivative of $\xi_{n,t}^{(k)}$ with respect to $t$.
\begin{equation}
\begin{split}
\partial_t \xi_{n,t}^{(k)} = \; &\int_0^t \rd s_2 \dots
\int_0^{s_{n-1}} \rd s_n  \, B^{(k)} \cU_0^{(k+1)} (t-s_2) \dots
\cU_0^{(k+n-1)} (s_{n-1} -s_n) B^{(k+n-1)} \\ &\hspace{3cm} \times
\cU_0^{(k+n)} (s_n) \gamma_0^{(k+n)} \, \\ &+i \sum_{j=1}^k \int_0^t
\rd s_1 \int_0^{s_1} \rd s_2 \dots \int_0^{s_{n-1}} \rd s_n \left[
\;
\Delta_j , \;\; \cU_0^{(k)} (t-s_1) B^{(k)} \dots \right. \\
&\left. \hspace{3cm} \times \cU_0^{(k+n-1)} (s_{n-1} -s_n)
B^{(k+n-1)} \cU_0^{(k+n)} (s_n) \gamma_0^{(k+n)} \right] \, .\,
\end{split}
\end{equation}
Hence, in Fourier space,
\begin{equation}\label{eq:hierxi}
\begin{split}
i\partial_t \xi_{n,t}^{(k)} &(\bq_k ; \bq'_k) = \sum_{j=1}^k \left(
q_j^2 -(q'_j)^2 \right) \xi_{n,t}^{(k)} (\bq_k; \bq'_k) + i \left(
B^{(k)} \xi_{n-1,t}^{(k+1)} \right) (\bq_k ; \bq'_k)\,,
\end{split}
\end{equation}
which proves the first equation in (\ref{eq:xi-theta}). Next we
show the second equation in (\ref{eq:xi-theta}).

Using (\ref{eq:KGamma2}), we compute the time derivative of
$\theta_{n,t}^{(k)}$:
\begin{equation}\label{dertheta}
\begin{split}
\partial_t \theta_{n,t}^{(k)} (\bq_k ; \bq'_k) =
&-i\sum_{\Gamma \in \cF_{n,k}} \int \rd \br_{n+k} \rd\br'_{n+k}
 \gamma^{(n+k)}_0
(\br_{n+k}; \br'_{n+k})
\\ &
\hspace{1cm} \times \frac{1}{(n+k)!} \sum_{\pi_2 \in S_{n+k}}
\prod_{e \in R_1 (\Gamma)} (-i\tau_e)  e^{-it \tau_e (q_{\pi_1
(e)}^{\sharp_e})^2 } \, \delta (q^{\sharp_e}_{\pi_1 (e)} -
r^{\sharp_e}_{\pi_2 (e)})
\\ &\hspace{1cm} \times \int \int_\bR \prod_{e\in E_2 (\Gamma)} \rd \alpha_e
\rd p_e  \prod_{e\in R_2 (\Gamma)} \delta (p_e - q_{\pi_1
(e)}^{\sharp_e}) \prod_{e \in L_2 (\Gamma)} \delta (p_e - r_{\pi_2
(e)}^{\sharp_e}) \\ &\hspace{1cm} \times \,
 e^{-it \sum_{e\in R_2(\Gamma)} \tau_e(\alpha_e+i\tau_e\eta_e)}
\left(  \sum_{e \in R_1 (\Gamma)}\tau_e (q_{\pi_1 (e)}^{\sharp_e})^2
+\sum_{e \in R_2 (\Gamma)} \tau_e(\a_e + i\tau_e\eta_e) \right) \\
&\hspace{1cm} \times \prod_{e \in E_2 (\Gamma)} \frac{1}{\a_e -
p_e^2 + i\tau_e\eta_e} \prod_{v \in V(\Gamma)} \delta
\Big(\sum_{e\in v} \pm \alpha_e \Big) \delta\Big(\sum_{e\in v}\pm
p_e \Big) \, .
\end{split}
\end{equation}
(Strictly speaking, this calculation is formal, since after the
differentiation the $d\alpha_e$ integral is not absolutely
convergent. We will remark on this issue at the end.) We write
\begin{equation} \sum_{e\in R_2 (\Gamma)} \tau_e(\a_e + i
\tau_e\eta_e) = \sum_{e\in R_2 (\Gamma)} \tau_e (\a_e - p_e^2 + i
\tau_e \eta_e) + \sum_{e \in R_2 (\Gamma)} \tau_e p_e^2 \,.
\end{equation}
Because of the delta-functions $\prod_{e\in R_2 (\Gamma)} \delta
(p_e - q_{\pi_1 (e)}^{\sharp_e})$, we also have
$$
     \sum_{e \in R_1 (\Gamma)} \tau_e (q_{\pi_1 (e)}^{\sharp_e})^2 +
  \sum_{e \in R_2 (\Gamma)} \tau_e p_e^2 =
   \sum_{e \in R (\Gamma)} \tau_e (q_{\pi_1 (e)}^{\sharp_e})^2
   = \sum_{j=1}^k \left( q_j^2 -(q'_j)^2 \right)\,.
$$
{F}rom the last two equations we obtain
\begin{equation}\label{eq:theta}
\begin{split}
i\partial_t \theta_{n,t}^{(k)} (\bq_k ; \bq'_k) = &\sum_{\Gamma \in
\cF_{n,k}}  \sum_{j=1}^k \left( q_j^2 -(q'_j)^2 \right) \int \rd
\br_{n+k} \rd\br'_{n+k} \, K_{\Gamma,t} (\bq_k , \bq'_k;
\br_{n+k} , \br'_{n+k}) \gamma^{(n+k)}_0 (\br_{n+k}; \br'_{n+k}) \\
&+ \sum_{\Gamma \in \cF_{n,k}}  \int \rd \br_{n+k} \rd\br'_{n+k}
 \gamma^{(n+k)}_0
(\br_{n+k}; \br'_{n+k}) \\
&\hspace{1cm} \times \frac{1}{(n+k)!} \sum_{\pi_2} \, \prod_{e \in
R_1 (\Gamma)} (-i\tau_e) e^{-it \tau_e (q_{\pi_1 (e)}^{\sharp_e})^2
}\;
\delta (q_{\pi_1 (e)}^{\sharp_e} -r_{\pi_2(e)}^{\sharp_e}) \\
&\hspace{1cm} \times \int \int_\bR\prod_{e\in E_2 (\Gamma)} \rd
\alpha_e \rd p_e  \prod_{e\in R_2 (\Gamma)} \delta (p_e - q_{\pi_1
(e)}^{\sharp_e}) \prod_{e \in L_2
(\Gamma)} \delta (p_e - r_{\pi_2 (e)}^{\sharp_e}) \\
&\hspace{1cm} \times \,  e^{-it \sum_{e\in
R_2(\Gamma)}\tau_e(\alpha_e+i\tau_e\eta_e)} \left(\sum_{e\in R_2
(\Gamma)}
 \tau_e (\a_e - p_e^2 + i \tau_e \eta_e)\right) \\
&\hspace{1cm} \times \prod_{e \in E_2 (\Gamma)} \frac{1}{\a_e -
p_e^2 + i\tau_e\eta_e} \prod_{v \in V(\Gamma)} \delta
\Big(\sum_{e\in v}\pm \alpha_e \Big) \delta\Big(\sum_{e\in v}\pm p_e
\Big) \, .
\end{split}
\end{equation}

For a given ${\bar e} \in R_2 (\Gamma)$ let ${\bar v}={\bar v}({\bar
e}) \in V(\Gamma)$ be the only vertex such that ${\bar e}\in {\bar
v}$ (by definition of $R_2 (\Gamma)$ there is such vertex).
 Then the second term on the r.h.s. of (\ref{eq:theta})
can be rewritten as (we use that $L_2(\Gamma)\cap
R_2(\Gamma)=\emptyset$):
\begin{equation}\label{eq:theta2}
\begin{split}
\sum_{\Gamma \in \cF_{n,k}} & \sum_{{\bar e} \in R_2 (\Gamma)}
\tau_{{\bar e}}
 \int \rd \br_{n+k} \rd\br'_{n+k}
 \gamma^{(n+k)}_0
(\br_{n+k}; \br'_{n+k}) \\
& \times \frac{1}{(n+k)!} \sum_{\pi_2}
 \prod_{e \in R_1 (\Gamma)}
 (-i\tau_e) e^{-it \tau_e (q_{\pi_1 (e)}^{\sharp_e})^2 }\;
\delta (q_{\pi_1 (e)}^{\sharp_e} -r_{\pi_2(e)}^{\sharp_e}) \\
& \times \int\int_\bR \prod_{ e\in{\bar v}} \rd\alpha_e \rd p_e \;
e^{-it\tau_{{\bar e}}(\a_{{\bar e}}+i\tau_{\bar e}\eta_{\bar e}) }
 \;\delta
\left( \sum_{e \in {\bar v}}\pm \a_e \right)
\, \delta \left( \sum_{e \in {\bar v}}\pm p_e \right) \\
&\times \int \int_\bR\prod_{e \in E_2 (\Gamma)\atop e \not\in {\bar
v}} \rd \alpha_e \rd p_e \prod_{e \in R_2 (\Gamma)\atop e\neq {\bar
e}} \delta (p_e - q_{\pi_1 (e)}^{\sharp_e}) \prod_{e \in L_2
(\Gamma)} \delta (p_e - r_{\pi_2 (e)}^{\sharp_e}) \\
& \times \,  e^{-it \sum_{e\in
R_2(\Gamma),e\neq {\bar e}}\tau_e(\alpha_e+i\tau_e\eta_e)} \\
& \times \prod_{e \in E_2 (\Gamma)\atop e\neq {\bar e}}
\frac{1}{\a_e - p_e^2 + i\tau_e\eta_e} \prod_{v \in V(\Gamma)\atop
v\neq {\bar v}} \delta \Big(\sum_{e\in v} \pm\alpha_e \Big)
\delta\Big(\sum_{e\in v} \pm p_e \Big) \, .
\end{split}
\end{equation}

Next we perform the $\a_{{\bar e}}$ integration. Using that, by
definition of the family $\boeta$, $\eta_{\bar e} = \sum_{\bar e
\neq e \in \bar v} \eta_e$, and the fact that \be
   \sum_{e\in \bar v}\pm \alpha_e =0 \;\;\Longleftrightarrow \;\;
   \tau_{\bar e}\alpha_{\bar e} = \sum_{e\in v, e\neq\bar e}
  \tau_e\alpha_e
\label{pmalpha} \ee from the definition of $\tau_e$,
 we obtain
\begin{equation}\label{eq:intabare}
\int_\bR \rd \a_{{\bar e}}\; e^{-it\tau_{{\bar e}}(\a_{{\bar
e}}+i\tau_{\bar e}\eta_{\bar e})    } \;\delta \left( \sum_{e \in
{\bar v}}\pm \a_e \right) = e^{-it \sum_{\bar e \neq e \in \bar v}
\tau_e (\a_e + i \tau_e \eta_e)} \, .
\end{equation}
Multiplying this contribution with the factor $\exp (-it \sum_{e\in
R_2(\Gamma),e\neq {\bar e}}\tau_e(\alpha_e+i\tau_e\eta_e))$ from
(\ref{eq:theta2}) and using (\ref{pmalpha}),
 we obtain $\exp ( -i t \sum_{e
\in R} \tau_e (\alpha_e + i \tau_e \eta_e)) $ with $R = \{ e \in R_2
(\Gamma) : e \neq \bar e \} \cup \{ e \in \bar v: e \neq \bar e\}$.
The set $R \cup R_1$ can be interpreted as the set of roots of a new
graph, with $k+1$ root-pairs and with $n-1$ vertices as follows.

\begin{figure}
\begin{center}
\epsfig{file=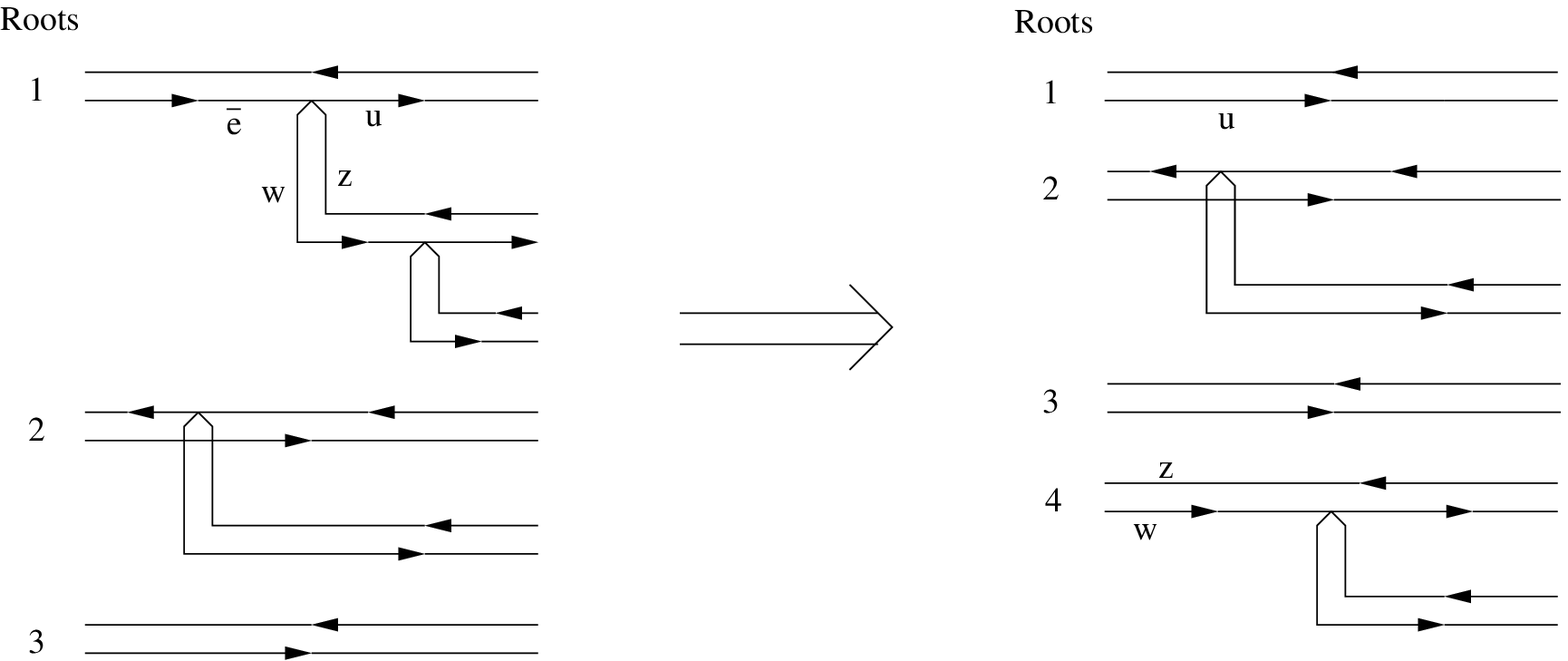,scale=.75}
\end{center}
\caption{The map $(\Gamma, \bar e) \to \wt
\Gamma$}\label{fig:wtGamma}
\end{figure}

Given $\Gamma \in \cF_{n,k}$ and $\bar e \in R_2 (\Gamma)$, we
define a new graph $\wt \Gamma = \wt \Gamma (\Gamma , \bar e) \in
\cF_{n-1, k+1}$ as follows: the roots of $\wt \Gamma$ are given by
$R(\wt \Gamma) = \{ e \in R(\Gamma) : e \neq \bar e \} \cup \{ e \in
\bar v : e \neq \bar e \}$, that is we remove the edge $\bar e$, and
we add all other edges adjacent to the vertex $\bar v$ to the set of
roots (recall that $\bar v \in V(\Gamma)$ is the unique vertex to
which $\bar e$ is adjacent). The label $\wt \pi_1$ of the roots in
$\wt \Gamma$ is defined so that the two unmarked son-edges of $\bar
e$ at the vertex $\bar v$ become the $(k+1)$-th pair of roots of
$\wt \Gamma$, while the marked son-edge at $\bar v$ inherits the
label of $\bar e$, and all the other roots keep their label. The
vertices and the edges of the new graph $\wt \Gamma$ are $V(\wt
\Gamma) = V(\Gamma) \setminus \{ \bar v \}$ and, respectively,
$E(\wt \Gamma) = E(\Gamma) \setminus \{ \bar e\}$. Finally, the
leaves of $\wt \Gamma$ are the same as the leaves of $\Gamma$, that
is $L(\Gamma) = L (\wt \Gamma)$. Graphically the map from $(\Gamma,
\bar e)$ (with $\Gamma \in \cF_{n,k}$ and $\bar{e} \in R_2
(\Gamma)$) to $\wt \Gamma \in \cF_{n-1,k+1}$ corresponds to the
cancellation of the edge $\bar e$ and of the vertex $\bar v$ and to
moving the two new roots (the two unmarked son-edges at $\bar v$)
with their full tree graphs to the bottom of the graph (see Fig.
\ref{fig:wtGamma}).

Note that the map from $\Gamma \in \cF_{n,k}$ and $\bar e \in R_2
(\Gamma)$ to $\wt \Gamma \in \cF_{n-1,k+1}$ is surjective but not
injective: in fact for every $\wt \Gamma \in \cF_{n-1, k+1}$ there
are $2k$ possible choices of $\Gamma \in \cF_{n,k}$ and $\bar e \in
R_2 (\Gamma)$, because the last $(k+1)$-th pair of roots of $\wt
\Gamma$ can be attached to any of the first $2k$ roots. This implies
that the sum over $\Gamma \in \cF_{n,k}$ and $\bar e \in R_2
(\Gamma)$ in (\ref{eq:theta2})
 can be replaced by a sum over $\wt \Gamma \in
\cF_{n-1,k+1}$ and a sum over the first $k$ pair of roots of $\wt
\Gamma$ plus a binary choice between the two paralell root edges.

With all these notations, (\ref{eq:theta2}) can be rewritten as
\begin{equation}\label{eq:afterder}
\begin{split}
\int & \rd \br_{n+k} \rd\br'_{n+k} \gamma^{(n+k)}_0 (\br_{n+k};
\br'_{n+k}) \sum_{\wt\Gamma \in \cF_{n-1,k+1}} \sum_{j=1}^{k} \int
\rd \wt \bq_{k+1} \rd \wt \bq'_{k+1} \, \prod^k_{\ell \neq j} \delta
(q_{\ell} - \wt q_{\ell}) \delta (q'_{\ell} - \wt q'_{\ell}) \\
&\times \Big[ \delta (q_j - \wt q_j - \wt q_{k+1} + \wt q'_{k+1})
\delta (q'_j - \wt q'_j) - \delta (q'_j - \wt q'_j - \wt q'_{k+1} +
\wt q_{k+1}) \delta (q_j
- \wt q_j) \Big] \\
& \times \frac{1}{(n+k)!} \sum_{\pi_2} \int \int_\bR\prod_{e\in E
(\wt\Gamma)} \rd \alpha_e \rd p_e \, \prod_{e \in R(\wt\Gamma)}
\delta (p_e - \wt q_{\pi_1 (e)}^{\sharp_e}) \prod_{e \in
L(\wt\Gamma)} \delta (p_e - r_{\pi_2 (e)}^{\sharp_e}) \\
\ &\times e^{-it \sum_{e\in R(\wt\Gamma)}
\tau_e(\alpha_e+i\tau_e\eta)} \prod_{e \in E(\wt\Gamma)}
\frac{1}{\a_e - p_e^2 + i\tau_e\eta_e} \prod_{v \in V(\wt\Gamma)}
\delta \Big(\sum_{e\in v}\pm \alpha_e \Big) \delta\Big(\sum_{e\in
v}\pm p_e \Big) \, .
\end{split}
\end{equation}
In the second line we rewrote the integration over the momenta
$p_e$, $e \in \bar v$, in (\ref{eq:theta2}), so that it is clear
that it describes exactly the action of the operator $iB^{(k)}$ (see
(\ref{eq:Bk})). {F}rom the last equation, together with
(\ref{eq:theta}), we obtain
\begin{equation}\label{eq:hiertheta}
i\partial_t \theta_{n,t}^{(k)} (\bq_k ; \bq'_k) = \sum_{j=1}^k
\left(q_j^2 - (q'_j)^2 \right) \theta_{n,t}^{(k)} (\bq_k; \bq'_k) +
i \left( B^{(k)} \theta_{n-1,t}^{(k+1)}\right) (\bq_k ; \bq'_k)
\end{equation}
which proves the second equation in (\ref{eq:xi-theta}). This
completes the proof of (\ref{eq:fullexp}) and (\ref{n0}).

\bigskip

\begin{figure}
\begin{center}
\epsfig{file=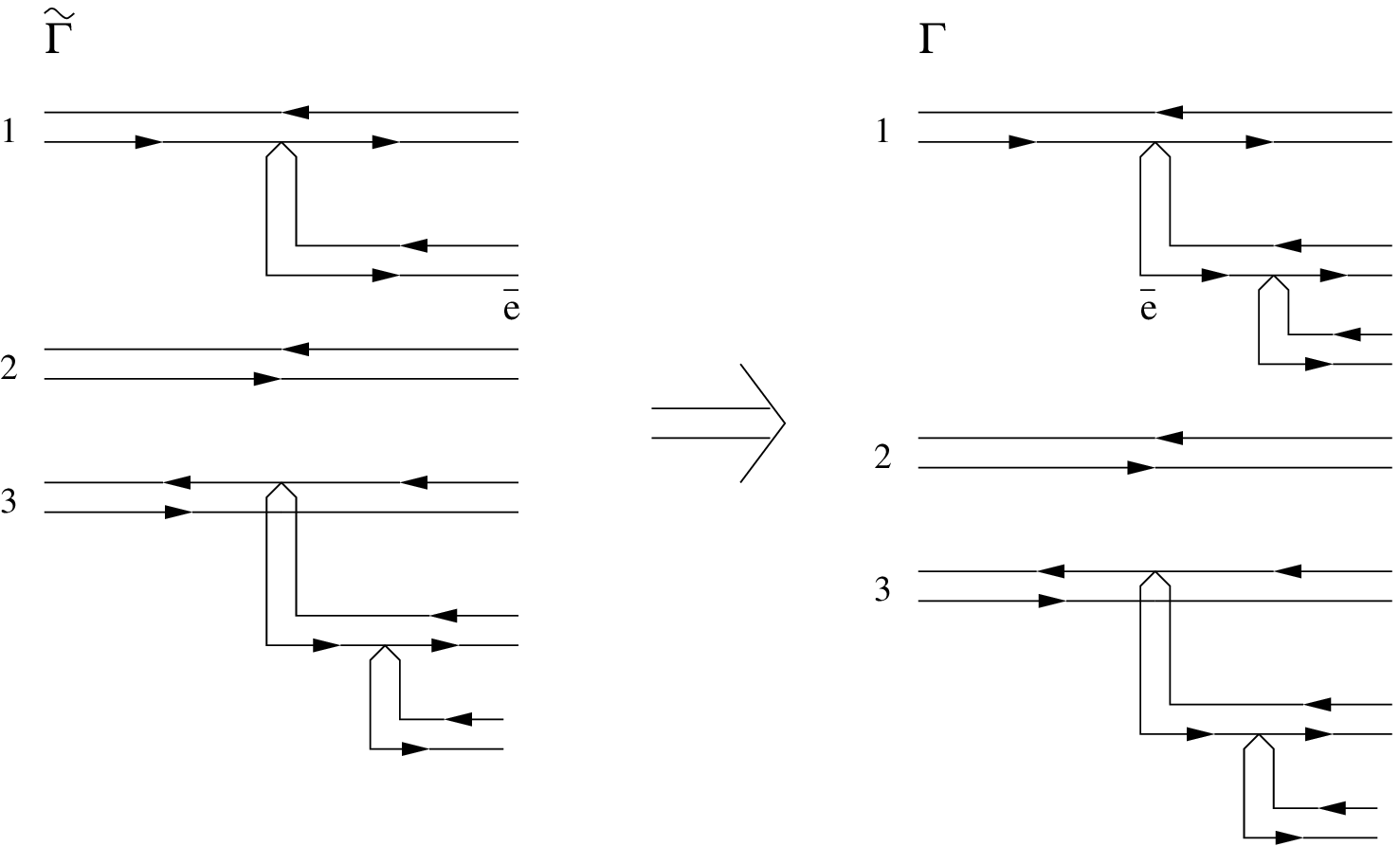,scale=.75}
\end{center}
\caption{The map $(\wt \Gamma, \bar e) \to
\Gamma$}\label{fig:Gamma}
\end{figure}

Let us now prove (\ref{eq:error}). The l.h.s. of (\ref{eq:error})
can be rewritten as
\begin{equation}
\begin{split}
\int_0^t &\rd s_1 \int_0^{s_1} \rd s_2 \dots \int_0^{s_{n-1}} \rd
s_n \; \cU_0^{(k)} (t-s_1) B^{(k)} \dots \cU_0^{(k+n-1)} (s_{n-1}
-s_n) B^{(k+n-1)} \gamma_{s_n}^{(k+n)} \\ &= \int_0^t \rd s \;
\Big[ \int_0^{t-s} \rd s_1  \int_0^{s_1} \dots \int_0^{s_{n-2}}
\rd s_{n-1} \; \cU_0^{(k)} (t-s-s_1) B^{(k)} \dots
\\ &\hspace{2cm} \times \cU_0^{(k+n-2)}
(s_{n-2} -s_{n-1}) B^{(k+n-2)} \cU_0^{(k+n-1)}
(s_{n-1}) \Big] \, B^{(k+n-1)} \gamma_{s}^{(k+n)} \; .
\end{split}
\end{equation}
{F}rom (\ref{eq:fullexp}), the last expression equals
\begin{equation}
\omega^{(k)}_t := \int_0^t \rd s \, \sum_{\wt \Gamma \in
\cF_{k,n-1}} K_{\wt \Gamma,t-s} B^{(k+n-1)} \gamma^{(k+n)}_s \, .
\end{equation}
Using (\ref{simpler}) and (\ref{eq:Bk}), we obtain
\begin{equation}\label{eq:KBG}
\begin{split}
\omega^{(k)}_t (\bq_k ; &\bq'_k) = -i \int_0^t \rd s \, \sum_{\wt
\Gamma \in \cF_{k,n-1}} \frac{1}{(n+k-1)!} \sum_{\pi_2 \in
S_{n+k-1}} \sum_{j=1}^{n+k-1} \\
&\times \int \rd \br_{n+k-1} \rd \br'_{n+k-1} \int  \int_\bR\prod_{e\in E
(\wt\Gamma)} \rd \alpha_e \rd p_e \, \prod_{e \in R(\wt\Gamma)}
\delta (p_e - q_{\pi_1 (e)}^{\sharp_e}) \prod_{e \in
L(\wt\Gamma)} \delta (p_e - r_{\pi_2 (e)}^{\sharp_e}) \\
\ &\times e^{-it \sum_{e\in R(\wt\Gamma)}
\tau_e(\alpha_e+i\tau_e\eta)} \prod_{e \in E(\wt\Gamma)}
\frac{1}{\a_e - p_e^2 + i\tau_e\eta_e} \prod_{v \in V(\wt\Gamma)}
\delta \Big(\sum_{e\in v}\pm \alpha_e \Big) \delta\Big(\sum_{e\in
v}\pm p_e \Big) \\ &\times \int \rd \bp_{n+k} \rd \bp'_{n+k} \,
\left( \prod_{\ell \neq j} \delta (r_{\ell} - p_{\ell}) \delta
(r'_{\ell} - p'_{\ell}) \right) \gamma^{(n+k)}_s (\bp_{n+k};
\bp'_{n+k}) \\ &\times \Big[ \delta (r'_j - p'_j) \delta (r_j -
(p_j + p_{n+k} -p'_{n+k})) - \delta (r_j -p_j) \delta (r'_j -
(p'_j + p'_{n+k} -p_{n+k})) \Big] \;,
\end{split}
\end{equation}
where the sum over $j$ and the last two lines correspond to the
action of the operator $B^{(n+k-1)}$ on the density
$\gamma^{(k+n)}_s$. Note that $j$ actually labels the leaf-pairs
of $\wt\Gamma$. Fixing $j=1,\dots ,k+n-1$ and one of the two terms
in the square bracket on the last line of (\ref{eq:KBG}) is
equivalent to choosing one of the leaves of $\wt \Gamma$. For a
given $\wt \Gamma$ and $\bar e \in L(\wt\Gamma)$, we can define a
new graph $\Gamma \in \cF_{n,k}$ by splitting the edge  $\bar e$
with a new vertex and attaching two new leaf edges to this vertex
(see Fig. \ref{fig:Gamma}).
The same graph in $\cF_{n,k}$ can clearly be obtained starting
from different graphs $\wt\Gamma \in \cF_{n-1,k}$. More precisely,
if we denote by $M(\Gamma)$ the set of maximal vertices of
$\Gamma$ ($v \in M(\Gamma)$ if $v\in V(\Gamma)$ and there is no
$\tilde v \in V(\Gamma)$ with $v \prec \tilde v$), then we find
all possible $\wt \Gamma$ by removing a vertex $\bar v  \in
M(\Gamma)$ (and deleting the son-edges of $\bar v$). It is hence
clear that, in (\ref{eq:KBG}) we can replace the sum over $\wt
\Gamma \in \cF_{n-1,k}$, the sum over $j \in \{ 1, \dots ,n+k-1\}$
and the binary choice of one of the two terms in the last line, by
a sum over all $\Gamma \in \cF_{n,k}$ and over all $\bar v \in
M(\Gamma)$.

In order to rewrite (\ref{eq:KBG}) in terms of the new Feynman
graph $\Gamma \in \cF_{n,k}$, we observe from (\ref{eq:KBG}), that
the son-edges of $\bar v$ will have a momentum like all the other
edges, and that momentum conservation holds at $\bar v$ (see the
last line of (\ref{eq:KBG})), but they will not have any
$\a$-variable, any $\eta$-variable, and any propagator. The
labelling $\pi_2$ of the leaves of $\wt \Gamma$ induces a
labelling with $\{ 1,\dots , n+k-1\}$ of the leaves of $\Gamma$,
with the exception of the two unmarked son-edges of the chosen
$\bar v \in M(\Gamma)$: these two edges are always labelled by the
number $n+k$. Of course, because of the permutation symmetry of
$\gamma^{(n+k)}$, we can restore a full symmetry of the
leaf-variables; to this end we replace the sum over $\pi_2 \in
S_{n+k-1}$ by a sum over $\pi_2 \in S_{n+k}$ and we replace the
factor $(n+k-1)!$ by $(n+k)!$. We conclude that
\begin{equation}
\begin{split}
\omega^{(k)}_t (\bq_k ; &\bq'_k) = -i \int_0^t \rd s \, \sum_{
\Gamma \in \cF_{n,k}} \frac{1}{(n+k)!} \sum_{\pi_2 \in S_{n+k}}
\sum_{\bar v \in M(\Gamma)} \sigma_{\bar v} \int \rd \bp_{n+k}\, \rd
\bp'_{n+k}
\\ &\times \int \prod_{e \in E(\Gamma)}\rd p_e \prod_{e\in E
(\Gamma)\backslash S_{\bar v}} \rd \alpha_e  \prod_{e \in
R(\Gamma)} \delta (p_e - q_{\pi_1 (e)}^{\sharp_e}) \prod_{e \in
L(\Gamma)} \delta (p_e - p_{\pi_2 (e)}^{\sharp_e}) e^{-it
\sum_{e\in R(\Gamma)} \tau_e(\alpha_e+i\tau_e\eta)} \\
&\times \prod_{e \in E(\Gamma)\backslash S_{\bar v}} \frac{1}{\a_e
- p_e^2 + i\tau_e\eta_e} \prod_{\bar v \neq v \in V(\Gamma)}
\delta \Big(\sum_{e\in v}\pm \alpha_e \Big) \prod_{v \in
V(\Gamma)} \delta\Big(\sum_{e\in v}\pm p_e \Big) \gamma_s^{(n+k)}
(\bp_{n+k}; \bp'_{n+k})
\end{split}
\end{equation}
where $S_{\bar v}$ is the set of son-edges of $\bar v$ and we recall that
$\sigma_{\bar v} = \sum_{e \in S_{\bar v}} \tau_e$. {F}rom
(\ref{Lsimpler}) we obtain (\ref{eq:error}).

\bigskip

Finally we comment on how to make the formal differentiation in
(\ref{dertheta}) rigorous. In the definition (\ref{eq:KGamma2}) we
could have defined $K_{\Gamma, t, \boeta, \e}$ instead of
$K_{\Gamma, t, \boeta}$ with introducing a regularizing factor
$\exp(-\e\sum_{e\in E_2(\Gamma)} \alpha^2_e)$ in the integrals and
similarly we would have defined $\theta_{n,t,\e}^{(k)}$ in
(\ref{def:theta}). Then the time derivative of
$\theta_{n,t,\e}^{(k)}$ can be computed by differentiating the
integrand. Note that before differentiation, the $\rd\alpha_e$
integrals in (\ref{eq:KGamma2}) are absolutely convergent and  the
convergence is
 uniform  in $\e>0$ and $t\in[0,T]$ for any fixed $T$.
Therefore  we obtain \be
    \lim_{\e\to0+0} \theta_{n,t,\e}^{(k)} = \theta_{n,t}^{(k)}
\label{eq:epsconv} \ee for any fixed $n,k$ and uniformly for
$t\in[0,T]$. In particular, the relation
 $\xi_{n,t}^{(k)} = \lim_{\e\to0+0}\theta_{n,0,\e}^{(k)}=0$ for $n\ge
1$ still holds.  The identity (\ref{1res}) will not
hold any more but \be
 \lim_{\e\to 0+0}\int_{-\infty}^{\infty} \frac{\rd \a_e \; e^{-\e\alpha_e^2}
\; e^{-it \tau_e \a_e}}{\a_e -p_e^2 +i \tau_e \eta} = (-i\tau_e)
e^{-it\tau_e p_e^2} e^{-\eta t} \, \label{1resuj} \ee still holds,
therefore $\xi_{0,t}^{(k)} = \lim_{\e\to 0+0}
\theta_{0,t,\e}^{(k)}$. In the integral (\ref{eq:intabare}) we pick
up a factor
\[\exp \Big[-\e
(\sum_{\bar e \neq e\in \bar v} \pm \alpha_e)^2 \Big]
\]
 which
converges to 1 in the limit. The integrations in
(\ref{eq:afterder}) are again absolutely convergent with or
without the regularizing factors. We therefore conclude the
following version of (\ref{eq:hiertheta}) for $\e>0$:
\begin{equation}\label{eq:hierthetareg}
i\partial_t \theta_{n,t,\e}^{(k)} (\bq_k ; \bq'_k) = \sum_{j=1}^k
\left(q_j^2 - (q'_j)^2 \right) \theta_{n,t,\e}^{(k)} (\bq_k;
\bq'_k) + i \left( B^{(k)} \theta_{n-1,t,\e}^{(k+1)}\right) (\bq_k
; \bq'_k) +o(1)
\end{equation}
as $\e\to0+0$. Integrating back this system of differential
equations, comparing the result with the solution to
(\ref{eq:hierxi}) and using that the difference in the initial
conditions vanish as $\e\to0$, we obtain that
$$
    \theta_{n,t,\e}^{(k)}=  \xi_{n,t}^{(k)} +o(1)
$$
for any fixed $n,k$ and uniformly on $t\in [0, T]$ for any fixed
$T>0$. Combining it with (\ref{eq:epsconv}), we obtain the
rigorous proof that  the differentiation in (\ref{dertheta}) is
allowed for our purposes. $\;\;\;\Box$

\subsection{Bounds for Amplitudes of Feynman Graphs}
\label{sec:boundgraph}

For brevity, we introduce the notation
$$
   \langle J^{(k)}, K_{\Gamma,t} \gamma^{(n+k)} \rangle: =
 \int \rd \bq_k \rd \bq'_k \rd \br_{n+k} \rd \br'_{n+k} \,
J^{(k)} (\bq_k ; \bq'_k) \, K_{\Gamma,t} (\bq_k , \bq'_k;
\br_{n+k}, \br'_{n+k}) \, \gamma^{(n+k)} (\br_{n+k}, \br'_{n+k})
$$
for an operator  $J^{(k)}\in \cK_{k}$ with  kernel $J^{(k)} (\bq_k
; \bq'_k)$ expressed in momentum space. We also define $\la
J^{(k)}, L_{\Gamma,t} \gamma^{(n+k)} \ra$ similarly. In the next
two theorems we show how to bound $\la J^{(k)}, K_{\Gamma, t}
\gamma^{(n+k)} \ra$ and $\la J^{(k)}, L_{\Gamma, t}
\gamma^{(n+k)}\ra$ (for $\Gamma \in \cF_{n,k}$ and for an
observable $J^{(k)}$ decaying sufficiently fast in momentum space)
in terms of the $\cH_{n+k}$-norm of $\gamma^{(n+k)}$. By Theorem
\ref{thm:graphs}, this will allow us to control the Duhamel
expansion (\ref{eq:duhamel}) of any solution of the infinite
hierarchy (\ref{eq:BBGKYint2}). Recall that the contribution $\la
J^{(k)},K_{\Gamma, t} \gamma^{(n+k)} \ra$ shows up in the analysis
of the fully expanded terms in (\ref{eq:duhamel}), while $\la
J^{(k)}, L_{\Gamma, t} \gamma^{(n+k)}\ra$ shows up in the error
term of (\ref{eq:duhamel}).

\begin{theorem}\label{thm:bounds}
Fix $k \geq 1$. For any $n \geq 1$,  suppose $\gamma^{(n+k)}$
is non-negative and symmetric with respect to permutations (in the
sense of (\ref{symm2})). Assume $0 < t \leq 1$. Choose $J^{(k)}
\in \cK_{k}$ that is symmetric with respect to permutations and whose
kernel satisfies
\begin{equation}\label{eq:Jbound}
|J^{(k)} (\bp_{k} ; \bp'_k)| \leq C \prod_{j=1}^k \frac{1}{\la p_j
\ra^3 \la p'_j \ra^3}\, .
\end{equation}
Then we have
\be \label{eq:JKG} \Big|  \langle J^{(k)},
K_{\Gamma,t} \gamma^{(n+k)} \rangle \Big| \leq C^{n} \, \tr \;
(1-\Delta_1) \dots (1-\Delta_{n+k}) \gamma^{(n+k)}\, \ee for every
$n \geq 0$.
\end{theorem}

\begin{theorem}\label{thm:boundsL}
Fix $k \geq 1$. For any $n \geq 1$,  suppose $\gamma^{(n+k)}$
is non-negative and symmetric with respect to permutations (in the
sense of (\ref{symm2})). Assume $0 < t \leq 1$. Choose $J^{(k)}
\in \cK_{k}$  symmetric with respect to permutations and with a
kernel satisfying (\ref{eq:Jbound}). Then, for every $n \geq 10+
k/2$, we have \be\label{eq:JLG} \Big| \langle J^{(k)},
L_{\Gamma,t} \gamma^{(n+k)} \rangle \Big| \leq C^{n} \,
t^{\frac{n}{4}} \tr \; (1-\Delta_1) \dots (1-\Delta_{n+k})
\gamma^{(n+k)}\, . \ee
\end{theorem}

{\it Remark.} The integer $k \geq 1$ is fixed.  The constant $C$ in
(\ref{eq:Jbound}) depends on $k$, and so do the constants on the
r.h.s. of (\ref{eq:JKG}) and (\ref{eq:JLG}). The restriction $t\leq
1$ plays no significant role in the theorems; it simplifies the
proof in a trivial manner. It is also possible to obtain a $t$-power
in (\ref{eq:JKG}) but we will not need it. Theorem \ref{thm:bounds}
will be applied to control the fully expanded terms in the Duhamel
series (\ref{eq:duhamel}), which, by Theorem \ref{thm:graphs}, can
be expressed in terms of the kernels $K_{\Gamma,t}$. For  our proof
of the uniqueness of the solution to the BBGKY hierarchy, it will be
sufficient to show that these terms are finite. They involve only
the initial condition, therefore these terms will be  identical for
any solution with the same initial data. On the other hand, Theorem
\ref{thm:boundsL}  will be applied to control the last  term in
(\ref{eq:duhamel}) involving a density matrix at an intermediate
time $s_n$. In this case it is not enough to show the finiteness of
the contributions; we also need to prove that they are small. It is
for this reason that in Theorem \ref{thm:boundsL} we need to extract
a time dependence and we will use the apriori bound
\eqref{eq:tbound} and  that $C^n t^{n/4}\to 0$ as $n\to\infty$ if
$t$ is small.
 Note also that the power
$n/4$ in (\ref{eq:JLG}) is not optimal and we do not aim at the
optimal $t$-dependence,  but we remark that this issue is
related to exploiting an additional smoothing effect of the free
Schr\"odinger evolution.

\bigskip

Before proving these theorems, we point out that the estimates
\eqref{eq:JKG}, (\ref{eq:JLG}) can be viewed as  Strichartz type
inequalities in the many particle setting. Recall that the
Strichartz inequality states  that
$$
        \int_0^t  \|  e^{is\Delta}f\|_p^r \; \rd s \leq C\| f \|_2^r,
   \qquad f\in L^2(\bR^3),
$$
for $r,p$ satisfying $\frac{2}{r} + \frac{3}{p}=\frac{3}{2}$ and
$2\leq r\leq \infty$. This inequality implies that
$$
        \int_0^t  \|  e^{is\Delta}f\|_p \; \rd s \leq C
t^{1-\frac{1}{r}}\| f \|_2\;,
$$
which  means that the free evolution  smoothes out possible
singularities of $f$  at the expense of reducing the $t$-power.

Another form of the Strichartz inequality asserts that
$$
   \Big\| \int_0^t  \rd s \, e^{i (t-s) \Delta}
    f_s\Big\|_{ L^{r}_{t} L^{p}_{x} }
   \le  C \| f_t\|_{L^{r'}_{t}
   L^{p'}_{x}}\, ,  \qquad f_t= f_t(x)\; ,
$$
where $ L^{r}_{t} L^{p}_{x}$ denotes the space $L^r(\bR;
L^p(\bR^3))$ and the positive exponents $p,r,p',r'$ satisfy
\begin{equation}
\frac 1 p + \frac 2 {3 r}  = \frac 1 2  ,  \quad    r\ge 2 \; ;
\qquad \frac 1 {p'} + \frac 2 {3 r'}  = \frac 7 6,    \quad r'\le
2\,.
\end{equation}
Once again, this estimate quantifies the smoothing effect of  the
free evolution  operator.  The price of reducing the $t$-power is
now expressed in terms of the change from the $L^{r}$ norm to the
$L^{r'}$ norm in the $t$ variable.

The kernels $K_{\Gamma,t}$ and $L_{\Gamma,t}$, though defined in
Green function form, actually have representations in terms of
$n$-fold time integrations like the formulae appearing in
\eqref{eq:fullexp} and \eqref{eq:error} (with the operator $B^{(m)}$
defined in \eqref{def:Bker}). If we replace the $\delta$-function
(which came from the two-body interaction) in the definition of
$B^{(m)}$ by a smooth function, the correct $t$ dependence in
(\ref{eq:JLG}) would be $t^n$ (at least for small $t$). The estimate
(\ref{eq:JLG}) states that the $\delta$  interaction is allowed if
we give up some power in $t$ --- in the same spirit as in the
Strichartz inequality.

Each integration step on the left hand side of  \eqref{eq:fullexp}
and \eqref{eq:error} actually involves a time integration and a
space integration via the partial trace in $B^{(m)}$. It would
therefore be natural to perform an iterative estimate involving
subsequent one-particle space-time dispersive bounds. Unfortunately,
we were unable to find an appropriate one-particle scheme to
implement this approach. Our method is much more complicated and it
involves tracking several singularity structures of the density
matrices in the integration step.

One reason to use the Feynman diagram representation is to obtain
estimates with correct  $n$ dependence. {F}rom the summation in the
definition of $B^{(m)}$ (see \eqref{def:B}), the number of terms on
the l.h.s of (\ref{eq:error}) is  $2^nk(k+1) \cdots (n+k-1) \sim
n!$. This factorial can be exactly compensated by the multiple time
integration, \be
    \int_0^t\rd s_1 \int_0^{s_1} \rd s_2 \ldots \int_0^{s_{n-1}}\rd s_n
    = \frac{t^n}{n!},
\label{ss} \ee but only if the $L^1$-norm is used in time. Higher
$L^r$-norms in time result in a partial loss of the $1/n!$ in
\eqref{ss} and thus the summation of these estimates over $n$ will
not converge  for any  $t$.

For this reason, we developed a new
method, based on the expansions (\ref{eq:fullexp}) and
(\ref{eq:error}) in terms of Feynman graphs.
Our graphical representation,  among its other merits,
reduces the number of terms in the expansion from $n!$ to $C^n$
(see \eqref{number}).
 This combinatorial reduction
stems from  combining  graphs like b) and c) on Fig.~\ref{fig:order}.
The dispersive properties of $e^{it\Delta}$ are now captured by the
decay properties of the integrands
in the kernels $K_{\Gamma,t}$, $L_{\Gamma,t}$ (see
(\ref{simpler}), (\ref{Lsimpler})). These multiple integrations
can be successively performed and we thus obtain
the bounds (\ref{eq:JKG}), (\ref{eq:JLG}). Our representation
treats all smoothing effects simultaneously and thus exploits an
additional decay which we were not able to obtain with one-particle
methods. However, we do not know if it is possible to design a one-particle
inequality similar to the Strichartz inequality to give a short proof for
 the estimates \eqref{eq:JKG} and (\ref{eq:JLG}) .

\subsection{Proof of  Theorem \ref{thm:bounds}}

To better explain how the dispersive properties of the free
evolution are used in our approach, we first discuss
the main ideas involved in the proof of (\ref{eq:JKG}) on a
heuristic level;
similar ideas apply to the proof of
(\ref{eq:JLG}). We have to bound  all
integrals over the three dimensional momentum variables $p_e$ and
over the one-dimensional variables $\a_e$ appearing in the
definition of the kernel $K_{\Gamma,t}$
(see
(\ref{simpler})). Notice that, because of the
singularity of the $\delta$-potential in position space, here we
face a large momentum problem: we have to make sure that
all integrals over $p_e$ (and $\a_e$) are convergent in the large
$p_e$ (respectively, large $\a_e$) regime. To this end we will
develop an integration scheme dictated by the structure of the
Feynman graph $\Gamma$.

We will start by
integrating over the variables $p_e$ and $\a_e$ associated with the
leaves of $\Gamma$. The momenta on the leaves are exactly
the variables of the density $\gamma^{(n+k)}$. The
factor $\tr
(1-\Delta_1) \dots (1-\Delta_{n+k}) \gamma^{(n+k)}$ on the r.h.s. of
(\ref{eq:JKG}) implies that each leaf carries
 a decaying factor $|p_e|^{-(2+\lambda)}$ (for
large $p_e$), for any $\lambda <1/2$. In the rigorous proof
we will have stronger restrictions on the value of $\lambda$. The
idea then is that we integrate over all the $p$ and $\a$ variables,
starting from the leaves and then moving towards the roots. At
each step we propagate the momentum decay from the son-edges to the
father edge of a certain vertex of $\Gamma$ by integrating out the
variables of the son-edges.

\begin{figure}
\begin{center}
\epsfig{file=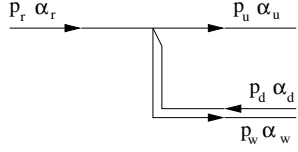,scale=.75}
\end{center}
\caption{Integration scheme: a typical vertex}\label{fig:verte}
\end{figure}

A typical step in the integration scheme is as follows: choose a
vertex $v \in V(\Gamma)$ such that we already have demonstrated a
decay $|p|^{-(2+\lambda)}$ in the momenta $p_u$, $p_d$, $p_w$ of the
three son edges of $v$ (denoted by $u,d$ and $w$, see Figure
\ref{fig:verte}). This means that all the momentum- and
$\a$-variables of the edges which are to the right of $u,d$ and $w$
in the graph have already been integrated out.
Then we first perform the integration over the three $\a$-variables of the
son edges. By power counting, we obtain formally
\begin{equation*} \int \rd \a_u \rd \a_d \rd \a_w \frac{\delta
(\a_r = \a_u +\a_d - \a_w)}{\langle \a_u - p_u^2 \rangle \langle
\a_d - p_d^2 \rangle \langle \a_w -p_w^2 \rangle} \leq
\frac{\const}{\langle \a_r -p_u^2 - p_d^2 + p_w^2 \rangle}
\end{equation*}
(up to logarithmic factors). Then we integrate over the momenta of
the son-edges by using their decay factor
 and we obtain, again by simple power counting,
\begin{equation*}
\int \frac{\rd p_u \rd p_d \rd p_w}{|p_u|^{2+\lambda}
|p_d|^{2+\lambda} |p_w|^{2+\lambda}} \, \frac{\delta (p_r = p_u +p_d
-p_w)}{\langle \a_r -p_u^2 - p_d^2 + p_w^2 \rangle} \leq
\frac{\const}{ |p_r|^{2 +\lambda}}
\end{equation*}
Here the power counting requires that  $3 \, (2+\lambda) + 2 - 6 > 2 + \lambda$ which holds  for any $\lambda
>0$. Thus the same decay in the large momentum regime propagated
from the son-edges to the father edge. This procedure can then be
iterated until we reach the roots of $\Gamma$. At that point we can
complete the integration scheme by using the smoothness (momentum
decay) of the observable $J^{(k)}$.

In this formal computation, the dispersive
nature  of the free evolution is expressed via the decay
 in $p$ and $\a$ of the resolvent $\langle \a - p^2 \rangle^{-1}$.
The decay in both  variables is critical to complete
the integration scheme. The rigorous proof of Theorems
\ref{thm:bounds} and \ref{thm:boundsL} is more involved than this simple
model calculation
 because the singularity structure
of  $\langle \a - p^2 \rangle^{-1}$ is spherical and it cannot
be described by a simple power counting alone.
We will have to consider edges of various types,
characterized by different decay properties,
 and we still have to close the iteration scheme.

\begin{proof}[Proof of Theorem \ref{thm:bounds}]
{F}rom the definition of $K_{\Gamma,t}$ in (\ref{eq:KGamma2}), we
have
\begin{equation}\label{eq:JKG1}
\begin{split}
 \langle J^{(k)}, K_{\Gamma,t} &\gamma^{(n+k)} \rangle =
\frac{1}{(n+k)!} \sum_{\pi_2 \in S_{n+k}} \int \rd \bq_k \rd \bq'_k
\rd \br_{n+k} \rd \br'_{n+k} \, J^{(k)} (\bq_k ; \bq'_k) \,
\gamma^{(n+k)} (\br_{n+k}, \br'_{n+k}) \\ &\times \prod_{e \in R_1
(\Gamma) = L_1(\Gamma)} (-i\tau_e) e^{-it \tau_e (q_{\pi_1
(e)}^{\sharp_e})^2
} \, \delta (q^{\sharp_e}_{\pi_1 (e)} - r^{\sharp_e}_{\pi_2 (e)}) \\
& \times \int \prod_{e \in E_2 (\Gamma)} \rd \alpha_e \rd p_e
\prod_{e \in R_2 (\Gamma)} \delta (p_e - q^{\sharp_e}_{\pi_1 (e)})
\prod_{e \in L_2 (\Gamma)} \delta (p_e - r^{\sharp_e}_{\pi_2(e)})
\;e^{-it \sum_{e\in R_2 (\Gamma)} \tau_e(\alpha_e+i\tau_e\eta_e)}
\\ &\times \prod_{e \in E_2 (\Gamma)} \frac{1}{\a_e - p_e^2 +
i\tau_e\eta_e} \prod_{v \in V(\Gamma)} \delta \Big(\sum_{e\in v}
\pm \alpha_e \Big) \delta\Big(\sum_{e\in v}\pm p_e \Big) \, .
\end{split}
\end{equation}
Because of the permutation symmetry of $\gamma^{(n+k)}$, the
integral has the same value for every choice of $\pi_2$. Hence,
instead of averaging, we fix $\pi_2 \in S_{n+k}$.  We define the
sets
\be
Q_1 : = \{ e \in L(\Gamma): \tau_e =1\}\; , \quad\mbox{ and }
\quad Q_2 = \{ e \in L(\Gamma): \tau_e =-1\} ;
\label{def:Q}
\ee
 that is $Q_1$ is the set of outward
leaves and $Q_2$ is the set of inward leaves. Clearly $L(\Gamma) =
Q_1 \cup Q_{2}$ and $|Q_1|=|Q_2|=n+k$. We use the notation
$\gamma^{(n+k)} (\{ (p_e;p_{e'}) \}_{e \in Q_1})$ to stress the
fact that, because of the permutation symmetry, the density
$\gamma^{(n+k)}$ only depends on the set of pairs $(p_e;p_{e'})$
of momenta associated with the (paired) leaves of $\Gamma$, and
not on the order of the pairs. Integrating over the variables
$\bq_k, \bq'_k$ and $\br_{n+k}, \br'_{n+k}$ and using all the
delta-functions, the absolute value of (\ref{eq:JKG1}) can be
estimated by
\begin{equation}\label{eq:JKG2}
\begin{split}
C e^{t \sum_{e \in R(\Gamma)} \eta_e} \int  &\prod_{e \in
E(\Gamma)} \rd p_e \prod_{e \in E_2 (\Gamma)} \rd \a_e \, \prod_{e
\in E_2 (\Gamma)} \frac{1}{|\a_e - p_e^2 + i\tau_e \eta_e|}
\prod_{e \in R(\Gamma)} \frac{1}{\la p_e \ra^3} \\ &
\hspace{2cm}\times \prod_{v \in V(\Gamma)} \delta (\sum_{e\in v}
\pm \a_e) \, \delta (\sum_{e\in v} \pm p_e) \; \Big|\gamma^{(n+k)}
( \{ (p_e;p_{e'}) \}_{e \in Q_1}) \Big| \, ,
\end{split}
\end{equation}
where the factor $\prod_{e\in R(\Gamma)} \la p_e \ra^{-3}$ comes
from estimating the observable $J^{(k)}$ using the assumption
(\ref{eq:Jbound}).

Since the observable $J^{(k)}$ is symmetric w.r.t. permutations,
and since $K_{\Gamma,t}$ preserves the symmetry, to compute the
quantity on the l.h.s. of (\ref{eq:JKG1}) we can replace the
density $\gamma^{(n+k)}$ by its restriction onto the subspace
$L^2_s (\bR^{3(n+k)})$ consisting of all permutation symmetric
functions in $L^2 (\bR^{3(n+k)})$. Hence, $\gamma^{(n+k)}$ can be
written as $\gamma^{(n+k)} = \sum_{j} \lambda_j |\psi_j \rangle
\langle \psi_j |$, with $\psi_j \in L^2_s (\bR^{3(n+k)})$ such
that $\| \psi_j \| =1$ for all $j$, with $\lambda_j \geq 0$ for
all $j$ (by the non-negativity of $\gamma^{(n+k)}$), and with
$\sum_j \lambda_j < \infty$. Hence it is enough to prove the bound
(\ref{eq:JKG}) for $\gamma^{(n+k)}$ being a one-dimensional
projection. In the following we therefore  assume that
$\gamma^{(n+k)} (\bp_{n+k} ; \bp'_{n+k}) = \psi (\bp_{n+k})
\overline{\psi} (\bp'_{n+k})$.

We again use the notation $\psi (\{ p_e \}_{e \in
Q_1})$ to indicate that $\psi$ is a function of the set of the
momenta associated with leaves in $Q_1$, and not of their order.
Moreover we choose $\eta_e = 1/t$, for all $e \in L(\Gamma)$: this
implies that $\eta_e \geq 1/t$ for every $e \in E(\Gamma)$, and
$\sum_{e \in R(\Gamma)} \eta_e = (2n+1)/t$. With a weighted
Schwarz inequality, we can then bound (\ref{eq:JKG2}) by
\begin{equation}\label{eq:long}
\begin{split}
C &e^{2n+1} \int \prod_{e \in E(\Gamma)} \rd p_e \, \prod_{e \in
E_2 (\Gamma)} \frac{\rd \a_e}{|\a_e - p_e^2 + \frac{i}{t}|} \,
\prod_{e \in R(\Gamma)} \frac{1}{\la p_e \ra^3}\, \prod_{v \in
V(\Gamma)}
\delta (\sum_{e\in v} \pm \a_e) \delta (\sum_{e\in v} \pm p_e) \\
& \hspace{1.5cm}\times \left( \frac{ \prod_{e \in Q_1} p_e^2
}{\prod_{e \in Q_2} p_e^2} |\psi ( \{ p_e \}_{e \in Q_1} )|^2 +
\frac{ \prod_{e \in Q_2} p_e^2 }{\prod_{e \in Q_1} p_e^2} |\psi (
\{ p_e \}_{e \in Q_2} )|^2 \right) \, \\ \leq \; &C^{n} \int \rd
\bp_{n+k} \, p_1^2 \dots p_{n+k}^2 |\psi (\bp_{n+k})|^2
\\ &\times \left( \sup_{\{ p_e \}_{e \in Q_1}} \int \prod_{e \in
E(\Gamma) \backslash Q_1} \rd p_e \prod_{e \in E_2 (\Gamma)}
\frac{\rd \a_e}{\la \a_e - p_e^2 \ra} \prod_{e \in R(\Gamma)}
\frac{1}{\la p_e \ra^3} \, \prod_{e \in Q_2}
\frac{1}{p_e^2}\right.  \prod_{v \in V(\Gamma)} \delta (\sum_{e\in
v} \pm \a_e)\, \delta (\sum_{e\in v} \pm p_e) \;
\\ &+ \sup_{\{ p_e \}_{e \in Q_2}} \int \prod_{e \in
E(\Gamma)\backslash Q_2 } \rd p_e \, \prod_{e \in E_2 (\Gamma)}
\frac{\rd \a_e}{\la \a_e - p_e^2 \ra} \prod_{e \in R(\Gamma)}
\frac{1}{\la p_e \ra^3} \, \prod_{e \in Q_1} \frac{1}{ p_e^2}
\left. \prod_{v \in V(\Gamma)} \delta (\sum_{e\in v} \pm \a_e) \,
\delta (\sum_{e\in v} \pm p_e) \right) \\ =: &\; A +  B
\end{split}
\end{equation}
where we used that, since we assumed that $t \leq 1$, $|\a_e
-p_e^2 +i/t|^{-1} \leq \la \a_e -p^2_e \ra^{-1}$. In the
contribution $A$, resulting from the first term in the
parenthesis, we have taken the supremum over all momenta $p_e$
associated with the leaves in  $Q_1$. We will refer to this
estimate as {\it freezing} these momenta and the corresponding
legs $e\in Q_1$ will be called {\it dead-edges}, the rest are
called {\it live-edges}.  In the contribution $B$ we froze all
momenta of the leaves $Q_2$.

In order to bound these integrals, we will successively
 integrate over all $\a$ variables and over
all non-frozen momenta starting from the leaves, until we are left
with an integral involving only the momenta of the roots. Finally,
we integrate out the root momenta: at this point we will also make
use of the decay factor $\prod_{e \in R(\Gamma)} \la p_e \ra^{-3}$
we gained from the test-function $J^{(k)}$.

The large $p_e$ and large $\alpha_e$ regimes are critical for the
convergence of our integrals. The decay of the non-frozen
leaf-momenta, $|p_e|^{-2}$, alone
 is not sufficient to render  these integrals finite;
for the intermediate edges even such decay is not available. The
propagators provide extra decays, $\langle \alpha
-p_e\rangle^{-1}$, but they may disappear (even with a possible
logarithmic divergence) in the $\rd\alpha_e$-integrals. On the
other hand, the delta functions of course help since they  reduce
the effective number of integrals. Finally, the test-function
provides a strong decay for the root variables (\ref{eq:Jbound}).
Due to the complexity of this structure, it requires a carefully
designed successive integration scheme, combined with appropriate
bounds, to show that these multiple integrals are actually
convergent. The precise bound will then easily follow along the
same lines. Unfortunately, the scheme is complicated by fact that
in certain estimates (namely when Lemma \ref{lm:plane} is applied)
mild local point singularities arise. This is unavoidable even if
one uses weights $\langle p_e \rangle^2$ instead of $p_e^2$  in
the Schwarz inequality in (\ref{eq:long}). Therefore some care is
needed to avoid accumulation of local divergences. We suggest the
reader to neglect this issue at the first reading and concentrate
only on the large momentum regime of the estimates in the proof of
(\ref{eq:JKG}).

Next we illustrate the successive integration scheme for
(\ref{eq:long}):
 we consider the term $A$, where the momenta of
the edges in $Q_1$ (recall $Q_1$ is the set of outward leaves) are
frozen. The analysis of the $B$ is analogous.

Since the delta functions always relate variables within the
connected components of $\Gamma$, the integrations can be done
independently in each connected component (tree) of $\Gamma$. The
order of integration is prescribed by the tree structure: we start
from the leaf-variables and proceed toward the root.

The key step is what we call {\it integrating out a vertex}. It
consists in integrating over the $\a$ variables and the  momenta
of the live son-edges of this vertex. The integral will be
estimated in terms of the $\a$-variable and the momentum variable
 of the father-edge and in terms of
the frozen momenta of the dead edges from the set $\{ \ell \in
Q_1: v \text{ lies on the route from $\ell$ to its root} \}$. A
vertex $v$ will be integrated only when all vertices $v'$ with
$v'\succ v$ have already been integrated out.

More precisely, we define an increasing sequence of subsets of the
vertices $V(\Gamma)$, $V_1(\Gamma)\subset V_2(\Gamma)
\subset\ldots\subset V(\Gamma) $,
 where $V_m(\Gamma)$ contains all vertices that have  been integrated out
after the first $m$ integration steps, in particular
$|V_m(\Gamma)|=m$. In the $(m+1)$-th integration step we integrate
out one of the maximal vertices in the set $V(\Gamma) \setminus
V_m(\Gamma)$. The maximality is considered with respect  to the
ordering defined by the restriction of $\prec$ onto  $V(\Gamma)
\setminus V_m(\Gamma)$. After $m=|V(\Gamma)|$ integration steps
all vertices have been integrated out.

The process of integrating out each last vertex $v$ (a vertex
whose father-edge is a root) in the $2k-|R_1 (\Gamma)|$
non-trivial connected components of $\Gamma$ is a little bit
different. In this case we integrate simultaneously over the
$\a$-variable of the son-edges and of the father-edge, and, like
in the other vertices, we integrate over the momenta of the
son-edges. Here we will estimate the integrals in terms of the
momentum of the father-edge, and of the momenta of the dead leaves
in the connected component we are considering. As a result, after
integrating out all vertices of $\Gamma$, we will be left with an
integral over the the momenta associated with the roots: the
integrand will depend on the root-momenta, on the dead momenta and
on the observable $J^{(k)}$.

Along the procedure we keep track of the available decay factors
for each edges. Every edge carries its own propagator, $\langle
\alpha_e-p_e^2\rangle^{-1}$, and we will focus on the additional
decay factors. We see from (\ref{eq:long}) that the every momentum
associated with a live leaf carries a decaying factor $1/|p_e|^2$.
We will show that when we integrate out a vertex,   a similar
polynomial decaying factor can be propagated to the father-edge.
Because of the propagators $\langle \a_e - p_e^2 \rangle^{-1}$, we
will actually be able to gain more decay in the momenta of the
father-edges, which will be available for the next integration,
when the momentum of such a father-edge will be integrated out as
a son-edge of the next vertex. The additional decay is typically
in the form of point singularity with a power higher than 2 (i.e.
of the form $|p_e-a|^{-2-\kappa}$, $\kappa>0$, with a possible
shift $a$ depending only on dead-momenta), but sometimes a
so-called {\it spherical decay}  in the form $\la \a_e +\beta_j +
(p_e - b_j)^2 \ra^{-1}$ arises, where the shifts $b_j, \beta_j$
depend only on the dead-edge momenta.

We will therefore distinguish different types of edges, according
to the momentum-decay they carry. For $e \in E_2 (\Gamma)$, we
define the set of vertices
$$
V_e = \{ v \in V(\Gamma) : e \text{ lies on the route
 from $v$ to its root} \},
$$
and the set
of dead leaves
\be
D_e = \{ \ell \in Q_1 : e \text{ lies on the route
from $\ell$ to its root}\} .
\label{def:De}
\ee
 We choose $\lambda >0$ and $\e >0$
sufficiently small (the correct conditions will be specified later
on). Then we have the following type of edges:
\begin{itemize}
\item[i)] \underline{$d$-edges}: these are the dead edges, over
whose momenta we do not integrate. Dead-edges are always leaves;
note that one companion of each pair of leaves is dead, the other
is live. \item[ii)] \underline{$2$-edges}: they carry a factor
\begin{equation}\label{eq:ii)}
\frac{1}{|p_e|^2} \, .
\end{equation}
These edges are exactly the live leaves. \item[iii)]
\underline{$(2+\lambda)$-edges and $(2+2\lambda)$-edges}: they
carry a sum of decaying factors
\begin{equation}\label{eq:iii)}
\sum_{j=1}^{\nu (e)} \frac{1}{|p_e -a_j|^{2+\lambda}}, \quad
\text{respectively } \quad  \sum_{j=1}^{\nu (e)} \frac{1}{|p_e
-a_j|^{2+2\lambda}} \end{equation} where $a_j$ are linear
combinations of the momenta of the dead-edges lying in $D_e$. Here
the number of terms, $\nu (e)$ is bounded by $\nu(e) \leq
C^{|V_e|}$, for a universal constant $C$.

\item[iv)] \underline{$(2+s+\kappa)$-edges} with
$\kappa=0,\lambda, 2\lambda$: they carry a sum of decaying factors
of the form:
\begin{equation}\label{eq:s}
\sum_{j=1}^{\nu (e)} \frac{1}{|p_e-a_j|^{2+\kappa}} \frac{1}{\la
\a_e +\beta_j + (p_e - b_j)^2 \ra^{1-\e}}\; .  \end{equation} Here
$a_j$, $b_j$,
 are linear combinations of the momenta of the dead-edges
in $D_e$, the numbers $\beta_j$ are quadratic functions of the
same momenta. The number of terms, $\nu (e)$, is bounded by $\nu
(e) \leq C^{|V_e|}$, for a universal constant $C$ (the symbol $s$
in $(2+s+\kappa)$ refer to the  ``spherical decay'' $\la \a +\beta
+(p-a)^2\ra^{-1+\e}$). If $e$ is a root edge, then it can still be
of the type $(2+s)$, $(2+s+\lambda)$ or $(2+s+2\lambda)$, but in
this case, in (\ref{eq:s}) we replace $\a_e$ with $p_e^2$.
\end{itemize}
The summations in (\ref{eq:iii)}) and (\ref{eq:s})  reflect
different  cases that originate from the fact that the three son-edges
of a vertex do not play fully symmetric roles. The precise number
and the possible relations among $a_j, b_j, \beta_j$ of these
terms play no role in the procedure since all our estimates will
be uniform in these shift variables. The reader can therefore
safely neglect the complicated structure of (\ref{eq:iii)}) and
(\ref{eq:s}) at the first reading. The only important issue is the
type of singularity and the power $\kappa$; these information are
carried in the shorthand notation $2$, $2+\kappa$, $2+s+\kappa$.

We will show that every time three edges of one of these types
meet as son-edges at a vertex, the father-edge will be again of
one of these types after integrating out this vertex.
 We will prove the following transitions, to
determine the type of the father-edge after integration, given the
types of the son-edges.
\begin{align}
\label{eq:cases} 1) \;& (d,d,2+\kappa) \to 2+s+\kappa, \quad
\text{for }
   \kappa=0,\lambda,2\lambda \nonumber \\
2) \;& (d,2+\kappa_1,2+\kappa_2) \to 2+2\lambda, \quad \text{for
} \kappa_{1,2} =0,\lambda,2\lambda \nonumber \\
3) \;& (2+\kappa_1,2+\kappa_2,2+\kappa_3) \to 2+2\lambda, \quad
\text{for } \kappa_{1,2,3} = 0,\lambda,2\lambda \quad \text{with }
\kappa_1 + \kappa_2 + \kappa_3 \geq 3\lambda \\
4) \;& (2+ 2\lambda, 2,2) \to 2+\lambda \nonumber \\
5) \;& (2+s+\kappa_1, 2+ \kappa_2,2+\kappa_3) \to 2+2\lambda \quad
\text{for } \kappa_{1,2,3} = 0, \lambda, 2\lambda  \nonumber
\end{align}
The short notation $(A, B, C)\to D$ means that $A, B, C$ type
decays on the son-edges yields a $D$-type decay on the father-edge
after integrating out the vertex. The order of $A, B, C$ is
irrelevant. For example, transition 2) is a short-hand writing of
the following estimate: \be\begin{split}
    \sup_{\alpha_r}\sum_{i=1}^{\nu(e_p)} \sum_{j=1}^{\nu(e_q)}
 \int \int_\bR  & \frac{\delta (\a_r = \a_p +\a_{q} - \a_{q'})
\; \rd \a_p \rd \a_q \rd\alpha_{q'} \delta(r=p+q-q')\rd p \rd
q}{\la \a_p - p^2 \ra \,|p- a_i|^{2+\kappa_1} \; \la \a_q - q^2
\ra \,  \la \a_{q'} - (q')^2 \ra\,
 |q'- b_j|^{2+\kappa_1} } \\
 & \leq C \sum_{k=1}^{\nu(e_r)}
 \frac{1}{ | r - c_k|^{2+2\lambda}}
\end{split}\ee
(see Fig.  \ref{fig:alfaqdead} for the notation, where we choose
the dashed line with momentum $q'$ to be the dead-edge for
definiteness). Here $c_k$ are linear combinations of $a_i, b_j$
variables and of the dead edge momentum~$q'$.
\begin{figure}
\begin{center}
\epsfig{file=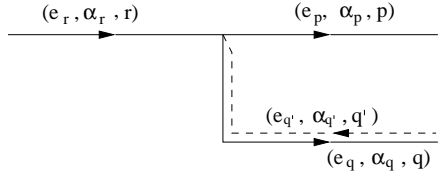,scale=.75}
\end{center}
\caption{The vertex of transition 2)}\label{fig:alfaqdead}
\end{figure}

\medskip

The transitions (\ref{eq:cases}) have to be combined with the
following set of rules, to see that they indeed form a closed
system along the successive vertex integration. The rules are as
follows:

\begin{itemize}
\item[a)] At every vertex, there are at most two $d$-edges. This
is clear, because $d$-edges are always outward pointing leaves,
and it is impossible to have a vertex with three son-edges having
the same orientation. \item[b)] At every vertex, there are at most
two $2$-edges: this follows analogously to a), because all
$2$-edges are inward pointing leaves. \item[c)] There is no vertex
with son-edges $(2+\lambda,2,2)$. Note that the $(2+\lambda)$-edge
can only result as the father-edge of a vertex with son-edges
$(2,2,2+2\lambda)$, and thus (since the $2$-edges always point
inward) it must have inward orientation. But all three son-edges
of a vertex cannot have the same orientation. \item[d)] There is
no vertex with son-edges $(2+\lambda,2+\lambda,2)$. As we have
seen in b) and c), all $2$-edges and $(2+\lambda)$-edges have
inward orientation and all three son-edges of a vertex cannot have
the same orientation.
\end{itemize}

Combining these rules with the transitions in (\ref{eq:cases}),
and observing that the spherical denominator $\la \a + \beta
+(p-a_j)^2 \ra^{-1+\e}$ can be always estimated by one (i.e. $+s$
can always be removed  from $2+s+\kappa$ for free), we see that we
have a closed system starting solely from $d$-edges and $2$-edges.
Along the integration, every edge in $\Gamma$ will become one of
the types described in i)-iv) above and each  vertex integration
corresponds to one of the steps 1)--5).

\medskip

Before proving the transitions 1)-5) rigorously, let us indicate
their validity by a simple power counting argument. All integrals
are locally convergent if $\lambda <1/3$, so it is sufficient to
focus on their large momentum (short distance) behavior.  The
integration variables $p_e$ are momentum variables with dimension
[{\it length}]$^{-1}$.
 Because of the propagators $\la \a_e -p_e^2 \ra^{-1}$,
the $\alpha_e$ variables have the same dimension as $p_e^2$, i.e.
[{\it length}]$^{-2}$. Hence the three propagators associated with
the three son-edges always have the dimension [{\it
length}]$^{6}$.

In the transition 1), we have effectively two $\a$ integrations
and no momentum integration. In fact, there are three
$\a$-variables and one live momentum associated with the three
son-edges of the vertex under consideration; but, because of the
$\delta$-functions $\delta (\sum_{e\in v} \pm \a_e)$ and $\delta
(\sum_{e\in v} \pm p_e)$, we only have to perform two $\a$
integration. Each  $\rd \a$-integration carries the dimension
[{\it length}]$^{-2}$. Since the momentum decay factor associated
with the $2+\kappa$-edge has the dimension [{\it
length}]$^{2+\kappa}$, the result of the integral has the
dimension [{\it length}]$^{6-2\cdot 2 +(2+\kappa)}$=[{\it
length}]$^{4+\kappa}$. The r.h.s. of 1), on the other hand, has
the dimension [{\it length}]$^{4+\kappa -2\e}$. Therefore $\e
>0$ guarantees that in the relevant short distance regime, the
r.h.s is indeed bigger than the l.h.s., as the corresponding
exponent on the r.h.s is smaller than the exponent on the left. We
lose some decay in our estimates to compensate for logarithmic
factors.

In the transition 2), we have
effectively two $\a$-integrations and one momentum integration
($\rd p_e$ has the dimension [{\it length}]$^{-3}$): hence the left side
of 2) has the dimensions [{\it length}]$^{3+\kappa_1+\kappa_2}$, while
the r.h.s. has the dimension [{\it length}]$^{2+2\lambda}$. For
$\lambda$ small enough, the exponent on the right is smaller than
the exponent on the left.

In 3),4), and 5) we have two $\a$- and
two momentum integrations. The dimension of the left side of 3) is
[{\it length}]$^{2+\kappa_1+\kappa_2+\kappa_3}$; the right side has the
dimension [{\it length}]$^{2+2\lambda}$. This explains where the
condition $\kappa_1+\kappa_2+\kappa_3 \geq 3\lambda$ comes from.
The transition 4) is similar. As for 5), the left side has the
dimension [{\it length}]$^{4+\kappa_1+\kappa_2+\kappa_3}$, and the right
side has the dimension [{\it length}]$^{2+2\lambda}$: again, for
$\lambda$ small enough, the exponent on the left is larger than
the exponent on the right.

\medskip

Next, we give a rigorous proof of the transitions
(\ref{eq:cases}). Choose a vertex $v \in V(\Gamma)$ and assume we
already performed the integration over all $v' \succ v$. We start
by performing the integration over the $\a$-variables associated
with the son-edges of the vertex $v$. Let $e_r$ denote the father
edge of the vertex $v$ (see  Fig. \ref{fig:alfa}). We distinguish
two cases according to whether $e_r$ is the root or not.

Suppose first that $e_r$ is not a root (that is $v$ is not the
last vertex left).
\begin{figure}
\begin{center}
\epsfig{file=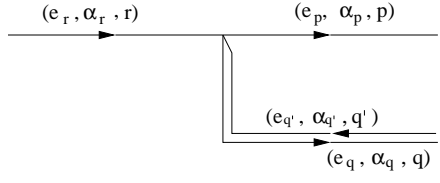,scale=.75}
\end{center}
\caption{Integrating out a vertex}\label{fig:alfa}
\end{figure}
If there is no spherical denominator among the son-edges, that is,
if the son-edges are all of the type $d$, $2$, $2+\lambda$ or
$2+2\lambda$, then we estimate the $\a$-integrations, for a fixed
and small enough $\e
>0$, by (with the notation of Fig. \ref{fig:alfa}):
\begin{equation}\label{eq:al}
\begin{split}
\int \rd \a_p \rd \a_q \rd \a_{q'} &\delta (\a_r = \a_p + \a_{q} -
\a_{q'}) \frac{1}{\la \a_p - p^2 \ra \, \la \a_q - q^2 \ra \, \la
\a_{q'} - (q')^2 \ra} \\ &= \int \rd \a_p \rd \a_q \frac{1}{\la
\a_p - p^2 \ra \, \la \a_q - q^2 \ra \, \la \a_{r} - \a_p -\a_q +
(q')^2 \ra}\\
&\lesssim \frac{1}{\la \a_r - p^2 - q^2 + (q')^2 \ra^{1-\e}} \, ,
\end{split}
\end{equation}
where we applied Lemma \ref{lm:pre1} twice, once in the $\a_q$ and
once in the $\a_p$-integration (after estimating $\la \a \ra^{-1}
\leq |\a|^{-1+ \delta}$, for a sufficiently small $\delta
>0$). On the other hand, if one of the son-edges (say the $e_p$
edge in Fig. \ref{fig:alfa}) is of the type $2+s$, $2+s+\lambda$
or $2+s+2\lambda$, then we use the bound
\begin{equation}\label{eq:al2}
\begin{split}
\int \rd &\a_p \rd \a_q \rd \a_{q'}  \frac{\delta (\a_r = \a_p +
\a_{q} - \a_{q'})}{\la \a_p - p^2 \ra \, \la \a_p + \beta +
(p-a)^2 \ra^{1-\e}
\la \a_q - q^2 \ra \, \la \a_{q'} - (q')^2 \ra} \\
&= \int \rd \a_p \rd \a_q \frac{1}{\la \a_p - p^2 \ra \, \la \a_p
+ \beta + (p-a)^2 \ra^{1-\e}\,  \la \a_q - q^2 \ra \, \la \a_{r} -
\a_p -\a_q + (q')^2 \ra}
\\ &\lesssim \int \frac{\rd \a_p}{\la \a_p \ra \, \la \a_p
+p^2 + \beta + (p-a)^2 \ra^{1-\e} \, \la \a_{r} - \a_p - p^2 - q^2
+ (q')^2 \ra}\\
&\lesssim \frac{1}{\la \beta + p^2 + (p-a)^2 \ra^{1-\e}} \int \rd
\a_p \left(\frac{1}{\la \a_p \ra} + \frac{1}{\la \a_p + \beta +p^2
+ (p-a)^2 \ra} \right) \frac{1}{\la \a_r - \a_p - p^2 - q^2 +
(q')^2 \ra} \\ &\lesssim \frac{1}{\la \wt\beta + (p- \wt a)^2
\ra^{1-\e}} \left( \frac{1}{\la \a_r - p^2 - q^2 + (q')^2
\ra^{1-\e}} + \frac{1}{\la \a_r + \beta + (p-a)^2 - q^2 + (q')^2
\ra^{1-\e}} \right)\,.
\end{split}
\end{equation}
where $\wt \beta$ and $\wt a$, like $\beta$ and $\a$ depend only
on the frozen momenta associated to the dead leaves in $D_{e_r}$
(see the definition (\ref{def:De})).

Suppose now that $e_r$ is a root, that is there
is no vertex $\tilde v$ with $\tilde v \prec v$. In this case we
also integrate over the $\a$ variable associated with the
father-edge $e_r$. We use
\begin{equation}\label{eq:al1root}
\begin{split}
\int \rd \a_r \rd \a_p \rd \a_q \rd \a_{q'} &\delta (\a_r = \a_p +
\a_{q} - \a_{q'}) \frac{1}{\la \a_r - r^2 \ra \la \a_p - p^2 \ra
\, \la \a_q - q^2 \ra \, \la \a_{q'} - (q')^2 \ra} \\ &= \int \rd
\a_r \rd \a_p \rd \a_q \frac{1}{\la \a_r -r^2 \ra \la \a_p - p^2
\ra \, \la \a_q - q^2 \ra \, \la \a_{r} - \a_p -\a_q +
(q')^2 \ra}\\
&\lesssim \frac{1}{\la r^2 - p^2 - q^2 + (q')^2 \ra^{1-\e}} \, ,
\end{split}
\end{equation}
if there is no spherical singularity in the son-edges, and
\begin{equation}\label{eq:al2root}
\begin{split}
\int \rd & \a_r \rd \a_p \rd \a_q \rd \a_{q'}  \frac{\delta (\a_r
= \a_p + \a_{q} - \a_{q'})}{\la \a_r - r^2 \ra \la \a_p - p^2 \ra
\, \la \a_p + \beta + (p-a)^2 \ra^{1-\e}
\la \a_q - q^2 \ra \, \la \a_{q'} - (q')^2 \ra} \\
&\lesssim \frac{1}{\la \wt\beta + (p- \wt a)^2 \ra^{1-\e}} \left(
\frac{1}{\la r^2 - p^2 - q^2 + (q')^2 \ra^{1-\e}} + \frac{1}{\la
r^2 + \beta + (p-a)^2 - q^2 + (q')^2 \ra^{1-\e}} \right)\,.
\end{split}
\end{equation}
if there is a spherical denominator in the son-edges.

\begin{figure}
\begin{center}
\epsfig{file=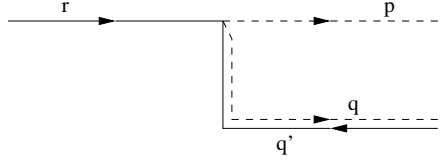,scale=.75}
\end{center}
\caption{A vertex with two dead edges}\label{fig:2dead}
\end{figure}
Next we have to estimate the momenta integrations. Let us first
consider the case $1)$ in (\ref{eq:cases}). Since we have two dead
edges and one momentum delta-function, effectively no integration
needs to be done. With the notation of Fig. \ref{fig:2dead}, where
dashed lines indicate dead-edges, we obtain, using the result of
(\ref{eq:al}),
\begin{equation}\label{eq:one}
\begin{split}
\int \frac{\rd q'}{|q'-a|^{2+\kappa}} \frac{\delta (r =
p+q-q')}{\la \a_{r} - p^2 - q^2 + (q')^2 \ra^{1-\e}} &=
\frac{1}{|r-b|^{2+\kappa}} \frac{1}{\la \a_r + \beta + (r- c)^2
\ra^{1-\e}}
\end{split}
\end{equation}
where $\kappa$ can assume the values $0, \lambda,2\lambda$, and
$\beta$, $b$, $c$ depend on $a$ and on the frozen momenta $p,q$.
If $e_r$ is a root, then, according to (\ref{eq:al1root}), we
replace $\a_r$ by $r^2$ in (\ref{eq:one}). This proves 1) (note
that the two dead edges always have the same orientation: with the
notation of Fig. \ref{fig:2dead}, it is impossible, for example,
that $p$ and $q'$ are both dead).

The transition 2) is proven, using the result of (\ref{eq:al}) (or
(\ref{eq:al1root}), if the father-edge is a root) in Proposition
\ref{prop:type2} (where $\kappa_1,\kappa_2$ can assume the values
$0,\lambda$ and $2\lambda$), under the conditions $\lambda <1/6$
and $\e <1/3$. Note that the vertices of the type 2) involve two
integrals, but because of the $\delta$-function from the momentum
conservation, one integral is a trivial substitution.

The transitions 3) and 4) are proven in Proposition
\ref{prop:type3}, using the result of the $\a$-integration
(\ref{eq:al}) (or (\ref{eq:al1root}) if the father-edge is a root)
under the condition that $\lambda < 1/6$ and $\e < \lambda /2$.
Finally, the transition 5) is shown, using the result of
(\ref{eq:al2}) (or (\ref{eq:al2root}), if the father-edge is a
root), in Proposition \ref{prop:type3sphe} (with $\lambda$
replaced by $2\lambda$), under the condition that $\kappa_1 +
\kappa_2 + \kappa_3 \leq 2\lambda$ and that $\lambda < 1/10$ and
$\e <\lambda$. If $\kappa_1 + \kappa_2 + \kappa_3
>2\lambda$, then we can drop the
denominator with the spherical singularity, and use
the transition 4) to prove 5). Therefore, assuming that $0 <
\lambda < 1/10$ and $0 < \e < \lambda/2$, we have proven the
transitions 1)-5) in (\ref{eq:cases}).

\medskip

After integrating over all vertices in the graph $\Gamma$ we are
left with an integral over the momenta of the roots in $R(\Gamma)
\backslash Q_1$ (a trivial root can be dead). Recall that we
already integrated over the $\a$-variable associated to the roots
(we performed this integration together with the integration over
the $\a$-variables of the son-edges of the roots in
(\ref{eq:al1root}) and (\ref{eq:al2root})). Estimating all the
spherical denominators left by one, we obtain from (\ref{eq:long})
that
\begin{equation}\label{eq:finalA}
\begin{split}
A \leq C^n &\left( \int \rd \bp_{n+k} \, p_1^2 \dots p_{n+k}^2 \,
|\psi (\bp_{n+k})|^2 \right) \\ &\times \sup_{\{ p_e \}_{e\in
Q_1}} \;  \prod_{e \in R(\Gamma) \cap Q_1} \; \frac{1}{\la p_e
\ra^3} \prod_{e \in R(\Gamma) \backslash Q_1}  \left( \sum_{j =
1}^{\nu (e)} \int \frac{\rd p_e}{|p_e - a_{e,j}|^{2+\kappa_e} \la
p_e \ra^3}\right) \,
\end{split}
\end{equation}
where, for every $e$, $\kappa_e =0,\lambda$ or $2\lambda$, where
$a_{e,j}$ are linear combinations of the momenta associated to
dead leaves in $D_e$, and where the number of terms $\nu (e)$ is
bounded by $\nu (e) \leq C^{|V_e|}$. Since the integrals are
uniformly bounded in the dead-momenta, and since $\sum_{e \in
R(\Gamma)\backslash Q_1} |V_e| = n$, it immediately follows that
\begin{equation}
A \leq C^n \left( \int \rd \bp_{n+k} \, p_1^2 \dots p_{n+k}^2 \,
|\psi (\bp_{n+k})|^2 \right) \, .
\end{equation}
In the same way we can prove that the term $B$ in (\ref{eq:long})
satisfies the same bound. Hence we conclude that
\begin{multline}
\Big| \int \rd \bq_k \rd \bq'_k \rd \br_{n+k} \rd \br'_{n+k} \,
J^{(k)} (\bq_k ; \bq'_k) \, K_{\Gamma,t} (\bq_k , \bq'_k;
\br_{n+k}, \br'_{n+k}) \, \gamma^{(n+k)} (\br_{n+k}, \br'_{n+k})
\Big| \\ \leq C^n \tr \; (1-\Delta_1) \dots (1-\Delta_{n+k})
\gamma^{(n+k)}\,
\end{multline}
for every $t \leq 1$.
\end{proof}

\subsection{Proof of  Theorem \ref{thm:boundsL}}

Despite the obvious analogy, the bound (\ref{eq:JLG}) cannot be
directly reduced to (\ref{eq:JKG}). The reason is that there are
three  propagators missing at the truncated vertex $\bar v$ that
appears in the definition (\ref{Lsimpler}). Their  missing decays
need to be propagated through the whole integration procedure until
the strong decay of the observable $J$ will compensate for them. For
this reason, edges carrying a momentum decay of the type i)-iv)
introduced in (\ref{eq:ii)})-(\ref{eq:s}) are not sufficient to
prove (\ref{eq:JLG}), and we need to introduce additional types of
decay. Of course this also requires to consider additional vertex
integrations, slightly different from the transitions 1)-5) in
(\ref{eq:cases}).

\begin{proof}[Proof of Theorem \ref{thm:boundsL}]
As in the proof of Theorem \ref{thm:bounds}, it is enough to
prove (\ref{eq:JLG}) for rank-one projectors,
$\gamma^{(n+k)} (\bp_{n+k}; \bp'_{n+k}) =
\psi (\bp_{n+k}) \overline{\psi} (\bp'_{n+k})$, for a $\psi \in
L^2_s (\bR^{3(n+k)})$; the general case follows then from the
expansion
$$
\gamma^{(n+k)} (\bp_{n+k};\bp'_{n+k} )= \sum_j
\lambda_j \psi_j (\bp_{n+k}) \overline{\psi_j} (\bp'_{n+k}),
\quad \mbox{with} \quad \psi_j \in L^2_s (\bR^{3(n+k)}), \;\;
\| \psi_j \| =1, \quad \forall j
$$
where the eigenvalues of
$\gamma^{(n+k)} $ satisfy $\lambda_j \geq 0$
for all $j$ and $\sum_j \lambda_j <
\infty$.

In order to prove (\ref{eq:JLG}) we start with the
expression (\ref{Lsimpler}). Recall that if there is only one
denominator containing $\alpha_e$, then the $\rd\alpha_e$ integral
would not be absolutely convergent, so this integration has to be
performed before taking the absolute value. This was the reason
behind distinguishing the roots of the trivial components,
$R_1(\Gamma)$, in (\ref{eq:KGamma2}). For the kernel $L_{\Gamma,
t}$, the same problem arises for the component containing $\bar
v\in M(\Gamma)$ (see (\ref{Lsimpler})), if this component has only
one vertex. This justifies the following definition.

For a given $\Gamma \in \cF_{n,k}$ and $\bar v \in M(\Gamma)$, we
define the set of edges $\wt E_2 (\Gamma,\bar v)$ as follows: if
there exists $\bar e \in R(\Gamma)$ such that $\bar e \in \bar v$,
then $\wt E_2 (\Gamma,\bar v): = E_2 (\Gamma) \backslash \{ \bar e
\}$. Otherwise $\wt E_2 (\Gamma,\bar v): = E_2 (\Gamma)$. In the
former case, starting from (\ref{Lsimpler}), we perform the
integration over the $\a_{\bar e}$ using (\ref{1res}) before we
take the absolute value; in other words, we treat $\bar e$ as the
trivial roots in $R_1 (\Gamma)$. This is necessary because, since
the son-edges of $\bar v$ do not carry a propagator, there is only
one denominator containing $\a_{\bar e}$ and the $\rd\alpha_e$
integral is not absolutely convergent. After performing the
integration over all $\a_e$ associated to $e \not \in \wt E_2
(\Gamma,\bar v)$, we take the absolute value of the integrand, and
we conclude that (\ref{eq:JLG}) is bounded, similarly to
(\ref{eq:long}), by (recall that we take here $\gamma^{(n+k)}
(\bp_{n+k}; \bp'_{n+k}) =\psi (\bp_{n+k})
\overline{\psi}(\bp'_{n+k})$)
\begin{equation}\label{eq:Llong2}
\begin{split}
C^{n} &\int \rd \bp_{n+k} \, \la p_1 \ra^2 \dots \la p_{n+k} \ra^2
|\psi (\bp_{n+k})|^2 \\ &\times \left( \sup_{\{ p_e \}_{e \in
Q_1}} \sum_{\bar v \in M(\Gamma)} \int \prod_{e \in E(\Gamma)
\backslash Q_1} \rd p_e \prod_{e \in \wt E_2 (\Gamma,\bar
v)\backslash S_{\bar v}} \frac{\rd \a_e}{|\a_e - p_e^2
+\frac{i}{t}|} \prod_{e \in R(\Gamma)} \frac{1}{\la p_e \ra^3} \,
 \right. \\
&\hspace{3cm} \times \prod_{e \in Q_2\backslash S_{\bar
v}}\frac{1}{p_e^2} \prod_{e\in Q_2 \cap S_{\bar v}} \frac{1}{\la
p_e \ra^2} \prod_{\bar v \neq v \in V(\Gamma)} \delta (\sum_{e\in
v} \pm \a_e) \prod_{v\in V(\Gamma)} \delta (\sum_{e\in v} \pm p_e)
\\ &+ \sup_{\{ p_e \}_{e \in Q_2}} \sum_{\bar v \in M(\Gamma)}
\int \prod_{e \in E(\Gamma)\backslash Q_2 } \rd p_e \prod_{e \in
\wt E_2 (\Gamma,\bar v)\backslash S_{\bar v}} \frac{\rd
\a_e}{|\a_e - p_e^2
+\frac{i}{t}|} \prod_{e \in R(\Gamma)} \frac{1}{\la p_e \ra^3} \\
& \hspace{3cm} \left.\times \prod_{e \in Q_1 \backslash S_{\bar
v}} \frac{1}{p_e^2} \, \prod_{e \in Q_1 \cap S_{\bar v}}
\frac{1}{\la p_e \ra^2} \prod_{\bar v \neq v \in V(\Gamma)} \delta
(\sum_{e\in v} \pm \a_e) \prod_{v \in V(\Gamma)} \delta
(\sum_{e\in v} \pm p_e) \right).
\end{split}
\end{equation}
Recall that $S_{\bar v}$ denotes the set of son-edges of $\bar v$
and that $Q_1$ and $Q_2$ denote the set of outward and,
respectively, inward leaves (\ref{def:Q}). Note that the weights
in the Schwarz inequality are somewhat different from the ones
used in (\ref{eq:long}): we take the decay factor $\la p \ra^{-2}$
instead of $|p|^{-2}$ in the son-edges of the vertex $\bar v$ (the
reason will be clear later on). Moreover, in contrast to
(\ref{eq:long}), we keep track of the $1/t$ factors in the
propagators to detect the short time behavior. To do so, we select
$\eta_e=t^{-1}$ for all leaf-edges, $e\in L(\Gamma)$,
 and notice that  for an arbitrary $e\in E(\Gamma)$
the value of $\eta_e$ is $t^{-1}$ times the number of edges in the
subgraph of descendents of $e$. In particular
$$
   \sum_{e\in R(\Gamma)} \eta_e = t^{-1}\cdot (3n)
$$
since the total number of non-root edges is $3n$. Therefore every
propagator in (\ref{Lsimpler}) carries at least the regularization
$t^{-1}$ (that is $\eta_e \geq t^{-1}$, for all $e \in
E(\Gamma)$); on the other side, the exponential prefactor is at
most $e^{3n}= C^n$ after taking absolute value. This justifies
(\ref{eq:Llong2}).

 To scale out the $t$ variables, we rescale $p_e$, for
all $e \in E (\Gamma)$, and we rescale $\a_e$, for $e \in \wt E_2
(\Gamma,\bar v) \backslash S_{\bar v}$ as
\[ p_e \to t^{-\frac{1}{2}} p_e \quad \text{for $e \in E(\Gamma)$ and}
\quad \a_e \to \a_e t^{-1} \quad \text{for $e\in \wt E_2
(\Gamma,\bar v)\backslash S_{\bar v}$.}
\] Then
\[ \frac{\rd \a_e}{|\a_e - p_e^2 + \frac{i}{t}|} \to \frac{\rd
\a_e}{|\a_e - p_e^2 + i|},  \qquad \delta ( \sum_{e \in v} \pm
\a_e ) \to t \, \delta (\sum_{e\in v} \pm \a_e), \qquad \delta (
\sum_{e \in v} \pm p_e ) \to  t^{3/2} \, \delta (\sum_{e\in v} \pm
p_e). \] {F}rom (\ref{eq:Llong2}) we can bound (\ref{eq:JLG}) by
(recall $t \leq 1$)
\begin{equation}\label{eq:Llong}
\begin{split}
C^{n} &t^{\frac{n-k}{2}-3} \int \rd \bp_{n+k} \,  \la p_1\ra^2
\dots \la p_{n+k} \ra^2 \, |\psi (\bp_{n+k})|^2 \\
&\times \left( \sup_{\{ p_e \}_{e \in Q_1}} \sum_{\bar v \in
M(\Gamma)} \int \prod_{e \in E(\Gamma) \backslash Q_1} \rd p_e
\prod_{e \in \wt E_2 (\Gamma,\bar v)\backslash S_{\bar v}}
\frac{\rd \a_e}{\la \a_e - p_e^2 \ra} \prod_{e\in R(\Gamma)}
\frac{1}{\la t^{-\frac{1}{2}} p_e \ra^{3}}  \right. \\
&\hspace{3cm} \times \prod_{e \in Q_2\backslash S_{\bar
v}}\frac{1}{p_e^2} \prod_{e\in Q_2 \cap S_{\bar v}} \frac{1}{\la
p_e \ra^2} \, \prod_{v \in V(\Gamma)\atop v\neq \bar v} \delta
(\sum_{e\in v} \pm \a_e)\, \prod_{v\in V(\Gamma)} \delta
(\sum_{e\in v} \pm p_e)
\\ &+ \sup_{\{ p_e \}_{e \in Q_2}} \sum_{\bar v \in M(\Gamma)}
\int \prod_{e \in E(\Gamma)\backslash Q_2 } \rd p_e \prod_{e \in
\wt E_2 (\Gamma, \bar v)\backslash S_{\bar v}} \frac{\rd \a_e}{\la
\a_e - p_e^2 \ra} \; \prod_{e\in R(\Gamma)} \frac{1}{\la
t^{-\frac{1}{2}} p_e \ra^3} \\ &\hspace{3cm}\times \prod_{e \in
Q_1\backslash S_{\bar v}}\frac{1}{p_e^2} \prod_{e\in Q_1 \cap
S_{\bar v}} \frac{1}{\la p_e \ra^2} \left. \, \prod_{ v \in
V(\Gamma) \atop v \neq \bar v} \delta (\sum_{e\in v} \pm \a_e)
\prod_{v \in V(\Gamma)} \delta (\sum_{e\in v} \pm p_e) \right) \\
=: &\; A_L + B_L \, .
\end{split}
\end{equation}
Note that here we rescale also the frozen momenta with $t^{-1/2}$:
of course this is allowed, since we take the supremum over them.
For $e \in Q_2 \cap S_{\bar v}$ in the term $A_L$ (and for $e \in
Q_1 \cap S_{\bar v}$ in $B_L$), we used that $\la t^{-1/2} p_e
\ra^{-2} \leq \la p_e \ra^{-2}$ (because $0 < t \leq 1$). We show
how to control $A_L$, where the outward leaves are dead; the proof
for $B_L$ is then analogous.

To estimate $A_L$ we proceed very
similarly as in our analysis of the contribution $A$ in
(\ref{eq:long}). But here we first have to get rid of the vertex
$\bar v$, whose son-edges do not have propagators. We distinguish
two cases.

In the first case, we assume that $\bar v$ is a vertex involving
two dead-edges (see Fig. \ref{fig:v-int}, part a)).
\begin{figure}
\begin{center}
\epsfig{file=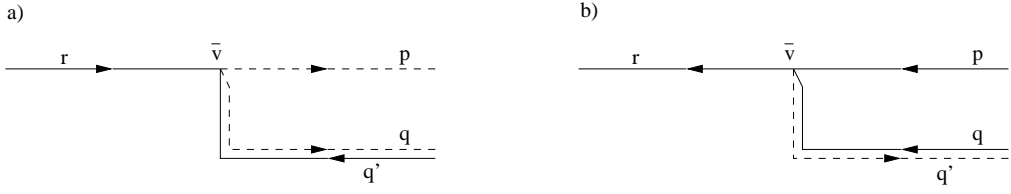,scale=.75}
\end{center}
\caption{Integrating out the vertex $\bar v$}\label{fig:v-int}
\end{figure}
Then the momentum integration at $\bar v$ (there is no
$\a$-integration here) gives
\begin{equation}\label{eq:v-int1}
\int \frac{\rd q'}{\la q' \ra^2} \delta (r- (p+q-q')) =
\frac{1}{\la r-p-q \ra^2} \leq \frac{1}{|r-p-q|^{1+2\lambda}}
\end{equation}
for $\lambda >0$ small enough. Edges carrying such a decay will be
called $(1+2\lambda)$-edges, similarly to the notation introduced
in i)-iv) above (see (\ref{eq:ii)})-(\ref{eq:s})). Note that in
the last inequality, we dropped part of the decay for large
momenta: this simplifies a little bit the classification of the
possible types of edges in $\Gamma$ (see the points v)-vii)
below). Here we used the fact that in (\ref{eq:Llong}) we chose
the decay $\la p_e \ra^{-2}$ instead of $|p_e|^{-2}$ for the
son-edges of the vertex $\bar v$ to avoid an irrelevant
complication in the short momentum regime.

In the second case, if $\bar v$ only involves one dead-edge (there
is always at least one dead edge adjacent to $\bar v$), then we
obtain (see Fig. \ref{fig:v-int}, part b))
\begin{equation}\label{eq:v-int2}
\int \frac{\rd p \rd q}{ \la p \ra^2 \la q \ra^2} \delta (r - (p
+q -q')) \lesssim \frac{C}{|r+q'|} \, .
\end{equation}
In this case, the father-edge of the $\bar v$-vertex will be
called a type $1$-edge.

After dealing with the vertex $\bar v$, we are faced with the
problem of integrating out the other $n-1$ vertices of $\Gamma$.
This is very similar to the problem we encountered when we proved
the bound (\ref{eq:JKG}): every non-trivial edge carries now a
propagator $\la \a_e -p_e^2\ra^{-1}$  exactly as in
(\ref{eq:long}). Here an edge is called trivial if it is a trivial
root or if it is a root adjacent to $\bar v$. The difference is
that now we start with more types of edges (in the analysis of
$K_{\Gamma,t}$, we started solely with $d$- and $2$-edges; here we
also have a $1$- or a $1+2\lambda$-edge): this implies that,
integrating out the other $n-1$ vertices, we will create new types
of edges, which were not defined in i)-iv) (see
(\ref{eq:ii)})-(\ref{eq:s}) in the proof of Theorem
\ref{thm:bounds}). Therefore, we have to supplement the
definitions i)-iv) with the following new types of edges.
\begin{itemize}
\item[v)] \underline{$1$-edges}: they carry a decaying factor:
\[
\frac{1}{|p_e - a|} \; ,
\]
 where $a$ is a linear combination of the dead
momenta in $D_e$. As we will see, there can be only one $1$-edge
along the integration procedure; it is the father-edge
 of $\bar v$, if it is not a root and if only one of
the son-edges of $\bar v$ is a dead edge (case b) in Fig.
\ref{fig:v-int}). In this case the two live son-edges have inward
orientation, so does the father-edge of $\bar v$, therefore the
$1$-edge is always inward.

\item[vi)]
\underline{$(1+\lambda)$-edges and $(1+2\lambda)$-edges}: they
carry a sum of decaying factors
\begin{equation}\label{eq:vi)}
\sum_{j=1}^{\nu (e)} \frac{1}{|p_e -a_j|^{1+\lambda}}, \quad
\text{respectively } \quad  \sum_{j=1}^{\nu (e)} \frac{1}{|p_e
-a_j|^{1+2\lambda}} \end{equation} where $a_j$ are linear
combinations of the momenta of the dead-edges lying in $D_e$. Here
the number of terms, $\nu (e)$ is bounded by $\nu(e) \leq
C^{|V_e|}$, for a universal constant $C$.
\item[vii)]
\underline{$(1+s+\kappa)$-edges} with $\kappa=0,\lambda,
2\lambda$: they carry a sum of decaying factors of the form:
\begin{equation}\label{eq:s1}
\sum_{j=1}^{\nu (e)} \frac{1}{|p_e-a_j|^{1+\kappa}} \frac{1}{\la
\a_e +\beta_j + (p_e - b_j)^2 \ra^{1-\e}}\; .  \end{equation} Here
$a_j$, $b_j$  are linear combinations of the momenta of the
dead-edges in $D_e$, the numbers $\beta_j$ are quadratic functions
of the same momenta. The number of terms, $\nu (e)$, is bounded by
$\nu (e) \leq C^{|V_e|}$, for a universal constant $C$ (the symbol
$s$ in $(1+s+\kappa)$ refer to the  ``spherical decay'' $\la \a
+\beta +(p-a)^2\ra^{-1+\e}$). If $e$ is a root edge, then it can
still be of the type $(1+s)$, $(1+s+\lambda)$ or $(1+s+2\lambda)$,
but in this case, in (\ref{eq:s1}) we replace $\a_e$ with $p_e^2$.
\end{itemize}
Note that in our classification of the edges of $\Gamma$, we
disregard the edges in $S_{\bar v}$: their only effect is to
produce a $(1+2\lambda)$- or a $1$-edge. Notice also that the new
types of edges appear only on the route from $\bar v$ to the root:
all other edges are still either $2$-, $(2+\lambda)$-,
$(2+2\lambda)$-, $(2+s)$-, $(2+s+\lambda)$-, or
$(2+s+2\lambda)$-edges. In order to take care of the new type of
edges we have to add to (\ref{eq:cases}) the following new
transitions:
\begin{align}\label{eq:casesL}
6)&\; (d,d,1+\kappa) \to 1+s+\kappa \quad \text{for } \kappa
=0,\lambda,2\lambda
\nonumber\\
7)&\; (d,1+\kappa_1,2+\kappa_2) \to 1+2\lambda \quad \text{for }
\kappa_{1,2} =0,\lambda,2\lambda \nonumber \\
8)&\; (1+\kappa_1 , 2+\kappa_2,2+\kappa_3) \to 1+2\lambda \quad
\text{for } \kappa_{1,2,3} =0,\lambda,2\lambda \quad \text{with }
\kappa_1 + \kappa_2 + \kappa_3 \geq 3\lambda \nonumber \\ 9)&\;
(1+2\lambda,2,2) \to 1 +\lambda \quad \\ 10)&\; (1,2+2\lambda,2)
\to 1+\lambda \quad \nonumber \\ 11)&\; (1+s+\kappa_1, 2+\kappa_2,
2+\kappa_3) \to 1+2\lambda \quad
\text{for } \kappa_{1,2,3} = 0,\lambda,2\lambda \nonumber \\
12)&\; (1+\kappa_1,2+s+\kappa_2, 2 + \kappa_3) \to 1+ 2\lambda
\quad \text{for } \kappa_{1,2,3} = 0,\lambda,2\lambda \nonumber
\end{align}
Moreover we have to add the following rules to the set a)-d)
introduced above.
\begin{itemize}
\item[e)] At every integration step, among the son-edges there can
only be one edge of the type v)-vii): this is clear because these
edges can only be found on the route from $\bar v$ to the root.
\item[f)] There is no vertex with son-edges of the type $(1,2,2)$,
$(1,2+\lambda,2)$, or $(1,2+\lambda,2+\lambda)$. This follows
because there cannot be three son-edges with the same orientation,
and because, from the previous observations (item b), c) and v)),
any $1$-edge, $2$-edge and $(2+\lambda)$-edge has always inward
orientation. \item[g)] There is no vertex with son-edges of the
type $(1+\lambda,2,2)$ or $(1+\lambda,2+\lambda,2)$. This follows
from the observation that the $1+\lambda$ edge, which can only
result from a transition 8) or 9), has always inward orientation
(because $1$- and $2$-edges have inward orientation).
\end{itemize}

Taking into account the fact that the spherical denominator $\la
\a + \beta + (p-a)^2 \ra^{-1+\e}$ can be always estimated by one
(i.e. $+s$ can always be removed from the decay characterization
of any edge), it is clear that the transitions 1)-12) together
with the rules a)-g) define a closed system, so that every edge
that arises in the successive vertex-integration of $\Gamma$ (with
the exception of the son-edges of $\bar v$) is of either one of
the types i)-vii) described above.

\medskip

Let us now prove the transitions 6)-12). The $\a$-integration can
be performed as in (\ref{eq:al}), (\ref{eq:al2}) (or in
(\ref{eq:al1root}) and (\ref{eq:al2root}), if we consider vertices
adjacent to root-edges). As for the momenta integration we proceed
as follows. The transition 6) can be shown similarly to the
transition 1) (see (\ref{eq:one})). The transition 7) follows,
using the result of (\ref{eq:al}) (or of (\ref{eq:al1root}), if
the father-edge is a root) to bound the $\a$-integration, from the
second part of Proposition \ref{prop:type2}, under the condition
that $0< \lambda <1/6$ and $\e < 1/3$. The transitions 8), 9) and
10) follow, again with the help of (\ref{eq:al}) or
(\ref{eq:al1root}), from Proposition \ref{prop:type3}, under the
assumption that $0 < \lambda <1/6$ and $\e <\lambda /2$. For
$\kappa_1+\kappa_2 +\kappa_3 \leq 2\lambda$, the transitions 11)
and 12) follow, using the result of (\ref{eq:al2}) (or
(\ref{eq:al2root}), if the father-edge is a root), by Proposition
\ref{prop:type3sphe} under the condition that $0 < \lambda < 1/10$
and $\e < \lambda$ (here we use this Proposition with $2\lambda$
instead of $\lambda$). If $\kappa_1 + \kappa_2+\kappa_3 >
2\lambda$, then we can drop the spherical denominator, and 11) and
12) follow from 8). Assuming that $0 < \lambda <1/10$ and $0 < \e
< \lambda /2$, this completes the proof of (\ref{eq:casesL}).

\medskip

Using the transitions (\ref{eq:cases}) and (\ref{eq:casesL}) we
can iteratively integrate over all vertices in $\Gamma$, until we
are left with an integral over the non-frozen root-momenta. {F}rom
(\ref{eq:Llong}) we obtain
\begin{equation}\label{eq:finalAL}
\begin{split}
A_L \leq &\; C^n t^{\frac{n-k-6}{2}} \left( \int \rd \bp_{n+k} \,
\la p_1 \ra^2 \dots \la p_{n+k} \ra^2 \, |\psi (\bp_{n+k})|^2
\right) \sup_{\{ p_e \}_{e \in Q_1}} \; \Bigg\{ \prod_{e \in
R(\Gamma) \cap Q_1} \; \frac{1}{\la t^{-\frac{1}{2}} p_e \ra^3} \\
&\times \sum_{\bar v \in M(\Gamma)} \prod_{  e \in R(\Gamma)
\backslash Q_1 \atop e\neq \bar e} \left( \sum_{j = 1}^{\nu (e)}
\int \frac{\rd p_e}{|p_e - a_{e,j}|^{2+\kappa_e} \la
t^{-\frac{1}{2}} p_e \ra^3}\right) \, \sum_{j=1}^{\nu (\bar e)}
\int \frac{\rd p_{\bar e}}{|p_{\bar e} -
a_{\bar{e},j}|^{1+\kappa_{\bar e}} \la t^{-\frac{1}{2}} p_{\bar e}
\ra^3} \Bigg\}\;,
\end{split}
\end{equation}
where we denote by $\bar e$ the unique root edge connected with
the vertex $\bar v$.
 As in (\ref{eq:finalA}), $\kappa_e
=0,\lambda$ or $2\lambda$, the $a_{e,j}$'s are linear combinations
of the momenta of the dead edges in $D_e$, and $\nu (e) \leq
C^{|V_e|}$. Note that the $(1+\kappa)$-type is inherited within
the tree containing $\bar v$, in particular the decay of the
corresponding root-edge is weaker than that of all other
root-edges.

Rescaling the momenta, we observe that
\begin{equation}
\sup_a \int \frac{\rd p}{|p-a|^{2+\kappa} \la t^{-1/2} p \ra^3}
\leq C t^{\frac{1-\kappa}{2}} \;, \quad \mbox{and}\quad
\sup_a \int \frac{\rd p}{|p-a|^{1+\kappa} \la t^{-1/2} p \ra^3}
\leq C t^{\frac{2-\kappa}{2}} \; .
\end{equation}
Since we assumed $t \leq 1$, since $|R(\Gamma)
\backslash Q_1| \geq k$, $|M(\Gamma)| \leq 2(n+k)$, and since
$\kappa_e \leq 2\lambda$, we obtain
\begin{equation}
A_L \leq (n+k) C^n t^{\frac{n-5}{2} - \lambda k} \; \tr \;
(1-\Delta_1) \dots (1-\Delta_{n+k}) \, \gamma^{(n+k)} \, .
\end{equation}
Since we assumed $n \geq 10 +k/2$, and $\lambda <1/10$, we find
\[ A_L \leq C^n \; t^{\frac{n}{4}} \; \tr \; (1-\Delta_1)
\dots (1-\Delta_{n+k}) \, \gamma^{(n+k)} \, .\] Since the same is
true for $B_L$ (see (\ref{eq:Llong})), this completes the proof of
 Theorem \ref{thm:boundsL}.
\end{proof}

\subsection{Proof of Theorem \ref{thm:unique}
}\label{sec:proofunique}

\bigskip

\begin{proof}[Proof of Theorem \ref{thm:unique}.] Suppose that $
\Gamma_{1,t} = \{ \gamma^{(k)}_{1,t} \}_{k \geq 1}$ and
$\Gamma_{2,t} = \{ \gamma^{(k)}_{2,t} \}_{k \geq 1}$ are two
solutions in $C([0,T], \cH)$ of the infinite hierarchy
(\ref{eq:BBGKYint2}),  such that, for $j =1,2$,
$\gamma_{j,t}^{(k)}$ is non-negative, symmetric w.r.t.
permutations and satisfies $\| \gamma_{j,t}^{(k)} \|_{\cH_k} \leq
C^k$, for all $k \geq 1$ and $t \in [0,T]$, and such that
$\gamma_{1,0}^{(k)} = \gamma_{2,0}^{(k)}$, for all $k\geq 1$. We
want to prove that $\Gamma_{1,t} = \Gamma_{2,t}$, for every $t \in
[0,T]$. To this end we will prove that, for every fixed $k \geq 1$,
$\gamma^{(k)}_{1,t} = \gamma^{(k)}_{2,t}$ for every $t \in [0,T]$
(as elements of $\cH_k$). By a
simple approximation argument it is then sufficient to prove that
\begin{equation}\label{eq:equal}
\tr \; J^{(k)} \, \left(\gamma_{1,t}^{(k)} - \gamma_{2,t}^{(k)}
\right) = 0
\end{equation}
for all $J^{(k)}$ in a dense subset of the dual space of $\cH_k$.
Since we assumed $\gamma^{(k)}_{1,t}$ and
$\gamma^{(k)}_{2,t}$ to be symmetric w.r.t. permutations, it is
enough to consider permutation symmetric observables $J^{(k)}$. We
will show (\ref{eq:equal}) for all permutation symmetric
$J^{(k)}$ with kernel $J^{(k)} (\bp_k ; \bp'_k)$ (in momentum
space) satisfying
\[ |J^{(k)} (\bp_k ; \bp'_k)| \leq C \prod_{j=1}^k \frac{1}{\la
p_j \ra^3 \la p'_j \ra^3} . \] For fixed $k \geq 1$ we can expand
$\gamma_{j,t}^{(k)}$ in a Duhamel-type expansion as in
(\ref{eq:duhamel}). With Theorem \ref{thm:graphs} we can identify
each term in the expansion (\ref{eq:duhamel}) as the sum of
contributions of Feynman graphs. We obtain, for any $n$, that
\begin{equation}
\gamma^{(k)}_{j,t} = \cU_0^{(k)} (t) \gamma_{j,0}^{(k)} +
\sum_{m=1}^{n-1} \sum_{\Gamma \in \cF_{m,k}} K_{\Gamma,t}
\gamma^{(k+m)}_{j,0} -i \sum_{\Gamma \in \cF_{n,k}} \int_0^t \rd s
\, L_{\Gamma,t-s} \gamma^{(k+n)}_{j,s}
\end{equation}
for $j=1,2$. Multiplying with the observable $J^{(k)}$ and taking
the trace we obtain \be\label{eq:duh} \tr \; J^{(k)}
\gamma^{(k)}_{j,t} = \la J^{(k)}, \, \cU_0^{(k)} (t)
\gamma^{(k)}_{j,0} \ra + \sum_{m=1}^{n-1} \sum_{\Gamma \in
\cF_{m,k}} \la J^{(k)}, K_{\Gamma,t}\gamma^{(m+k)}_{j,0}\ra -i
\sum_{\Gamma \in \cF_{n,k}} \int_0^t \rd s \; \la J^{(k)},
L_{\Gamma,t-s} \gamma^{(n+k)}_{j,s}\ra \ee for $j=1,2$. {F}rom
Theorem \ref{thm:bounds}, it follows that the terms in the sum
over $m$ are bounded in absolute value by $C^m \|
\gamma_{j,0}^{(m+k)} \|_{\cH_{m+k}}$, in particular they are
finite. Since $\gamma_{1,0}^{(k+m)} = \gamma_{2,0}^{(k+m)}$ for
every $m \geq 1$, when we take the difference between $\tr \,
J^{(k)} \gamma_{1,t}^{(k)}$ and $\tr \, J^{(k)}
\gamma_{2,t}^{(k)}$, the free evolution terms $\la J^{(k)},
\cU_0^{(k)} \gamma_{j,0}^{(k)} \ra$ and all the terms in the sum
over $m$ disappear and it only remains to bound the contributions
from the last term in (\ref{eq:duh}). {F}rom Theorem
\ref{thm:bounds}, and since $|\cF_{n,k}| \leq C^{n+k}$, we obtain,
under the assumption that $t \leq 1$ and $n \geq 10+ k/2$,
\begin{equation}
\begin{split}
\Big|\tr \, J^{(k)} \left( \gamma_{1,t}^{(k)} - \gamma_{2,t}^{(k)}
\right)\Big| &\leq C^n \, \int_0^t \rd s  (t-s)^{\frac{n}{4}}
\left( \| \gamma^{(n+k)}_{1,s} \|_{\cH_{k+n}} + \|
\gamma^{(n+k)}_{2,s} \|_{\cH_{k+n}} \right) \leq C^n \,
t^{\frac{n}{4}} \; ,
\end{split}
\end{equation}
where we used that, by assumption, $\sup_{s \in [0,T]} \|
\gamma^{(n+k)}_{j,s} \|_{\cH_{n+k}} \leq C^{n+k}$ for $j=1,2$.
Hence, if we choose $t < \min (1, (1/2C)^4)$ we conclude that
\begin{equation}
\Big|\tr \, J^{(k)} \left( \gamma_{1,t}^{(k)} - \gamma_{2,t}^{(k)}
\right)\Big| \leq 2^{-n}.
\end{equation}
Since $n \geq 1$ is arbitrary, this clearly proves
(\ref{eq:equal}) for every $t \leq \min (1,(1/2C)^4)$. The proof
can then be iterated to show that $\gamma_{1,t}^{(k)} =
\gamma_{2,t}^{(k)}$ for all $t \in [0,T]$.
\end{proof}

\section{Integrating Out a Vertex}
\label{sec:bounds}\setcounter{equation}{0}

In this section we prove some estimates used in the proofs of
Theorem \ref{thm:bounds} and Theorem \ref{thm:boundsL} to control
the momentum integration in the transitions 1)-5) in
(\ref{eq:cases}) and 6)-12) in (\ref{eq:casesL}).

\subsection{Preliminary Estimates}

We begin by proving some useful lemmas: they contain the
prototypes of integration we have to deal with when integrating
out a vertex.

\begin{lemma}\label{lm:pre1}
For every $\e,\lambda,\eta$ with $0\leq \e <\lambda<1$ and $0 <
\eta < \lambda -\e$ there exists a constant $C_{\lambda,\e,\eta}$
such that \be
    \int_{-\infty}^\infty
\frac{\rd \beta}{\langle \alpha - \beta\rangle^{1-\e} \;
|\beta|^\lambda} \leq \frac{C_{\lambda,\e,\eta}}{\langle \alpha
\rangle^{\lambda -\e -\eta}} \label{2den} \ee for all $\alpha \in
\bR$.
\end{lemma}
\begin{proof}
If $|\alpha|\leq 1$, then $\langle \alpha \rangle \sim 1$,
$\langle \alpha - \beta \rangle \sim \langle \beta \rangle$ and
(\ref{2den}) is trivial. For $|\alpha|\ge 1$ we split the integral
into two parts. For $|\beta|\leq |\alpha|/2$, we use
$$
     \frac{1}{\langle \alpha - \beta\rangle^{1-\e}} \lesssim
 \frac{1}{\langle
 \alpha - \beta\rangle^{1-\lambda+\eta} \langle \alpha
\rangle^{\lambda-\e-\eta}}
$$
and we obtain
\begin{equation}
\begin{split}
\int_{|\beta| \leq |\alpha|/2} \frac{\rd \beta}{ \langle \alpha -
\beta\rangle^{1-\e} |\beta|^{\lambda}} &\leq \frac{C}{ \langle
\alpha \rangle^{\lambda-\e-\eta}} \int_{|\beta| \leq |\alpha|/2}
\frac{\rd \beta}{\langle
 \alpha - \beta\rangle^{1-\lambda+\eta} |\beta|^{\lambda}} \\ &\leq \frac{C}{
 \langle \alpha \rangle^{\lambda-\e-\eta}} \left( \int_{|\beta| \leq 1/2}\frac{ \rd \beta}{|\beta|^{\lambda}}
 + \int \rd \beta \left(\frac{1}{\langle \beta
 \rangle^{1+\eta}}+ \frac{1}{\langle \alpha - \beta
 \rangle^{1+\eta}} \right)\right) \\ &\leq \frac{C}{
 \langle \alpha \rangle^{\lambda-\e-\eta}}
 \end{split}
 \end{equation}
where we used the Schwarz inequality and the fact that, if
$|\beta|\geq 1/2$, $|\beta| \sim \langle \beta \rangle$. For
$|\beta|\ge |\alpha|/2$, on the other hand, we use
$$
 \frac{1}{|\beta|^\lambda} \lesssim
 \frac{1}{|\alpha |^{\lambda-\e-\eta} | \beta |^{\e+\eta}}
\lesssim \frac{1}{\la\alpha \ra^{\lambda-\e-\eta} | \beta
|^{\e+\eta}}
$$
and conclude similarly.
\end{proof}

\begin{lemma}\label{lm:sphere}
For every $\e,\delta,\gamma$ with $0\le\e < 1$, $\delta <
(1/2)-\e$, $\delta >-1/2$, and $0 \leq \gamma < \min ( 1-\e;
1+2\delta; 1-2\delta-2\e)$, and for every $\eta
>0$ sufficiently small (depending on $\e,\delta,\gamma$), there
exists a constant $C = C_{\delta,\e,\gamma,\eta}$ with \be
  I=
 \int \frac{\rd p}{|p|^{2-2\delta} \, \langle \alpha - (p-a)^2
\rangle^{1-\e}}
   \leq \frac{C}{\langle a \rangle^{\gamma} \, \langle
 \alpha - a^2\rangle^{\frac{1}{2}-\frac{\gamma}{2}-\delta -\e -\eta}}
\label{spherical} \ee for all $\a \in \bR$ and $a \in \bR^3$.
\end{lemma}

\begin{proof}
We consider first the case $|a|<1$. Then $\la \alpha - a^2\ra \sim
\la \alpha \ra$ and $\la a \ra \sim 1$, so it is sufficient to
prove the estimate  (\ref{spherical}) when $a^2$ and $\la a \ra$
are removed from the r.h.s. of (\ref{spherical}). We now
distinguish two cases, depending on the size of $|\alpha|$.

If $|\alpha|\leq 10$, then the integral $I$ is comparable with
$$
 I\lesssim   \int \frac{\rd p}{|p|^{2-2\delta} \, \langle p-a
\rangle^{2-2\e}} \lesssim \int_{|p| \leq 1} \frac{\rd
p}{|p|^{2-2\delta}} + \int \rd p \, \left( \frac{1}{\langle p
\rangle^{4-2\delta-2\e}} + \frac{1}{\langle p-a \rangle^{4-2\delta
-2\e}} \right)
$$
which is uniformly bounded in $\a$ and $a$. Here we applied a
Schwarz inequality, and we used that, by assumption, $\delta
> -1/2$ and $\delta + \e < 1/2$. Since in this case $\langle \alpha \rangle
\simeq 1$, this proves (\ref{spherical}).

If $|\a|\ge 10$, then we split the $\rd p$-integration into three
different regimes. In the regime $p^2\leq |\alpha|/2$ we have
$\langle \alpha - (p-a)^2\ra\sim \la \alpha\ra$ and, putting
$y=p^2$, \be \label{eq:prelim1}
 \int_{|p|^2 \leq |\alpha| /2} \frac{\rd p}{|p|^{2-2\delta} \, \langle \alpha - (p-a)^2
\rangle^{1-\e}} \lesssim \frac{1}{\la\alpha\ra^{1-\e}}
\int_0^{|\alpha|/2} \rd y \, \frac{1}{|y|^{\frac{1}{2}-\delta}}
\lesssim \frac{1}{\la \alpha\ra^{{\frac{1}{2}-\e-\delta}}} \ee
because $\delta > -1/2$. In the regime
 $|\alpha|/2 \leq p^2\le 2|\alpha|$ we
have
\begin{equation}\label{eq:prelim2}
\begin{split}
\int_{|\a|/2 \leq p^2 \leq 2|\a|} &\;\frac{\rd p}{|p|^{2-2\delta}
\, \langle \alpha - (p-a)^2 \rangle^{1-\e}} \lesssim
\frac{1}{\la\alpha\ra^{1-\delta}} \int_{|\alpha| /2 \leq p^2\le
2|\alpha|} \frac{\rd p}{\langle \alpha - (p-a)^2 \rangle^{1-\e}}
\\ &\leq \frac{1}{\la \alpha \ra^{1-\delta}} \int_{|\alpha|/2 \leq
(q+a)^2\le 2|\alpha|}
 \frac{\rd q}{\langle \alpha - q^2
\rangle^{1-\e}} \leq \frac{1}{\la\alpha\ra^{1-\delta}}
\int_{|\alpha| /3 \leq q^2\le 3|\alpha|}
 \frac{\rd q}{\langle \alpha - q^2 \rangle^{1-\e}} \\ & \lesssim
 \frac{1}{\la\alpha\ra^{1-\delta}} \int_{|\a|/3}^{3|\a|}
 \frac{\rd y \, \sqrt{y}}{\la \a - y\ra^{1-\e}} \lesssim
 \frac{1}{\la\alpha\ra^{\frac{1}{2}-\delta}} \int_{|\a|/3}^{3|\a|}
 \frac{\rd y}{| \a - y |^{1-\e}} \lesssim \frac{1}{\la \a
 \ra^{\frac{1}{2}-\e-\delta}} \, .
 \end{split}
 \end{equation}
Finally, for $p^2\ge 2|\alpha|$, we have $\la \alpha -(p-a)^2\ra
\sim p^2$ and hence \be\label{eq:prelim3}
    \int_{p^2 \geq 2 |\a|} \frac{\rd p}{|p|^{2-2\delta}\, \langle \alpha - (p-a)^2
\rangle^{1-\e}} \lesssim \int_{p^2\ge 2|\alpha|}
 \frac{\rd p}{|p|^{4-2\delta-2\e}} \lesssim
\frac{1}{\la\alpha\ra^{\frac{1}{2}-\e-\delta}} \,. \ee Combining
(\ref{eq:prelim1}), (\ref{eq:prelim2}), and (\ref{eq:prelim3}), we
obtain (\ref{spherical}) for arbitrary $\gamma \geq 0$ and $\eta
>0$.

\medskip

Now we turn to the case $|a|\ge 1$. By rotational symmetry, we can
assume that $a=(|a|,0,0)$. After a change of variables and
introducing $\varrho:= p_2^2+p_3^2$ we find
$$
   I \lesssim \int_\bR dp_1 \int_0^\infty \frac{d\varrho}{
 | \varrho + p_1^2|^{1-\delta}
 \langle \alpha - a^2 -\varrho - p_1^2 + 2|a|p_1\rangle^{1-\e}} \,
 .
$$
We define the new variables:
$$
    u: = \varrho + p_1^2, \qquad  v: = \alpha - a^2 - u +
    2|a|p_1\,.
$$
The map $D \ni (u,v) \to (p_1, \varrho) \in \bR\times [0,\infty)$
is one-to-one if we choose
\[ D = \Big\{ (u,v)\in \bR^2 \; : \; u\ge\Big( \frac{\alpha -a^2-u-v}{2|a|}
\Big)^2\Big\} \,.\] Computing the Jacobian of this transformation,
we obtain
$$
   I \lesssim  \frac{1}{|a|}\int_D \; \frac{\rd u \rd v}{| u |^{1-\delta}
\la v\ra^{1-\e}}.
$$
Using the definition of the domain $D$, we get, for $0 \leq \gamma
\leq 1$,
\begin{equation}
I \lesssim \frac{1}{|a|} \int_D \frac{\rd u \rd
v}{|u|^{\frac{1}{2} + \frac{\gamma}{2}-\delta}
 \la v \ra^{1-\e} \Big| \frac{\alpha -a^2-u-v}{2|a|}\Big|^{1-\gamma}}
\lesssim \frac{1}{|a|^{\gamma}} \int_{\bR^2} \frac{\rd u \rd v}{|
u |^{\frac{1}{2}+\frac{\gamma}{2}-\delta} |\alpha
-a^2-u-v|^{1-\gamma}\la v\ra^{1-\e}} \, .
\end{equation}
Applying the bound (\ref{2den}) twice, to integrate first over $v$
and then over $u$, and using the assumptions that $0 \leq \gamma <
\min (1-\e; 1+2\delta; 1-2\delta-2\e)$ and that $\eta$ is small
enough, we find
\begin{equation}
I \lesssim \frac{1}{|a|^{\gamma}} \int_{\bR} \frac{\rd u }{| u
|^{\frac{1}{2} +\frac{\gamma}{2} - \delta} \, \la \alpha
-a^2-u\ra^{1-\gamma-\e- \frac{\eta}{2}} } \lesssim \frac{
C_{\delta, \e,\gamma, \e}   }{\la a \ra^{\gamma} \langle
 \alpha - a^2 \rangle^{\frac{1}{2}-\frac{\gamma}{2} -\delta -\e -\eta}}
 \,.
\end{equation}
Here we used that $|a| \geq 1$, to replace $|a|$ by $\la a \ra$.
\end{proof}

\begin{lemma}\label{lm:plane}
For any $\e,\delta,\eta$ with  $0\leq \e < 2\delta <1$, and
$0<\eta <2\delta-\e$, there exists a constant $C_{\delta,\eta,\e}$
such that \be
     I=\int \frac{\rd p}{|p|^{2+2\delta} \, \la \a -p\cdot a\ra^{1-\e}}
   \leq \frac{C_{\delta,\eta,\e}}{\la\alpha\ra^{2\delta-\e-\eta}
 |a|^{1-2\delta}}
\label{plane} \ee for every $a \in \bR^3$, $\a \in \bR$.
\end{lemma}
\begin{proof}
By rotational symmetry, we can assume $a=(|a|,0,0)$. Introducing
the variable $\varrho = p^2_2 + p_3^2$, we find
$$
    I \lesssim \int_\bR \frac{\rd p_1 }{\la\alpha-p_1|a|\ra^{1-\e}}
   \int_0^\infty\frac{\rd\varrho}{|p_1^2+\varrho|^{1+\delta}}
  \lesssim \int_\bR \frac{\rd p_1 }{\la\alpha-p_1|a|\ra^{1-\e} |p_1|^{2\delta}}
   =\frac{1}{|a|^{1-2\delta}} \int_\bR\frac{\rd y}{\la \alpha - y\ra^{1-\e}
|y|^{2\delta}}
$$
and we conclude by (\ref{2den}).
\end{proof}

\subsection{Momentum Integration}

\begin{figure}
\begin{center}
\epsfig{file=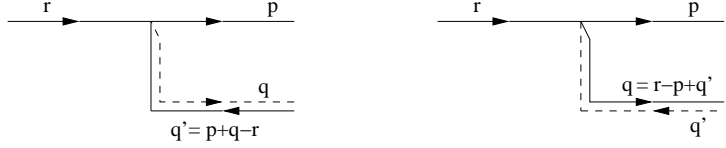,scale=.75}
\end{center}
\caption{The vertices of Prop. \ref{prop:type2}}\label{fig:prop7}
\end{figure}

Using the results of the last subsection, we provide here bounds
for the integration over the momenta carried by the son-edges of a
given vertex. The bounds in the next proposition are used to
integrate out vertices with one dead edge, i.e.
vertices of type  2) in
(\ref{eq:cases}) and  type 7) in (\ref{eq:casesL})).
These vertices
involve integration over two momenta; however, because of the
momentum delta-function, effectively we only need to integrate
over one momentum (see Fig. \ref{fig:prop7}).

\begin{proposition}\label{prop:type2}
Suppose $0<\lambda < 1/6$ and $0 \leq \eps < 1/3$. Let
$\kappa_1,\kappa_2 \geq 0$, with $\kappa_1 + \kappa_2 \leq
4\lambda$. Then there exists a constant $C = C
(\lambda,\e,\kappa_1,\kappa_2)$ such that
\begin{equation}\label{2a}
\sup_\a \int \frac{\rd p}{|p-a_1|^{2+\kappa_1}
|r-p-q-a_2|^{2+\kappa_2}} \frac{1}{\la \a - p^2 \pm (r-p-q)^2
\ra^{1-\e}} \leq C \sum_{j=1}^{\mu} \frac{1}{|r-b_j|^{2+2\lambda}}
\end{equation}
and
\begin{equation}\label{2b}
\sup_\a \int \frac{\rd p}{|p-a_1|^{1+\kappa_1}
|r-p-q-a_2|^{2+\kappa_2}} \frac{1}{\la \a - p^2 \pm (r-p-q)^2
\ra^{1-\e}} \leq C \sum_{j=1}^{\mu} \frac{1}{|r-b_j|^{1+2\lambda}}
\end{equation}
for any $r,q,a_1,a_2 \in \bR^3$. Here $b_j$ are linear
combinations of $a_1, a_2$ and of the frozen momentum $q$, and
$\mu$ is a universal integer constant.
\end{proposition}
{\it Remark. } The different signs in the propagator on the l.h.s.
of (\ref{2a}) are needed because of the two possible orientation
of the dead edge. The positive sign in  (\ref{2a}) corresponds to
the vertex on the left side of Fig. \ref{fig:prop7} since in this
case the two live son-edges, carrying the propagators $p^2$ and
$(p+q-r)^2$, have opposite orientation.  For the vertex on the
right, first replace  $q$ with $-q'$ in (\ref{2a}) and  then use
the negative sign, corresponding to the paralell orientations of
the live son-edges. The bound (\ref{2b}) is used when one of the
son-edges is a $1+\kappa$--edge, for $\kappa=0,\lambda$ or
$2\lambda$. The different signs in the propagator again take care
of the possible orientations of the dead edge and of the
$1+\kappa$--edge.

\begin{proof}[Proof of Proposition \ref{prop:type2}.]
We will make use of the following inequality to separate
denominators.
\begin{lemma}\label{lemma:ab}
 For arbitrary $\alpha,\beta >0$ and $0\leq \gamma
\leq \min (\alpha,\beta)$, there exists a constant
$C_{\alpha,\beta,\gamma}$ such that
\begin{equation}\label{eq:weighted}
\frac{1}{|a|^{\alpha} |b-a|^{\beta}} \leq
\frac{C_{\a,\beta,\gamma}}{|b|^{\gamma}} \left(
\frac{1}{|a|^{\alpha+\beta-\gamma}} +
\frac{1}{|b-a|^{\alpha+\beta-\gamma}} \right) \, .
\end{equation}
\end{lemma}
\begin{proof}[Proof of Lemma \ref{lemma:ab}.]
 Note that, since $|b| \leq |b-a| + |a|$,
we have $|b|^{\gamma} \leq C_{\gamma} ( |b-a|^{\gamma} +
|a|^{\gamma} )$ and thus
\begin{equation}
\frac{|b|^{\gamma}}{|a|^{\gamma} |b-a|^{\gamma}} \leq C_{\gamma}
\left(\frac{1}{|a|^{\gamma}} + \frac{1}{|b-a|^{\gamma}} \right)
\,.
\end{equation}
Hence
\begin{equation}
\begin{split}
\frac{1}{|a|^{\alpha} |b-a|^{\beta}} &= \frac{1}{|a|^{\gamma}
|b-a|^{\gamma}} \, \frac{1}{|a|^{\a-\gamma} |b-a|^{\beta-\gamma}}
\leq \frac{C_\gamma}{|b|^{\gamma}} \left( \frac{1}{|a|^{\gamma}} +
\frac{1}{|b-a|^{\gamma}} \right) \frac{1}{|a|^{\a-\gamma}
|b-a|^{\beta-\gamma}}\\
&\leq \frac{C_{\gamma}}{|b|^{\gamma}} \left( \frac{1}{|a|^{\alpha}
|b-a|^{\beta-\gamma}} + \frac{1}{|a|^{\a-\gamma} |b-a|^{\beta}}
\right)
\end{split}
\end{equation}
and (\ref{eq:weighted}) follows by a Schwarz inequality.
\end{proof}

Returning to the proof of Proposition \ref{prop:type2},
 we introduce a
parameter $\theta$ which can assume the values $1$ and $2$. In
this way we can prove (\ref{2a}) and (\ref{2b}) in parallel;  we
use $\theta = 2$ for the proof of (\ref{2a}), $\theta=1$ for the
proof of (\ref{2b}). According to the sign in the denominator, we
have to bound one of the two integrals
\begin{equation}\label{eq:I-II}
\begin{split}
\text{(I)} &:= \sup_\a \int \frac{\rd p}{|p-a_1|^{\theta+\kappa_1}
|r-p-q-a_2|^{2+\kappa_2}} \frac{1}{\la \a - 2p \cdot (r-q) \ra^{1-\e}} \\
\text{(II)} &:= \sup_\a \int \frac{\rd
p}{|p-a_1|^{\theta+\kappa_1} |r-p-q-a_2|^{2+\kappa_2}}
\frac{1}{\la \a - 2 (p-\frac{r-q}{2})^2 \ra^{1-\e}}\,.
\end{split}
\end{equation}
Here we shifted the $\a$ variable by a $p$-independent number:
this is of course allowed because we take the supremum over $\a$.

We consider first the integral (I). Using (\ref{eq:weighted}) we
obtain, for arbitrary $-1 < \gamma \leq \min (\kappa_1,\kappa_2)$,
\begin{equation*}
\begin{split}
\text{(I)} \lesssim \; &\frac{1}{|r-q-a_1-a_2|^{\theta + \gamma}}
\;
\\ &\times \sup_{\a} \int \rd p \, \left( \frac{1}{|p-a_1|^{2+\kappa_1
+\kappa_2 - \gamma}} + \frac{1}{|r-q-p-a_2|^{2+\kappa_1 +\kappa_2
- \gamma}} \right) \, \frac{1}{\la \a -2p\cdot (r-q) \ra^{1-\e}}
\\\lesssim \; &\frac{1}{|r-q-a_1-a_2|^{\theta+\gamma}}
\; \sup_{\a} \int \frac{\rd p}{|p|^{2+\kappa_1 +\kappa_2 -
\gamma}} \, \frac{1}{\la \a -2p\cdot (r-q) \ra^{1-\e}}\,.
\end{split}
\end{equation*}
Assuming that
\begin{equation}\label{eq:ekg}
\e < \kappa_1 + \kappa_2 -\gamma < 1
\end{equation}
we can find $\eta >0$ such that $0< \eta < \kappa_1 + \kappa_2
-\gamma -\e$ and we can apply Lemma \ref{lm:plane} to find
\begin{equation}
\begin{split}
\text{(I)} &\lesssim \frac{1}{|r-q-a_1-a_2|^{\theta+\gamma} \,
|r-q|^{1+ \gamma - \kappa_1 - \kappa_2}}\\ &\lesssim
\frac{1}{|r-q|^{\theta +1 +2\gamma-\kappa_1 -\kappa_2}}+
\frac{1}{|r-q-a_1-a_2|^{\theta +1 +2\gamma-\kappa_1-\kappa_2}} \, .
\end{split}
\end{equation}
Choosing $2\gamma = \kappa_1 + \kappa_2 - 1 + 2\lambda$ we can
bound (I) by the r.h.s. of (\ref{2a}) or of (\ref{2b}) (according
to whether $\theta =2$ or $\theta=1$), with $\mu =2$, $b_1=q$ and
$b_2 = q+a_1+a_2$: we only have to check that this choice of
$\gamma$ is compatible with the condition $-1 < \gamma \leq \min
(\kappa_1 , \kappa_2)$ and with (\ref{eq:ekg}). This follows from
the assumptions that $\kappa_1 + \kappa_2 \leq 4\lambda$, $\lambda
< 1/6$ and $0 \leq \e < 1/3$.

\medskip

Next we consider the term (II) in (\ref{eq:I-II}). Using
(\ref{eq:weighted}) and changes of variables, we conclude that
\begin{equation}
\begin{split}
\text{(II)} \lesssim &\; \frac{1}{|r-q-a_1-a_2|^{\theta}}
\\ &\times \sup_{\a} \int \rd p \left( \frac{1}{|p-a_1|^{2+ \kappa_1
+\kappa_2}} + \frac{1}{|r-p-q-a_2|^{2+ \kappa_1 +\kappa_2}}
\right) \frac{1}{\la \a - 2 (p - \frac{r-q}{2})^2 \ra^{1-\e}}
\\ \lesssim \;&\frac{1}{|r-q-a_1-a_2|^{\theta}}
\\ &\times \sup_{\a} \int \frac{\rd p}{|p|^{2+\kappa_1+\kappa_2}}
 \left( \frac{1}{\la \a -2 (p +a_1 - \frac{r-q}{2})^2
 \ra^{1-\e}}+ \frac{1}{\la \a -2 (p +a_2 - \frac{r-q}{2})^2
 \ra^{1-\e}}\right)\,.
\end{split}
\end{equation}
Applying Lemma \ref{lm:sphere}, with $\gamma = 2\lambda$ and
$\delta=-(\kappa_1 + \kappa_2)/2$, we obtain for $0 < \eta < 1/2 -
\lambda -\e$,
\begin{equation}
\begin{split}
\text{(II)} &\lesssim \frac{1}{|r-q-a_1-a_2|^{\theta}} \left(
\frac{1}{|r-q -2a_1|^{2\lambda}} + \frac{1}{|r-q
-2a_2|^{2\lambda}} \right) \; \sup_{\a} \frac{1}{\la \a
\ra^{\frac{1+\kappa_1+\kappa_2}{2} - \lambda -\e-\eta}} \\
&\lesssim  \frac{1}{|r-q-a_1-a_2|^{\theta+2\lambda}} +
\frac{1}{|r-q-2a_1|^{\theta+2\lambda}} +
\frac{1}{|r-q-2a_2|^{\theta+2\lambda}} \, .
\end{split}
\end{equation}
Here we used that $\lambda <1/6$, $\e < 1/3$ and $\kappa_1 +
\kappa_2 \leq 4\lambda$. This completes the proof of (\ref{2a})
and (\ref{2b}).
\end{proof}

\medskip

\begin{figure}
\begin{center}
\epsfig{file=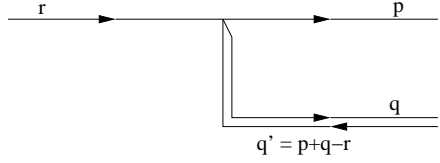,scale=.75}
\end{center}
\caption{The vertex of Prop. \ref{prop:type3}}\label{fig:prop8}
\end{figure}

In the next proposition we show how to integrate out vertices
where all the son-edges are live $(2+\kappa)$-- or
$(1+\kappa)$--edges, with $\kappa=0,\lambda$, or $2\lambda$. These
will include the vertex integrations of type 3) and 4) in
(\ref{eq:cases}) and type 8), 9) and 10) in (\ref{eq:casesL}).
After the $\alpha$-integrations, these vertices involve
integration over three momenta $p,q,q'$. Using the delta-function
$\delta (r-(p+q-q'))$, we are left with two effective integrations
which need to be controlled (see Fig. \ref{fig:prop8}).

\begin{proposition}\label{prop:type3}
Suppose $\kappa_1,\kappa_2,\kappa_3 \geq 0$ with $0< \kappa_1 +
\kappa_2 + \kappa_3 < 1$. Let $0 \leq \kappa < \kappa_1 + \kappa_2
+ \kappa_3$, and $\e < (\kappa_1 + \kappa_2 + \kappa_3 -
\kappa)/2$. Then there is a constant $C=C
(\kappa_1,\kappa_2,\kappa_3,\kappa,\eps)$ such that
\begin{multline} \label{3a}
\sup_\a \int \frac{\rd p \, \rd q}{|p-a_1|^{2+\kappa_1}
|q-a_2|^{2+\kappa_2}|r-p-q-a_3|^{2+\kappa_3}} \\
\times \frac{1}{\la \a - p^2 - q^2+ (r-p-q)^2\ra^{1-\e}} \leq C
\sum_{j=1}^{\mu} \frac{1}{|r-b_j|^{2+\kappa}}
\end{multline}
and
\begin{multline} \label{3b1}
\sup_\a \int \frac{\rd p \, \rd q}{|p-a_1|^{1+\kappa_1}
|q-a_2|^{2+\kappa_2} |r-p-q-a_3|^{2+ \kappa_3}} \\
\times \frac{1}{\la \a - p^2 - q^2+ (r-p-q)^2\ra^{1-\e}} \leq C
\sum_{j=1}^{\mu} \frac{1}{|r-b_j|^{1+\kappa}}
\end{multline}
and
\begin{multline} \label{3b2}
\sup_\a \int \frac{\rd p \, \rd q}{|p-a_1|^{2+\kappa_1}
|q-a_2|^{2+\kappa_2} |r-p-q-a_3|^{1+\kappa_3}}\\ \times
\frac{1}{\la \a - p^2 - q^2+ (r-p-q)^2\ra^{1-\e}} \leq C
\sum_{j=1}^{\mu} \frac{1}{|r-b_j|^{1+\kappa}}
\end{multline}
for all $r,a_1,a_2,a_3 \in \bR^3$. Here the $b_j$ are linear
combinations of $a_1,a_2,a_3$, and $\mu$ is a universal constant.
\end{proposition}
{\it Remark. } The bound (\ref{3b1}) is used when the edge
carrying momentum $p$ is a $(1+\kappa)$--edge, for
$\kappa=0,\lambda,2\lambda$. The bound (\ref{3b2}), on the other
hand, is used when the edge with momentum $q'=r-p-q$ is a
$(1+\kappa)$--edge (if the edge with momentum $q$ is a
$(1+\kappa)$--edge, then, after exchanging $p \leftrightarrow q$,
we can use (\ref{3b1})).

\begin{proof}
In this proof we will assume that $a_1=a_2=a_3=0$: the
generalization to $a_1,a_2,a_3 \neq 0$ can be obtained with
similar shifts as we did in the proof of Proposition
\ref{prop:type2}. Using
\[ \frac{1}{|p|^{\kappa_1} |q|^{\kappa_2} |r-p-q|^{\kappa_3}} \leq
 \left(\frac{1}{|p|^{\theta}} +
\frac{1}{|q|^{\theta}}+\frac{1}{|r-p-q|^{\theta}} \right)
\] with $\theta =\kappa_1 + \kappa_2 + \kappa_3$, and the
symmetry of the integrand w.r.t. the exchange $p \leftrightarrow
q$, the l.h.s. of (\ref{3a}) is bounded by
\begin{equation}\label{eq:3a1}
\begin{split}
2 \; \sup_\a &\int \frac{\rd p  \, \rd q}{|p|^{2+\theta}
|q|^{2}|r-p-q|^{2}}\frac{1}{\la \a - p^2 - q^2+
(r-p-q)^2\ra^{1-\e}} \\ &+ \sup_\a \int \frac{\rd p \, \rd
q}{|p|^2 |q|^2 |r-p-q|^{2+\theta}}\frac{1}{\la \a - p^2 - q^2+
(r-p-q)^2\ra^{1-\e}} \,  \\ \lesssim \; &\sup_\a \int \frac{\rd
q}{|q|^{2} |r-q|^2} \int \rd p \, \left( \frac{1}{|p|^{2+\theta}}
+ \frac{1}{|r-q-p|^{2+\theta}} \right) \frac{1}{\la \a - p^2 -
q^2+ (r-p-q)^2 \ra^{1-\e}}\\
\lesssim \; &\sup_{\a_1 } \int \frac{\rd q}{|q|^{2} |r-q|^{2}}
\int \frac{ \rd p}{|p|^{2+ \theta}}
\frac{1}{\la \a_1 - 2 r\cdot q -2 p \cdot (r-q)\ra^{1-\e}} \\
&+ \sup_{\a_2 } \int \frac{\rd q}{|q|^{2} |r-q|^{2}} \int \frac{
\rd p}{|p|^{2+ \theta}} \frac{1}{\la \a_2 - 2(q - \frac{r}{2})^2 +
2 p\cdot (r-q) \ra^{1-\e}} \,
\end{split}
\end{equation}
where we applied the inequality (\ref{eq:weighted}) and then we
shifted the variable $q\to (r-p-q)$ to obtain the last term.
Applying Lemma \ref{lm:plane} we conclude that the l.h.s. of
(\ref{3a}) can be estimated by
\begin{equation}\label{eq:pands}
\begin{split}
\sup_\a \int &\frac{\rd p \, \rd q}{|p|^{2+\kappa_1}
|q|^{2+\kappa_2}|r-p-q|^{2+\kappa_3}} \frac{1}{\la \a - p^2 - q^2+
(r-p-q)^2\ra^{1-\e}} \\ \lesssim \; &\sup_{\a_1} \int \frac{\rd
q}{|q|^{2} |r-q|^{3 - \theta}} \frac{1}{\la \a_1 - 2 \, r\cdot q
\ra^{\theta - \eps-\eta}} \\ &+ \sup_{\a_2} \int \frac{\rd
q}{|q|^{2} |r-q|^{3 - \theta}} \frac{1}{\la \a_2 - 2(q -
\frac{r}{2})^2 \, \ra^{\theta - \eps-\eta}} \, ,
\end{split}
\end{equation}
for any sufficiently small $\eta >0$.
To bound the first term we
use again (\ref{eq:weighted}) and Lemma \ref{lm:plane}:
\begin{equation}\label{eq:3a12}
\begin{split}
\sup_{\a_1} \int &\frac{\rd q}{|q|^{2} |r-q|^{3 - \theta}}
\frac{1}{\la \a_1 - 2 \, r\cdot q \ra^{\theta-\e-\eta}}\\
&\lesssim \frac{1}{|r|^{2-\frac{\theta - \kappa}{2}}} \sup_{\a_1}
\int \rd q \, \left(\frac{1}{|q|^{3 - \frac{\theta +\kappa}{2}}} +
\frac{1}{|r-q|^{3-\frac{\theta+\kappa}{2}}} \right) \frac{1}{\la
\a_1 - 2 \, r\cdot q \ra^{\theta - \eps-\eta}}
\\ &\lesssim \frac{1}{|r|^{2 + \kappa}}\; \sup_{\a_1} \frac{1}{\la
\a_1 \ra^{\frac{\theta-\kappa}{2} -\e -2\eta}}  \lesssim
\frac{1}{|r|^{2+ \kappa}} ,
\end{split}
\end{equation}
because $\kappa < \theta$, $0 \leq 2\e < \theta -\kappa$ and
$\eta>0$ is arbitrarily small. Using (\ref{eq:weighted}) and Lemma
\ref{lm:sphere} (with $\gamma =\kappa$), the second term on the
r.h.s. of (\ref{eq:pands}) can be controlled by:
\begin{equation}
\begin{split}
\sup_{\a_2} \int &\frac{\rd q}{|q|^{2} |r-q|^{3 - \theta}}
\frac{1}{\la \a_2 - 2(q - \frac{r}{2})^2 \, \ra^{\theta -\eps-\eta}} \\
& \lesssim \frac{1}{|r|^{2}} \sup_{\a_2} \int \rd p \left(
\frac{1}{|q|^{3-\theta}} +
\frac{1}{|r-q|^{3-\theta}}\right)\frac{1}{\la \a_2 - 2(q -
\frac{r}{2})^2 \, \ra^{\theta - \eps -\eta}}
\\ &\lesssim \frac{1}{|r|^{2+\kappa}}\; \sup_{\a_3} \frac{1}{\la
\a_3 \ra^{\frac{\theta-\kappa}{2} - \e- 2\eta}} \lesssim
\frac{1}{|r|^{2+\kappa}} \, ,
\end{split}
\end{equation}
because $\theta > \kappa$ and $2\e < \theta -\kappa$, and $\eta$
is arbitrarily small. This completes the proof of (\ref{3a}).

\medskip

Next we prove (\ref{3b1}). With $\theta = \kappa_1 + \kappa_2$, we
have, similarly to (\ref{eq:pands}),
\begin{equation}\label{eq:3b11}
\begin{split}
\sup_\a \int &\frac{\rd p \, \rd q}{|p|
|q|^{2+\kappa_1}|r-p-q|^{2+\kappa_2}} \frac{1}{\la \a - p^2 - q^2+
(r-p-q)^2\ra^{1-\e}} \\ \lesssim \; &\sup_{\a} \int \frac{ \rd
p}{|p| |r-p|^{2}} \int \rd q  \left( \frac{1}{|q|^{2+\theta}} +
\frac{1}{|r-q-p|^{2+\theta}} \right) \frac{1}{\la \a - p^2 - q^2+
(r-p-q)^2\ra^{1-\e}} \\ \lesssim \; &\sup_{\a_1} \int \frac{ \rd
p}{|p||r-p|^{3-\theta}} \frac{1}{\la \a_1 - 2 \, p \cdot r
\ra^{\theta - \e -\eta}} + \sup_{\a_2} \int \frac{ \rd p}{|p|
|r-p|^{3-\theta}} \frac{1}{\la \a_2 - 2 \, (p - \frac{r}{2})^2
\ra^{\theta-\e-\eta}} \, ,
\end{split}
\end{equation}
for any small $\eta >0$. Using Lemma \ref{lm:plane}, the first
term can be handled as follows
\begin{equation}
\begin{split}
\sup_{\a_1} &\int \frac{ \rd p}{|p| |r-p|^{3-\theta}} \frac{1}{\la
\a_1 - 2 \, p \cdot r \ra^{\theta -\e-\eta}} \\ &\lesssim
\frac{1}{|r|^{1-\frac{\theta}{2}}} \; \sup_{\a_1} \int \rd p
\left( \frac{1}{|p|^{3-\frac{\theta}{2}}} +
\frac{1}{|r-p|^{3-\frac{\theta}{2}}} \right) \frac{1}{\la \a_1 - 2
\, p \cdot r \ra^{\theta - \e -\eta}}
\\ &\lesssim \frac{1}{|r|} \; \sup_{\a_1} \frac{1}{\la \a_1
\ra^{\frac{\theta}{2} - \e -2\eta}} \lesssim \frac{1}{|r|}
\end{split}
\end{equation}
because $\theta > 0$, $\e < \theta/2$, and $\eta$ is arbitrarily
small. As for the second term on the r.h.s. of (\ref{eq:3b11}),
from Lemma \ref{lm:sphere}, we find (using again that $0 < \e
<\theta/2$ and choosing $\gamma=0$)
\begin{equation}
\begin{split}
\sup_{\a_2} \int &\frac{ \rd p}{|p| |r-p|^{3-\theta}} \frac{1}{\la
\a_2 - 2 \, (p - \frac{r}{2})^2 \ra^{\theta - \e - \eta}} \\
&\lesssim \frac{1}{|r|} \; \sup_{\a_2} \int \rd p \left(
\frac{1}{|p|^{3-\theta}} + \frac{1}{|r-p|^{3-\theta}}\right)
\frac{1}{\la \a_2 - 2 \, (p - \frac{r}{2})^2
\ra^{\theta - \e -\eta}} \\
&\lesssim \frac{1}{|r|} \sup_{\a_3} \frac{1}{\la \a_3
\ra^{\frac{\theta}{2} -\e -2\eta}} \lesssim \frac{1}{|r|}\,.
\end{split}
\end{equation}
Exactly the same proof also works for (\ref{3b2}).
\end{proof}

Finally, in the next proposition, we show how to integrate out
vertices with three alive son-edges, one of which carries a spherical
denominator (that is one of the son-edges is a $(2+s+\kappa)$-- or a
$(1+s+\kappa)$--edge): these are vertices of the type 5) in
(\ref{eq:cases}) and of the type 11) or 12) in (\ref{eq:casesL}).

\begin{proposition}\label{prop:type3sphe}
Suppose $0< \lambda <1/5$ and $0 \leq \e < \lambda/2$. Let
$\kappa_1,\kappa_2, \kappa_3 \geq 0$, with $\kappa_1 + \kappa_2 +
\kappa_3 \leq \lambda$. Then there exists a constant
$C=C(\lambda,\e, \kappa_1,\kappa_2,\kappa_3)$ such that
\begin{multline}\label{3c}
\sup_{\a,\beta, c} \int \frac{\rd p \, \rd q}{|p-c_1|^{2+\kappa_1}
|q-c_2|^{2+\kappa_2} |r-p-q-c_3|^{2+\kappa_3}} \frac{1}{\la \beta +
(p-c)^2 \ra^{1-\e}} \\ \times \frac{1}{\la \a - p^2 \pm q^2
 \pm (r-p-q)^2 \ra^{1-\e}} \leq C \sum_{j=1}^{\mu}
\frac{1}{|r-b_j|^{2+\lambda}}
\end{multline}
and
\begin{multline}\label{3d1}
\sup_{\a,\beta, c} \int \frac{\rd p \, \rd q}{|p-c_1|^{1+\kappa_1}
|q-c_2|^{2+\kappa_2} |r-p-q-c_3|^{2+\kappa_3}} \frac{1}{\la \beta +
(p-c)^2 \ra^{1-\e}} \\ \times \frac{1}{\la \a - p^2 \pm q^2
 \pm (r-p-q)^2 \ra^{1-\e}} \leq C \sum_{j=1}^{\mu}
\frac{1}{|r-b_j|^{1+\lambda}}
\end{multline}
and
\begin{multline}\label{3d2}
\sup_{\a,\beta, c} \int \frac{\rd p \, \rd q}{|p-c_1|^{2+\kappa_1}
|q-c_2|^{1+\kappa_2} |r-p-q-c_3|^{2+\kappa_3}} \frac{1}{\la \beta +
(p-c)^2 \ra^{1-\e}} \\ \times \frac{1}{\la \a - p^2
\pm q^2 \pm (r-p-q)^2 \ra^{1-\e}} \leq C \sum_{j=1}^{\mu}
\frac{1}{|r-b_j|^{1+\lambda}}
\end{multline}
for every $c_1,c_2,c_3 \in \bR^3$. Here the $b_j$ are linear
combinations of $c_1,c_2,c_3$ and $\mu$ is a universal integer
constant. The bounds hold for all four possible choices of the two
signs.
\end{proposition}
{\it Remarks. } The bound (\ref{3d1}) is used if one of the
son-edges is a $(1+\kappa)$--edge and one of the other two
son-edges is a $(2+s+\kappa)$--edge, with $\kappa=0,\lambda$, or
$2\lambda$. The bound (\ref{3d2}), on the other hand, is used if
one of the son-edges is a $(1+s+\kappa)$--edge. The different
signs in the propagators are needed depending on the orientation
of the edge carrying the spherical denominator (in our notation
this is the edge with momentum $p$), and on which one of the two
terms on the r.h.s. of (\ref{eq:al2}) arising from the
$\a$-integration we are considering. Note that since one of the
son edges always has an opposite orientation with respect to the
other two, the sign combination $(--)$ in the propagators does not
arise in our applications. Although the estimates remain true also
in this case, we prove them only for the other combinations
$(++)$, $(+-)$ and $(--)$.

\begin{proof}
We prove the proposition in the case $c_1 = c_2 = c_3 = 0$: to
generalize the proof for $c_1,c_2,c_3 \neq 0$ one can proceed as
we did in Proposition \ref{prop:type2} (since we assume $c_1 = c_2
= c_3 =0$, in our proof we will only need one term on the r.h.s.
of (\ref{3c}), (\ref{3d1}) and (\ref{3d2}), that is we can take
$\mu =1$ and $b_1=0$; however, when $c_1,c_2,c_3$ do not vanish,
one need $\mu > 1$, as in Prop. \ref{prop:type2}). Without loss of
generality we can also assume that $\e >0$ (the l.h.s. of
(\ref{3c}), (\ref{3d1}) and (\ref{3d2}) is clearly increasing in
$\e$). We begin by proving (\ref{3c}) and (\ref{3d1}): to this end
we introduce the parameter $\theta$ which can assume the values
$1,2$ (to prove (\ref{3c}) we use $\theta=2$, to prove (\ref{3d1})
we use $\theta=1$). The possible combinations of the two signs
lead to the three contributions
\begin{equation}\label{eq:I-IV}
\begin{split}
\text{(I)} &:=\sup_{\a,\beta, c} \int \frac{\rd p \, \rd
q}{|p|^{\theta+\kappa_1} |q|^{2+\kappa_2} |r-p-q|^{2+\kappa_3}}
\frac{1}{\la \beta + (p-c)^2 \ra^{1-\e}} \frac{1}{\la \a -
\frac{1}{4} (p+r)^2 + (q+\frac{p-r}{2})^2 \ra^{1-\e}} \\
\text{(II)} &:= \sup_{\a,\beta, c} \int \frac{\rd p \, \rd
q}{|p|^{\theta + \kappa_1} |q|^{2+\kappa_2} |r-p-q|^{2+\kappa_3}}
\frac{1}{\la \beta + (p-c)^2\ra^{1-\e}}
\frac{1}{\la \a - (p -\frac{r}{2})^2 - q \cdot  (p-r) \ra^{1-\e}} \\
\text{(III)} &:= \sup_{\a,\beta, c} \int \frac{\rd p \, \rd
q}{|p|^{\theta+\kappa_1} |q|^{2+\kappa_2} |r-p-q|^{2+\kappa_3}}
\frac{1}{\la \beta + (p-c)^2\ra^{1-\e}} \frac{1}{\la \a -p\cdot r
+ q \cdot (p-r) \ra^{1-\e}}
\end{split}
\end{equation}
The first formula corresponds to the sign choice $(++)$, the
second and third ones are the choices $(+-$) and $(-+)$ (as
remarked above, we do not consider, in this proof, the case
$(--)$, because we do not need it in our applications). Recall
that $\theta =2$ is needed for the proof of (\ref{3c}) and $\theta
=1$ for the proof of (\ref{3d1}). To derive (\ref{eq:I-IV}), we
shifted the variable $\a$ by some number independent of $p$ and
$q$: this is clearly allowed, because we take the supremum over
$\a$.

We start by estimating the contribution (I). With $\kappa =
\kappa_1 + \kappa_2 + \kappa_3$, and using the assumption $\kappa
\leq \lambda$, we find, from (\ref{eq:weighted}),
\begin{equation}\label{eq:est0}
\begin{split}
\frac{1}{|p|^{\theta+\kappa_1} |q|^{2+\kappa_2}
|r-p-q|^{2+\kappa_3}} \lesssim \;
&\frac{1}{|p|^{\theta+\kappa}|r-p|^{2-2\lambda-\frac{\kappa}{2}+\e}}
\left( \frac{1}{|q|^{2+2\lambda+\frac{\kappa}{2}-\e}} +
\frac{1}{|r-p-q|^{2+2\lambda+\frac{\kappa}{2}-\e}} \right) \\ & +
\frac{1}{|p|^{\theta} |r-p|^{2-2\lambda+\frac{\kappa}{2}+\e}}
\left( \frac{1}{|q|^{2+2\lambda+\frac{\kappa}{2}-\e}} +
\frac{1}{|r-p-q|^{2+ 2\lambda + \frac{\kappa}{2}-\e}} \right) \, .
\end{split}
\end{equation}
This implies, with a simple shift of the $q$ variable, that
\begin{equation}\label{eq:I0}
\begin{split}
\text{(I)} \lesssim \; &\sup_{\a,\beta, c} \int \frac{\rd
p}{|p|^{\theta+\kappa} |r-p|^{2-2\lambda-\frac{\kappa}{2}+\e}}
\frac{1}{\la \beta +(p-c)^2\ra^{1-\e}} \\ &\hspace{2cm}\times \int
\frac{\rd q}{|q|^{2+2\lambda +\frac{\kappa}{2}-\e}} \frac{1}{\la
\a - \frac{1}{4} (p+r)^2 +
(q+\frac{p-r}{2})^2 \ra^{1-\e}} \\
&+\sup_{\a,\beta, c} \int \frac{\rd p}{|p|^{\theta}
|r-p|^{2-2\lambda+\frac{\kappa}{2}+\e}} \frac{1}{\la \beta
+(p-c)^2\ra^{1-\e}} \\ &\hspace{2cm}\times \int \frac{\rd
q}{|q|^{2+2\lambda +\frac{\kappa}{2}-\e}} \frac{1}{\la \a -
\frac{1}{4} (p+r)^2 + (q+\frac{p-r}{2})^2 \ra^{1-\e}}\,.
\end{split}
\end{equation}
To bound the $q$-integrals, we apply Lemma \ref{lm:sphere} with
$\gamma = 1-2\lambda -\frac{\kappa}{2} -\e$, $2\delta=
-2\lambda-\frac{\kappa}{2} + \e$ and $\eta=\e$; this is allowed,
because $1-2\lambda -\frac{\kappa}{2} -\e >0$ (since $\e <\lambda
/2$ and $\lambda < 1/5$), because $1-2\lambda-(\kappa/2)-\e < 1
-2\lambda -(\kappa/2) +\e$ (since we assumed $\e >0$), and because
$1-2\lambda - (\kappa/2) -\e < 1 + 2\lambda +(\kappa/2) -3\e$. We
obtain (using that $\kappa \leq \lambda <1/5$)
\begin{equation}\label{eq:I1}
\begin{split}
\text{(I)} \lesssim \; &\sup_{\a,\beta,c} \int \frac{\rd
p}{|p|^{\theta+\kappa} |r-p|^{3-4\lambda-\kappa}} \frac{1}{\la
\beta + (p-c)^2 \ra^{1-\e}} \frac{1}{\la \a - p\cdot r
\ra^{2\lambda + \frac{\kappa}{2} -2\eps}}
\\ &+\sup_{\a,\beta,c} \int \frac{\rd p}{|p|^{\theta}
|r-p|^{3-4\lambda}} \frac{1}{\la \beta + (p-c)^2 \ra^{1-\e}}
\frac{1}{\la \a - p\cdot r  \ra^{2\lambda + \frac{\kappa}{2}
-2\eps}} \, \\
\lesssim \; &\frac{1}{|r|^{\theta}} \sup_{ \a,\beta, c} \int
\frac{\rd p}{|p|^{3-4\lambda}} \frac{1}{\la \beta + (p- c)^2
\ra^{1-\e}} \frac{1}{\la \a - p\cdot r \ra^{2\lambda +
\frac{\kappa}{2} -2\eps}}
\end{split}
\end{equation}
where we used (\ref{eq:weighted}) and a simple shift of the
$p$-variable (and also of the variables $\a$ and $c$, over which
we take the supremum). To estimate this term we use a H\"older
inequality:
\begin{equation}\label{eq:3c11}
\begin{split}
\sup_{\a,\beta, c} \int &\frac{\rd p}{|p|^{3-4\lambda}}
\frac{1}{\la \beta + (p- c)^2 \ra^{1-\e}} \frac{1}{\la \a-
p\cdot r \ra^{2\lambda + \frac{\kappa}{2} - 2\eps}} \\
& \lesssim \sup_{\a, \beta,c} \, \left( \int \frac{\rd p}{|p|^{2+
\frac{1-5\lambda -(\kappa/2) +2\eps}{1-2\lambda -(\kappa/2)
+2\e}}} \frac{1}{\la \beta+ (p- c)^2 \ra}
\right)^{1-2\lambda - \frac{\kappa}{2} +2\e} \\
&\hspace{4cm}\times \left( \int \frac{\rd p}{|p|^{2 +
\frac{\lambda + (\kappa/2) -2\e}{2\lambda + (\kappa/2) -2\e}}}
\frac{1}{\la \a -  p\cdot r \ra} \right)^{2\lambda +
\frac{\kappa}{2} - 2\e}
\end{split}
\end{equation}
where we used that $(1-\e)/(1-2\lambda -(\kappa/2) +2\e) \geq 1$
(this follows from $\e < \lambda/2$) and that $\la \beta - (p-c)^2
\ra \geq 1$. The first integral is bounded uniformly in $\beta$
and $c$: this follows from Lemma \ref{lm:sphere} (with $\gamma
=0$), because $-1 < (1-5\lambda -(\kappa/2) +2\eps)/(1-2\lambda
-(\kappa/2) +2\e) < 1$ (this follows from $\kappa \leq \lambda
<1/5$). As for the second integral, since $0 < (\lambda +
(\kappa/2) -2\e)/(2\lambda + (\kappa/2) -2\e) <1$ (from $\e <
\lambda /2$), we can apply Lemma \ref{lm:plane}:
\begin{equation}\label{eq:3c112}
\sup_{\wt\a,\beta,c} \int \frac{\rd p}{|p|^{3-4\lambda}}
\frac{1}{\la \beta + (p-\tilde c)^2 \ra^{1-\e}} \frac{1}{\la
\wt\a- 2 p\cdot r \ra^{2\lambda + \frac{\kappa}{2} - 2\eps}}
\lesssim \frac{1}{|r|^{\lambda}} \,.
\end{equation}
The powers of $|p|$ in (\ref{eq:3c11}) were chosen exactly in order to get
this decay. With (\ref{eq:I1}), we conclude that (I) $\lesssim
|r|^{-\theta-\lambda}$.

\medskip

Next we consider the term (II) in (\ref{eq:I-IV}). Instead of
(\ref{eq:est0}) we use here the similar bound
\begin{equation}\label{eq:est1}
\begin{split}
\frac{1}{|p|^{\theta+\kappa_1} |q|^{2+\kappa_2}
|r-p-q|^{2+\kappa_3}} \lesssim \;
&\frac{1}{|p|^{\theta+\kappa}|r-p|^{2-2\lambda-\frac{\kappa}{2}}}
\left( \frac{1}{|q|^{2+2\lambda+\frac{\kappa}{2}}} +
\frac{1}{|r-p-q|^{2+2\lambda+\frac{\kappa}{2}}} \right) \\ & +
\frac{1}{|p|^{\theta} |r-p|^{2-2\lambda+\frac{\kappa}{2}}} \left(
\frac{1}{|q|^{2+2\lambda+\frac{\kappa}{2}}} + \frac{1}{|r-p-q|^{2+
2\lambda + \frac{\kappa}{2}}} \right) \, .
\end{split}
\end{equation}
With a shift of the $q$ variable, we obtain
\begin{equation}\label{eq:II0}
\begin{split}
\text{(II)} \lesssim &\;\sup_{\a,\beta,c} \int \frac{\rd
p}{|p|^{\theta+\kappa} |r-p|^{2-2\lambda - \frac{\kappa}{2}}}
\;\frac{1}{\la \beta + (p-c)^2 \ra^{1-\e}} \int \frac{\rd
q}{|q|^{2+2\lambda +\frac{\kappa}{2}}}
\; \frac{1}{\la \a - (p-\frac{r}{2})^2 - q \cdot (p-r) \ra^{1-\e}} \\
&+\sup_{\a,\beta,c} \int \frac{\rd p}{|p|^{\theta+\kappa}
|r-p|^{2-2\lambda - \frac{\kappa}{2}}} \;\frac{1}{\la \beta +
(p-c)^2 \ra^{1-\e}} \int \frac{\rd q}{|q|^{2+2\lambda
+\frac{\kappa}{2}}} \; \frac{1}{\la \a -
p\cdot r - q \cdot (p-r) \ra^{1-\e}}\\
&+\sup_{\a,\beta,c} \int \frac{\rd p}{|p|^{\theta}
|r-p|^{2-2\lambda + \frac{\kappa}{2}}} \;\frac{1}{\la \beta +
(p-c)^2 \ra^{1-\e}} \int \frac{\rd q}{|q|^{2+2\lambda
+\frac{\kappa}{2}}}
\; \frac{1}{\la \a - (p-\frac{r}{2})^2 - q \cdot (p-r) \ra^{1-\e}} \\
&+\sup_{\a,\beta,c} \int \frac{\rd p}{|p|^{\theta}
|r-p|^{2-2\lambda + \frac{\kappa}{2}}} \;\frac{1}{\la \beta +
(p-c)^2 \ra^{1-\e}} \int\frac{\rd q}{|q|^{2+2\lambda
+\frac{\kappa}{2}}} \; \frac{1}{\la  \a - p\cdot r - q \cdot (p-r)
\ra^{1-\e}}\,.
\end{split}
\end{equation}
Applying Lemma \ref{lm:plane} (with $\eta=\eps$) to bound the
$q$-integral in the four terms, we find
\begin{equation}
\begin{split}
\text{(II)} \lesssim &\;\sup_{\a,\beta,c} \int \frac{\rd
p}{|p|^{\theta+\kappa} |r-p|^{3-4\lambda -\kappa}} \; \frac{1}{\la
\beta + (p-c)^2 \ra^{1-\e}} \frac{1}{\la  \a -
(p-\frac{r}{2})^2 \ra^{2\lambda +\frac{\kappa}{2}-2\e}}\\
&+\;\sup_{\a,\beta,c} \int \frac{\rd p}{|p|^{\theta+\kappa}
|r-p|^{3- 4\lambda -\kappa}} \; \frac{1}{\la \beta + (p-c)^2
\ra^{1-\e}} \frac{1}{\la \a - p\cdot
r \ra^{2\lambda +\frac{\kappa}{2}-2\e}}\\
&+\sup_{\a,\beta,c} \int \frac{\rd p}{|p|^{\theta}
|r-p|^{3-4\lambda}} \, \frac{1}{\la \beta + (p-c)^2 \ra^{1-\e}}
\frac{1}{\la \a - (p-\frac{r}{2})^2 \ra^{2\lambda
+\frac{\kappa}{2}-2\e}}\\
&+\sup_{\a,\beta,c} \int \frac{\rd p}{|p|^{\theta}
|r-p|^{3-4\lambda}} \, \frac{1}{\la \beta + (p-c)^2 \ra^{1-\e}}
\frac{1}{\la \a - p\cdot r \ra^{2\lambda
+\frac{\kappa}{2}-2\e}}\,.
\end{split}
\end{equation}
Using that $\kappa \leq \lambda <1/5$, it follows from
(\ref{eq:weighted}) and from a shift of the $p$ variable (and of
the $\a$ and $c$ variable as well), that
\begin{equation}\label{eq:II2}
\begin{split}
\text{(II)} \lesssim \; & \frac{1}{|r|^{\theta}} \;
\sup_{\a,\beta, c} \int \frac{\rd p}{|p|^{3-4\lambda}} \,
\frac{1}{\la \beta + (p- c)^2 \ra^{1-\e}} \frac{1}{\la \a -
(p-\frac{r}{2})^2 \ra^{2\lambda
+\frac{\kappa}{2}-2\e}}\\
&+\frac{1}{|r|^{\theta}} \; \sup_{\a,\beta, c} \int \frac{\rd
p}{|p|^{3-4\lambda}} \, \frac{1}{\la \beta + (p- c)^2 \ra^{1-\e}}
\frac{1}{\la \a - p\cdot r \ra^{2\lambda
+\frac{\kappa}{2}-2\e}}\,.
\end{split}
\end{equation}
The second term is identical to the r.h.s. of (\ref{eq:I1}), and
can be estimated in the same way. As for the first term on the
r.h.s. of the last equation, we apply a H\"older inequality (we
use here the same exponents as in (\ref{eq:3c11}) but here we
divide the powers of $|p|$ in a different way):
\begin{equation}\label{eq:3c12}
\begin{split}
\sup_{\a,\beta,c} \int &\frac{\rd p}{|p|^{3-4\lambda}}
\frac{1}{\la \beta + (p- c)^2 \ra^{1-\e}} \frac{1}{\la \a-
(p-\frac{ r}{2})^2 \ra^{2\lambda + \frac{\kappa}{2} - 2\eps}} \\
& \lesssim \sup_{\a, \beta,c} \, \left( \int \frac{\rd p}{|p|^{2+
\frac{1-4\lambda }{1-2\lambda -\frac{\kappa}{2} +2\e}}}
\frac{1}{\la \beta + (p- c)^2 \ra} \right)^{1-2\lambda -
\frac{\kappa}{2} +2\e}  \left( \int \frac{\rd p}{|p|^2}
\frac{1}{\la \a - (p-\frac{ r}{2})^2\ra } \right)^{2\lambda +
\frac{\kappa}{2} - 2\e} \, .
\end{split}
\end{equation}
Since $-1 < (1-4\lambda )/(1-2\lambda -(\kappa/2) +2\e) < 1$ (as
follows from $\lambda <1/5$), the first integral is bounded
uniformly in $\beta$ and $c$ by Lemma \ref{lm:sphere}. To estimate
the second integral we use again Lemma \ref{lm:sphere}, with
$\gamma = \lambda / (2\lambda + (\kappa/2) -2\eps)$ (this is
allowed because $\lambda / (2\lambda + (\kappa/2) -2\eps) <1$ for
$\eps < \lambda /2$). We conclude that
\begin{equation}
\sup_{\a,\beta, c} \int \frac{\rd p}{|p|^{3-4\lambda}}
\frac{1}{\la \beta - (p- c)^2 \ra^{1-\e}} \frac{1}{\la \a-
(p-\frac{ r}{2})^2 \ra^{2\lambda + \frac{\kappa}{2} - 2\eps}}
\lesssim \frac{1}{\la r \ra^{\lambda}} \leq
\frac{1}{|r|^{\lambda}} \, .
\end{equation}
{F}rom (\ref{eq:II2}) it follows that (II) $\lesssim
|r|^{-\theta-\lambda}$.

\medskip

Next we consider the term (III) in (\ref{eq:I-IV}). With
(\ref{eq:est1}), we find
\begin{equation}
\begin{split}
\text{(III)} \lesssim &\;\sup_{\a,\beta,c} \int \frac{\rd
p}{|p|^{\theta+\kappa} |r-p|^{2-2\lambda - \frac{\kappa}{2}}}
\;\frac{1}{\la \beta + (p-c)^2 \ra^{1-\e}} \int \frac{\rd
q}{|q|^{2+2\lambda +\frac{\kappa}{2}}}
\; \frac{1}{\la \a - p\cdot r+ q\cdot (p-r) \ra^{1-\e}} \\
&+\sup_{\a,\beta,c} \int \frac{\rd p}{|p|^{\theta+\kappa}
|r-p|^{2-2\lambda - \frac{\kappa}{2}}} \;\frac{1}{\la \beta +
(p-c)^2 \ra^{1-\e}} \int \frac{\rd q}{|q|^{2+2\lambda
+\frac{\kappa}{2}}} \; \frac{1}{\la \a -
(p-\frac{r}{2})^2 + q \cdot (p-r) \ra^{1-\e}}\\
&+\sup_{\a,\beta,c} \int \frac{\rd p}{|p|^{\theta}
|r-p|^{2-2\lambda + \frac{\kappa}{2}}} \;\frac{1}{\la \beta +
(p-c)^2 \ra^{1-\e}} \int \frac{\rd q}{|q|^{2+2\lambda
+\frac{\kappa}{2}}}
\; \frac{1}{\la \a - p\cdot r + q \cdot (p-r) \ra^{1-\e}} \\
&+\sup_{\a,\beta,c} \int \frac{\rd p}{|p|^{\theta}
|r-p|^{2-2\lambda + \frac{\kappa}{2}}} \;\frac{1}{\la \beta +
(p-c)^2 \ra^{1-\e}} \int\frac{\rd q}{|q|^{2+2\lambda
+\frac{\kappa}{2}}} \; \frac{1}{\la \a - (p-\frac{r}{2})^2 + q
\cdot (p-r) \ra^{1-\e}}\,.
\end{split}
\end{equation}
These terms can be estimated as we did with the four terms on the
r.h.s. of (\ref{eq:II0}) (the different sign in front of the
factor $q\cdot (p-r)$ plays no role in our bounds).

\medskip

As for (\ref{3d2}), similarly to (\ref{eq:I-IV}) we find the three
contributions
\begin{equation}\label{eq:I'-IV'}
\begin{split}
\text{(I')} &:=\sup_{\a,\beta, c} \int \frac{\rd p \, \rd
q}{|p|^{2+\kappa_1} |q|^{1+\kappa_2} |r-p-q|^{2+\kappa_3}}
\frac{1}{\la \beta + (p-c)^2 \ra^{1-\e}} \frac{1}{\la \a -
\frac{1}{4} (p+r)^2 + (q+\frac{p-r}{2})^2 \ra^{1-\e}} \\
\text{(II')} &:= \sup_{\a,\beta, c} \int \frac{\rd p \, \rd
q}{|p|^{2 + \kappa_1} |q|^{1+\kappa_2} |r-p-q|^{2+\kappa_3}}
\frac{1}{\la \beta + (p-c)^2\ra^{1-\e}}
\frac{1}{\la \a - (p -\frac{r}{2})^2 - q \cdot  (p-r) \ra^{1-\e}} \\
\text{(III')} &:= \sup_{\a,\beta, c} \int \frac{\rd p \, \rd
q}{|p|^{2+\kappa_1} |q|^{1+\kappa_2} |r-p-q|^{2+\kappa_3}}
\frac{1}{\la \beta + (p-c)^2\ra^{1-\e}} \frac{1}{\la \a -p\cdot r
+ q \cdot (p-r) \ra^{1-\e}} \,.
\end{split}
\end{equation}
The analysis of (I')--(III') is then very similar to the one of
(I)--(III): the only difference is that instead of using
(\ref{eq:est0}) and (\ref{eq:est1}), we employ
\begin{equation}\label{eq:est2}
\begin{split}
\frac{1}{|p|^{2+\kappa_1} |q|^{1+\kappa_2} |r-p-q|^{2+\kappa_3}}
\lesssim \;
&\frac{1}{|p|^{2+\kappa}|r-p|^{1-2\lambda-\frac{\kappa}{2}+\e}}
\left( \frac{1}{|q|^{2 + 2\lambda + \frac{\kappa}{2}-\e}} +
\frac{1}{|r-p-q|^{2+2\lambda+\frac{\kappa}{2}-\e}} \right) \\ & +
\frac{1}{|p|^{2} |r-p|^{1-2\lambda+\frac{\kappa}{2}+\e}} \left(
\frac{1}{|q|^{2+2\lambda+\frac{\kappa}{2}-\e}} +
\frac{1}{|r-p-q|^{2+ 2\lambda + \frac{\kappa}{2}-\e}} \right) \,
\end{split}
\end{equation}
to bound the term (I'), and
\begin{equation}\label{eq:est3}
\begin{split}
\frac{1}{|p|^{2+\kappa_1} |q|^{2+\kappa_2} |r-p-q|^{2+\kappa_3}}
\lesssim \;
&\frac{1}{|p|^{2+\kappa}|r-p|^{1-2\lambda-\frac{\kappa}{2}}}
\left( \frac{1}{|q|^{2+2\lambda+\frac{\kappa}{2}}} +
\frac{1}{|r-p-q|^{2+2\lambda+\frac{\kappa}{2}}} \right) \\ & +
\frac{1}{|p|^{2} |r-p|^{1-2\lambda+\frac{\kappa}{2}}} \left(
\frac{1}{|q|^{2+2\lambda+\frac{\kappa}{2}}} + \frac{1}{|r-p-q|^{2+
2\lambda + \frac{\kappa}{2}}} \right) \,
\end{split}
\end{equation}
to bound (II') and (III') (we use (\ref{eq:est2}) when the
propagator is quadratic in $q$, (\ref{eq:est3}) when it is
linear).
\end{proof}

\appendix

\section{Some Technical Bounds}\label{app}
\setcounter{equation}{0}

In this Appendix we collect some simple information that are used
throughout the paper.

\begin{lemma}\label{lm:scatt}
Let $a_N$ be the scattering length of $\frac{1}{N} V_N$ with $V_N$
given in (\ref{def:VN}) and let $b_0:= \int V_N = \int V$. We
assume additionally that $V$ is radially symmetric. Then for any
$0<\beta<1$ \be
   \lim_{N\to\infty} N a_N = \frac{b_0}{8\pi}  \; .
\label{Nalim}
\ee
\end{lemma}

\begin{proof}
Let $R$ be the radius of the support of $V$, i.e.
$V(x) =0$ for $|x| \geq R$.
 An upper bound for $a_N$ can be obtained by the inequality (see,
 for example \cite{SR})
\begin{equation}\label{eq:scatt0}
8\pi a_N \leq  \int \frac{1}{N} V_N= \frac{b_0}{N}\,.
\end{equation}
To derive a lower bound for $a_N$, we recall that the scattering
length can be computed through the integral
\begin{equation}\label{eq:scatt1}
8\pi a_N = \int \frac{1}{N} V_N f \;,
\end{equation}
where $f(x)$ is the radial symmetric solution of the zero energy
equation
\[ \left( -\Delta + \frac{1}{2N} V_N
\right) f = 0 \] with $f(x) \to 1$ as $|x| \to \infty$. It is
easy to show that
\begin{equation}\label{eq:scatt3}
f(x) \geq \left\{ \begin{array}{ll}  1-\frac{a_N}{|x|} \quad
&\text{for } |x| \geq a_N  \\
0 \quad &\text{for } |x| \leq a_N
\end{array} \right. \, .
\end{equation}
This can be proven as follows. We write $f(x) = g (|x|)/|x|$. Then
$g$ satisfies the one-dimensional equation \[ -g'' (r)
+\frac{N^{3\beta}}{2N} V(N^{\beta}r) g(r) =0 \, .\] Simple
arguments show that $g(r) \geq 0$ for all $r \geq 0$. Moreover,
for $r \geq R N^{-\beta}$, we have $g(r) = r -a_N$ (it is easy to
see that this definition of the scattering length $a$ agrees with
(\ref{eq:scatt1})). For $r <R N^{-\beta}$ we have
\begin{equation}
\begin{split}
g(RN^{-\beta}) - g(r) &= \int_r^{RN^{-\beta}} \rd s \, g' (s) \\
&=\int_r^{RN^{-\beta}} \rd s \, \left( g' (RN^{-\beta}) -
\int_s^{RN^{-\beta}} \rd \tau \, g'' (\tau) \right)\\
&=(RN^{-\beta} -r) - \int_r^{RN^{-\beta}} \rd s \,
\int_s^{RN^{-\beta}} \rd \tau \, \frac{N^{3\beta -1}}{2} \,
V(N^{\beta} \tau) g (\tau),
\end{split}
\end{equation}
because $g' (RN^{-\beta}) =1$. Since $g(RN^{-\beta}) = RN^{-\beta}
- a_N$, we obtain
\begin{equation}
r-g(r) = a_N -\int_r^{RN^{-\beta}} \rd s \, \int_s^{RN^{-\beta}}
\rd \tau \, N^{3\beta -1} V(N^{\beta} \tau) g (\tau)\,.
\end{equation}
{F}rom the non-negativity of the potential and from $g \geq 0$, it
follows that $r-g(r) \leq a_N$ for all $r < RN^{-\beta}$, and
hence $f(x) = g(|x|)/|x| \geq 1 - a_N/|x|$. Eq. (\ref{eq:scatt3})
now follows because $f(x) \geq 0$ for all $x \in \bR^3$.

Inserting (\ref{eq:scatt3}) into (\ref{eq:scatt1}), and using the
bound (\ref{eq:scatt0}), we conclude that
\begin{equation}
\begin{split}
8\pi N a_N &\geq \int_{|x| \geq a_N} \rd x \, N^{3\beta}
V(N^{\beta} x) \left( 1 - \frac{a_N}{|x|} \right) \\ & \geq
\int_{|x| \geq b_0/8 \pi N} \rd x \, N^{3\beta} V(N^{\beta} x) -
\frac{b_0}{8\pi N} \int
\frac{\rd x }{|x|} \, N^{3\beta} V(N^{\beta}x) \\
& \geq b_0 - \int_{|x| \leq b_0/8\pi N} \rd x \, N^{3\beta}
V(N^{\beta} x) - \frac{b_0}{8\pi} N^{\beta- 1} \int \frac{\rd x
}{|x|} \, V(x) \\
& \geq b_0 - \frac{4\pi}{3} \left(\frac{b_0 N^{\beta-1}}{8\pi
}\right)^3 \| V \|_{\infty} - \frac{b_0}{8\pi} N^{\beta- 1} \| V
\|_{\infty} \int_{|x| \leq R} \frac{\rd x }{|x|}\,.
\end{split}
\end{equation}
This, together with (\ref{eq:scatt0}), implies that
\begin{equation}
b_0 - C_1 N^{3\beta -3} -C_2 N^{\beta-1} \leq 8\pi N a_N \leq
b_0\, ,
\end{equation}
for two $N$-independent constants $C_1, C_2$. Hence, for $0 <
\beta <1$, we obtain (\ref{Nalim}).
\end{proof}

\bigskip

In the next lemma we prove that solutions of the  nonlinear
Schr\"odinger equation which are in the space $H^1 (\bR^3)$ at
time $t=0$, have $H^1$ norm uniformly bounded in time.

\begin{lemma}\label{lm:GPEbound}
Suppose $\ph \in H^1 (\bR^3)$, and let $\ph_t$ be the solution of
the  nonlinear Schr\"odinger equation
\begin{equation}\label{eq:GPE2} i\partial_t \ph_t = -\Delta \ph_t
+ b_0 |\ph_t|^2 \ph_t \,
\end{equation} with $b_0 >0$.
Then \begin{equation}\label{eq:GPbound}
 (\ph_t , (1- \Delta) \ph_t
) = \int \rd x \, \left( |\ph_t (x)|^2 + |\nabla \ph_t (x)|^2
\right) \leq C
\end{equation} for all $t \in \bR$. Hence, with $\gamma_t^{(k)} (\bx_k
;\bx'_k) = \prod_{j=1}^k \ph_t (x_j) \overline{\ph_t} (x'_j)$, we
have \begin{equation}\label{eq:GPHbound}
 \tr \; |S_1 \dots S_k
\gamma^{(k)}_t S_k \dots S_1| \leq C^k \end{equation} The constant
$C$ only depends on $b_0$ and on the $H^1$-norm of $\ph$.
\end{lemma}
\begin{proof}
The $L^2$-norm of $\ph_t$ is conserved in time. Also the energy \[
E(\ph) = \int \rd x \, |\nabla \ph (x)|^2 + \frac{b_0}{2} \int \rd
x \, |\ph (x)|^4 \] is conserved. By the Sobolev inequality, we
have
\[ \int \rd x |\nabla \ph (x)|^2
\leq  E(\ph) \leq C \| \ph \|_{H^1}^4 \, ,\] for a constant $C$
only depending on $b_0$. Hence
\[ \int |\nabla \ph_t (x)|^2 \leq E (\ph_t) = E (\ph) \leq C \|
\ph \|_{H^1}^4 .\]
\end{proof}

The following lemma is useful to bound the pair interaction $V_N
(x)$ in terms of the kinetic energy.

\begin{lemma}\label{lm:sob}
Let $V \in L^1 (\bR^3) \cap L^{3/2} (\bR^3)$. Put $V_N (x) =
N^{3\beta} V(N^{\beta} x)$. Then
\[ V_N (x_1 -x_2) \leq \mbox{(const.)}
\| V \|_{L^1} \, (1 -\Delta_1) (1-\Delta_2)
\,
\] and
\[ V_N (x_1 -x_2) \leq \mbox{(const.)}
 N^{\beta} \| V \|_{L^{3/2}} (1-\Delta_1) \, \]
hold with  universal constants.
\end{lemma}
\begin{proof}
For a proof of these results, see Lemma 5.2 in \cite{ESY}.
\end{proof}

\medskip

Finally, we give a proof of Lemma \ref{lm:sobsob} that is a slight
modification of the proof of Proposition 8.1 in \cite{ESY}.

\begin{proof}[Proof of Lemma \ref{lm:sobsob}]
By the positivity of
$\gamma^{(k+1)}$, it is enough to prove (\ref{eq:gammaintbound})
for the special case $\gamma^{(k+1)} (\bx_{k+1} , \bx'_{k+1}) = f
(\bx_{k+1}) \overline{f}(\bx'_{k+1})$. We can then bound the
l.h.s. of (\ref{eq:gammaintbound}) by the sum
\begin{equation}\label{eq:delta0}
\begin{split}
\Big| &\int \rd {\bf x}_{k+1} \rd {\bf x}_{k+1}' \, J^{(k)} (\bx_k
;\bx'_k) \, \left(\delta_{\alpha_1} (x'_{k+1} - x_{k+1}) - \delta
(x'_{k+1} - x_{k+1}) \right) \delta_{\alpha_2} (x_j -x_{k+1}) \, f
(\bx_{k+1}) \overline{f} (\bx'_{k+1}) \Big|
\\ &+ \Big| \int \rd {\bf x}_{k+1} \rd {\bf x}_{k+1}' \, J^{(k)}
(\bx_k ; \bx'_k) \, \delta (x'_{k+1} - x_{k+1}) \left(
\delta_{\alpha_2} (x_j -x_{k+1}) - \delta (x_j - x_{k+1})\right)\,
f (\bx_{k+1}) \overline{f} (\bx'_{k+1}) \Big| \, .
\end{split}
\end{equation}
The first term can be bounded by
\begin{equation}\label{eq:delta1}
\begin{split}
\Big| \int \rd {\bf x}_{k+1} &\rd {\bf x}_{k+1}' \, J^{(k)} (\bx_k
; \bx'_k) \big(\delta_{\alpha_1} (x'_{k+1} - x_{k+1}) - \delta
(x'_{k+1} - x_{k+1}) \big) \delta_{\alpha_2} (x_j -x_{k+1}) \, f
(\bx_{k+1}) \overline{f} (\bx'_{k+1}) \Big|
\\ \leq  \int &\rd \bx_{k+1} \rd \bx'_k \,
|J^{(k)} (\bx_k ; \bx'_k) | \, \delta_{\alpha_2} (x_j -x_{k+1}) \,
|f (\bx_{k+1})| \\ &\hspace{2cm} \times  \Big|
 \int \rd x'_{k+1} \, \delta_{\alpha_1} (x_{k+1} - x'_{k+1})
\big[ f (\bx'_k , x_{k+1})- f
(\bx'_k , x'_{k+1})\big] \Big| \, .
\end{split}
\end{equation}
We use the estimate $\delta_{\alpha_1}(x) \leq \frac{C}{|B|} \cdot
{\bf 1}_B(x)$ where $B: = \{ x\; : \; |x|\leq \alpha_1\}$. A
standard Poincar\'e-type inequality (see, e.g. Lemma 7.16 in
\cite{GT}) yields that
 \be\label{pineq}
\Big | \int \rd x'_{k+1} \, \delta_{\alpha_1} (x_{k+1} - x'_{k+1})
\, \big[ f (\bx'_k , x_{k+1}) - f (\bx'_k, x'_{k+1}) \big] \Big|
\le C \int_{|y| \le \alpha_1}
 \frac {|\nabla_{k+1} f (\bx'_k , x_{k+1} + y)|} { |y|^2  } \; \rd y
\ee for any $\bx'_k$ and $x_{k+1}$.
Inserting this inequality on
the r.h.s. of (\ref{eq:delta1}) and applying a Schwarz inequality
we get
\begin{equation}
\begin{split}
\Big| \int \rd &{\bf x}_{k+1} \rd {\bf x}_{k+1}' \, J^{(k)} (\bx_k
; \bx'_k) \, \left(\delta_{\alpha_1} (x'_{k+1} - x_{k+1}) - \delta
(x'_{k+1} - x_{k+1}) \right) \delta_{\alpha_2}
(x_j -x_{k+1}) \, f (\bx_{k+1}) \overline{f} (\bx'_{k+1}) \Big| \\
\leq \; & C \int \rd \bx_{k+1} \rd \bx'_k
 \rd y \, \frac{{\bf 1} (|y| \leq \alpha_1)}{|y|^2} \,
\delta_{\alpha_2} (x_j -x_{k+1}) |J^{(k)} (\bx_k ; \bx'_k) | \,
\\ &\hspace{2cm} \times \left( |f (\bx_k
,x_{k+1})|^2 + |\nabla_{k+1} f (\bx'_k , x_{k+1} +y)|^2 \right) \\
\leq \; &C  \, \alpha_1  \left( \sup_{\bx_k} \int \rd \bx'_{k}\,
|J^{(k)} (\bx_k ; \bx'_k)|\right) \; \int \rd \bx_{k+1} \,
\delta_{\alpha_2} (x_j -x_{k+1}) |f (\bx_k ,x_{k+1})|^2
\\ & + C \left( \sup_{\bx'_k,x_{k+1}} \int \rd \bx_k \, \delta_{\alpha_2}
(x_j -x_{k+1}) |J^{(k)} (\bx_k ; \bx'_k)| \right) \\
&\hspace{2cm} \times \int \rd \bx'_k \rd x_{k+1}  \rd y \,
\frac{{\bf 1} (|y| \leq \alpha_1)}{|y|^2} \, |\nabla_{k+1} f (\bx'_k
, x_{k+1} +y)|^2\,.
\end{split}
\label{A7}
\end{equation}
 In the first term we apply Lemma \ref{lm:sob} and in the
second term we shift the $x_{k+1}$ variable, and then we compute
the $y$-integral. Moreover we use that
\begin{equation}
\sup_{\bx_k} \int \rd \bx'_{k}\, |J^{(k)} (\bx_k ; \bx'_k)| \leq
C^k \, \tri J^{(k)} \tri_j
\end{equation}
and
\begin{equation}
\sup_{\bx'_k,x_{k+1}} \int \rd \bx_{k} \, \delta_{\alpha_2} (x_j
-x_{k+1}) \, |J^{(k)} (\bx_k ; \bx'_k)| \leq C^k  \tri J^{(k)}
\tri_j
\end{equation}
for a universal constant $C$ (recall the definition of the norm
$\tri J^{(k)} \tri_j$ from (\ref{eq:Jnorm})). Thus, we find
\begin{equation*}
\begin{split}
\Big| \int \rd &{\bf x}_{k+1} \rd {\bf x}_{k+1}' \, J^{(k)} (\bx_k
; \bx'_k) \, \left(\delta_{\alpha_1} (x'_{k+1} - x_{k+1}) - \delta
(x'_{k+1} - x_{k+1}) \right) \delta_{\alpha_2}
(x_j -x_{k+1}) \, f (\bx_{k+1}) \overline{f} (\bx'_{k+1}) \Big| \\
\leq \; & C^k \alpha_1 \tri J^{(k)} \tri_j \, \tr \; (1- \Delta_j)
(1-\Delta_{k+1}) \gamma^{(k+1)}\,.
\end{split}
\end{equation*}
In order to control the second term on the r.h.s. of
(\ref{eq:delta0}), we use that
\begin{equation}
\begin{split}
\int \rd {\bf x}_{k+1} &\rd {\bf x}_{k+1}' \, J^{(k)} (\bx_k ;
\bx'_k)  \delta (x'_{k+1} - x_{k+1}) \left( \delta_{\alpha_2} (x_j
-x_{k+1}) - \delta (x_j - x_{k+1})\right)\, f (\bx_{k+1})
\overline{f} (\bx'_{k+1})   \\ &= \int \rd \bx_k \rd \bx'_k \rd
x_{k+1}  \, \left( \delta_{\alpha_2} (x_j -x_{k+1}) - \delta (x_j
- x_{k+1})\right)\, J^{(k)} (\bx_k ; \bx'_k)\, f (\bx_{k},
x_{k+1}) \overline{f} (\bx'_{k}, x_{k+1}) \\ &= - \int \rd x_1 \dots
\rd x_{j-1} \rd x_{j+1} \dots \rd x_{k+1} \, \rd \bx'_k \,
\overline{f} (\bx'_{k}, x_{k+1}) \\ &\hspace{1cm} \times \left(
J^{(k)} (\hat{\bx}_k ; \bx'_k) \, f (\hat{\bx}_{k}, x_{k+1}) -
\int \rd x_j \, \delta_{\alpha_2} (x_j -x_{k+1}) J^{(k)} (\bx_k ;
\bx'_k) f(\bx_k,x_{k+1})
 \right) \; ,
\end{split}
\end{equation}
where we introduced the notation $\hat{\bx}_k = (x_1, \dots
,x_{j-1},x_{k+1},x_{j+1}, \dots ,x_k)$ (that is $\hat{\bx}_k$ is
the same as $\bx_k$, but with $x_j$ replaced by $x_{k+1}$). Using
again the generalization of the Poincar\'e inequality that led to
\eqref{pineq}, we obtain
\begin{equation}
\begin{split}
\Big| \int &\rd {\bf x}_{k+1} \rd {\bf x}_{k+1}' \, J^{(k)} (\bx_k
; \bx'_k)  \delta (x'_{k+1} - x_{k+1}) \left( \delta_{\alpha_2}
(x_j -x_{k+1}) - \delta (x_j - x_{k+1})\right)\, f (\bx_{k+1})
\overline{f} (\bx'_{k+1}) \Big|  \\ \leq \; &C \int \rd \bx_k \rd
\bx'_k \rd x_{k+1}  \frac{{\bf 1} (|x_j - x_{k+1}|\leq
\alpha_2)}{|x_j - x_{k+1}|^2} \, \Big|\nabla_j \big[ J^{(k)}
(\bx_k ; \bx'_k) \, f (\bx_{k}, x_{k+1}) \big]\Big|\, |f
(\bx'_{k}, x_{k+1})|
\\ \leq \; &C \int \rd \bx_k \rd \bx'_k \rd x_{k+1}
\frac{{\bf 1} (|x_j - x_{k+1}|\leq \alpha_2)}{|x_j - x_{k+1}|^2} \,
\\ & \times \left(|\nabla_j J^{(k)} (\bx_k ; \bx'_k)|  \, |f (\bx_{k},
x_{k+1})|\, |f (\bx'_{k}, x_{k+1})| + |J^{(k)} (\bx_k; \bx'_k)| \,
|\nabla_j f (\bx_k ,x_{k+1})| \, |f(\bx'_k,x_{k+1})| \right)
\\ \leq \; &C\int \rd \bx_k \rd \bx'_k \rd x_{k+1}
\frac{{\bf 1} (|x_j - x_{k+1}|\leq \alpha_2)}{|x_j - x_{k+1}|^2}
|\nabla_j J^{(k)} (\bx_k ; \bx'_k)|  \, \left( \kappa \, |f
(\bx_{k},
x_{k+1}))|^2 + \kappa^{-1} |f (\bx'_{k}, x_{k+1})|^2 \right) \\
&+C \int \rd \bx_k \rd \bx'_k \rd x_{k+1} \frac{{\bf 1} (|x_j -
x_{k+1}|\leq \alpha_2)}{|x_j - x_{k+1}|^2} |J^{(k)} (\bx_k;
\bx'_k)| \left( \kappa\, |\nabla_j f (\bx_k ,x_{k+1})|^2 +
\kappa^{-1}  |f(\bx'_k,x_{k+1})|^2 \right).
\end{split}
\end{equation}
In the terms proportional to $\kappa$ we drop the restriction
${\bf 1} (|x_j - x_{k+1}|\leq \alpha_2)$ and we apply the Hardy's
inequality to the $x_{k+1}$-integration. In the terms containing
$\kappa^{-1}$, on the other hand, we perform  the $x_j$
integration (after estimating $|J^{(k)}|$ and $|\nabla_j J^{(k)}|$
by their supremum).  We get
\begin{equation*}
\begin{split}
\Big| \int &\rd {\bf x}_{k+1} \rd {\bf x}_{k+1}' \, J^{(k)} (\bx_k
; \bx'_k)  \delta (x'_{k+1} - x_{k+1}) \left( \delta_{\alpha_2}
(x_j -x_{k+1}) - \delta (x_j - x_{k+1})\right)\, f (\bx_{k+1})
\overline{f} (\bx'_{k+1}) \Big|  \\ \leq \; & C \kappa \left(
\sup_{\bx_k} \int \rd \bx'_k (|J^{(k)} (\bx_k;\bx'_k)| + |\nabla_j
J^{(k)} (\bx_k ; \bx'_k)|) \right) \, \int \rd \bx_{k+1} \,
|(1-\Delta_{k+1})^{1/2} (1-\Delta_j)^{1/2} f (\bx_{k+1})|^2 \\
&+ C \kappa^{-1} \alpha_2 \left( \sup_{\bx'_k, x_j} \int \rd
x_1\dots \rd x_{j-1} \rd x_{j+1} \dots \rd x_{k} \, (|J^{(k)}
(\bx_k ; \bx'_k)| + |\nabla_j J^{(k)} (\bx_k ; \bx'_k)|) \right)
\\ &\hspace{1cm} \times \int \rd \bx'_k \rd x_{k+1} \, |f(\bx'_k,x_{k+1})|^2
\\ \leq \; & C^k (\kappa + \alpha_2 \kappa^{-1}) \tri J^{(k)} \tri_j \, \tr \;
(1-\Delta_j) (1-\Delta_{k+1}) \gamma^{(k+1)}.
\end{split}
\end{equation*}
Choosing $\kappa = \alpha_2^{1/2}$, we find
(\ref{eq:gammaintbound}).
\end{proof}

\end{document}